\documentclass[twocolumn, preprintnumbers ,amsmath, amssymb, aps,longbibliography]{revtex4}
\usepackage{color}
\usepackage{amsmath}
\usepackage{amssymb}  % More symbols
\usepackage{bbold}
\usepackage{float}
\usepackage{pifont}   % Ding symbols
\usepackage{graphicx} % Include figure files
\usepackage{dcolumn}  % Align table columns on decimal point
\usepackage{bm}       % bold math
\usepackage{multirow} % Table functions
\usepackage{hyperref}
\usepackage{url}
\usepackage{subcaption}
\usepackage{placeins}% For making floats not move around everywhere
\newcommand{\vect}[1]{\mathbf{#1}}
\newcommand{\ket}[1]{\left|{#1}\right\rangle}
\newcommand{\comm}[2]{\left[{#1},{#2}\right]}

\newcommand{\norm}[1]{\left\|{#1}\right\|}
\newcommand{\tr}{\mathrm{tr}}
\begin{document}
%\DeclareGraphicsRule{*}{png}{*}{}

%%%% User-defined commands %%%%
\renewcommand\floatpagefraction{0.88} %% default value: 0.5
\renewcommand\topfraction{0.88}       %% default value: 0.7

\author{Michael Vogl\footnote{These authors contributed equally to this work.}}
\author{Pontus Laurell$^*$}
\author{Aaron D. Barr}
\affiliation{Department of Physics, The University of Texas at Austin, Austin, TX 78712, USA}
\author{Gregory A. Fiete}
\affiliation{Department of Physics, Northeastern University, Boston, MA 02115, USA}
\affiliation{Department of Physics, The University of Texas at Austin, Austin, TX 78712, USA}
\affiliation{Department of Physics, Massachusetts Institute of Technology, Cambridge, MA 02139, USA}

%\title{Flow equation treatment of high frequency Floquet systems - a march into perturbatively inaccessible regimes}
\title{A flow equation approach to periodically driven quantum systems}
\date{\today}

\begin{abstract}

We present a theoretical method to generate a highly accurate {\em time-independent} Hamiltonian governing the finite-time behavior of a time-periodic system.  The method exploits infinitesimal unitary transformation steps, from which renormalization group-like flow equations are derived to produce the effective Hamiltonian.  Our tractable method has a range of validity reaching into frequency--and drive strength--regimes that are usually inaccessible via high frequency $\omega$ expansions in the parameter $h/\omega$, where $h$ is the upper limit for the strength of local interactions.  We demonstrate exact properties of our approach on a simple toy-model,  and test an approximate version of it on both interacting and non-interacting many-body Hamiltonians, where it offers an improvement over the more well-known Magnus expansion and other high frequency expansions. %The approximation is equivalent to a rotating frame approximation but much less restrictive because it can quite easily capture most features of a rotating frame approximation when it is impossible to calculate the rotating frame approximation by other means. 
For the interacting models, we compare our approximate results to those found via exact diagonalization.  While the approximation generally performs better globally than other high frequency approximations, the improvement is especially pronounced in the regime of lower frequencies and strong external driving. This regime is of special interest because of its proximity to the resonant regime where the effect of a periodic drive is the most dramatic.  Our results open a new route towards identifying novel non-equilibrium regimes and behaviors in driven quantum many-particle systems.
\end{abstract}

%\pacs{73.20.At, 73.21.Ac, 81.05.Uw}

\maketitle

%%%%%%%%%%%%%%%%
% Introduction %
%%%%%%%%%%%%%%%%

% Note: In PRX/PRL style, citations before period or comma, e.g. [2].
\section{Introduction}
Recent years have seen rapid progress in our understanding of dynamics and non-equilibrium phenomena in quantum systems  \cite{RevModPhys.83.863,RevModPhys.89.011004}. This has been a result of experimental advances in the ability to control cold atom  \cite{RevModPhys.89.011004,RevModPhys.80.885,RevModPhys.83.1523}
and condensed matter systems \cite{Basov2017,doi:10.1146/annurev-matsci-070813-113258,RevModPhys.83.471}, by developments in time-resolved laser techniques  \cite{doi:10.1080/00018732.2016.1194044,1402-4896-92-3-034004}, and by the fact that stepping into the time domain opens up new ways of ultrafast control of material properties \cite{PhysRevLett.113.057201,Mentink2015,Basov2017} and access to different phases of matter. These include photoinduced superconductivity \cite{Fausti2011,doi:10.1080/00107514.2017.1406623}, hidden orders \cite{Stojchevska2014}, and metastable states \cite{Gerasimenko2017}, but also entirely novel phases, such as time crystals \cite{10.1038/nature21413,10.1038/nature21426} and non-equilibrium topological phases \cite{PhysRevX.6.041001,PhysRevLett.116.250401}. 

In particular, there has been growing interest in periodically driven (or Floquet) \cite{Bukov2015,PhysRev.138.B979} many-body systems, which can bear a close resemblance to equilibrium systems \cite{10.1038/nphys4106}. The Floquet systems come in three established thermodynamic classes: integrable \cite{PhysRevLett.112.150401,PhysRevLett.118.260602,2018arXiv180500031W}, many-body localized (MBL) \cite{PhysRevLett.116.250401,Ponte2015,PhysRevLett.114.140401}, and generic interacting ones \cite{PhysRevE.90.012110}. The first two classes can avoid thermalization, allowing for a notion of a Floquet phase of matter at long stroboscopic times $t=nT$, where $T$ is the period of the Hamiltonian, $H(t+T)=H(t)$, and $n$ is an integer. The physics of these phases is captured by an effective, time-independent Floquet Hamiltonian $H_F$, given via the time evolution operator over one period $U(T)=\exp \left( -iH_F T\right)$. 

In non-interacting systems $H_F$ can be used to dynamically engineer interesting and topological band structures \cite{Wang2013a,PhysRevLett.106.220402,PhysRevB.96.125144,PhysRevLett.114.125301,PhysRevLett.116.176401,PhysRevB.95.035136,PhysRevB.97.035422,PhysRevX.6.021013,DAlessio2015,PhysRevX.7.041008}, most notably Floquet topological insulators \cite{PhysRevB.79.081406,Lindner2011,PhysRevB.84.235108,PhysRevLett.107.216601,PhysRevX.3.031005,PhysRevB.90.115423}. Our main interest, however, is interacting systems where $H_F$ can be engineered to drive phase transitions \cite{PhysRevLett.102.100403,PhysRevLett.95.260404}, or, in the case of Floquet-MBL systems, realize new phases without equilibrium analogs \cite{PhysRevLett.116.250401,PhysRevX.6.041001,PhysRevLett.118.030401,PhysRevLett.119.123601,2018arXiv180304490L}. Clean interacting Floquet systems are the least studied of the three classes, perhaps because they were expected to heat up to a featureless state with infinite effective temperature \cite{PhysRevB.94.020201,PhysRevX.4.041048}. However, it was recently theoretically discovered that under very general conditions they may remain in a prethermal state until exponentially long times $\tau^\star$  \cite{PhysRevLett.115.256803,PhysRevB.95.014112,PhysRevLett.116.120401,PhysRevX.7.011026,PhysRevE.93.012130}, which has been verified numerically in several models \cite{Machado2017,Weidinger2016}.

The existence of a prethermal regime is important because realistic systems usually contain integrability-breaking perturbations that support it, and because the thermalization (or more specifically, the energy absorption) time $\tau^\star$ can correspond to experimentally accessible time scales. The existence of such a regime also implies that there is interesting physics to be found at intermediate times $0<t<\tau^\star$\,\cite{PhysRevLett.115.256803,Kuwahara2016}, where one may use time-dependent perturbations to drive dynamical phase transitions \cite{PhysRevLett.110.135704,PhysRevLett.119.080501,Flaeschner2016,PhysRevB.97.085152}, control interactions \cite{PhysRevLett.107.210405,PhysRevB.95.024306}, or engineer phase transitions and topological phases \cite{PhysRevLett.115.205301,PhysRevLett.116.125301,Peronaci2017,PhysRevX.7.011018,PhysRevB.95.045102}. 

To understand the properties of a system in the prethermal regime, it is convenient to use a description in terms of the effective Hamiltonian, $H_F$. It is, however, notoriously difficult to calculate $H_F$ or the exact time-evolution operator $U(t)$ for interacting systems, so generally one uses an expansion technique to find an approximate, effective Hamiltonian in the high-frequency limit. These include the Magnus expansion \cite{Blanes2009,Feldm1984,Magnus1954}, rotating frames \cite{PhysRevB.95.014112}, and many more \cite{Bukov2015,PhysRevA.68.013820,PhysRevX.4.031027,PhysRevLett.115.075301,Eckardt2015,PhysRevB.93.144307,PhysRevB.94.235419,PhysRevLett.116.125301,PhysRevB.25.6622}. Unfortunately, these methods do not produce a cleanly convergent expansion series for general systems. Instead, they are asymptotic expansions, subject to an optimal cut-off order which prevents them (in principle) from reaching into the lower frequency regimes \cite{PhysRevB.95.014112,Blanes2009}. By this statement we do not mean to imply that methods such as ours may not be subject to their own cut-offs but that these cut-offs may differ \cite{PhysRevLett.120.200607}. Whether this is the case for the exact version of the flow equations in this paper is a matter that still has to be determined.

One of the more controlled descriptions of a system occurs in the quasiequilibrium regime, $W \ll \hbar\omega \ll \Delta$, where $W$ is the bandwidth of the system, $\omega$ the driving frequency ($\hbar$ Planck's constant), and $\Delta$ is the gap to the continuum of higher energy states. While this separation of energy scales is quite feasible in cold atom systems, it is harder to reach in solid state systems. Mott insulators are the most promising class of systems in this regard, but even there the range of frequencies is limited since we typically have $W\sim 1$eV and $\Delta \sim 1$eV, which are of the same order of magnitude. In addition, lower frequency regimes are required for certain topological phases \cite{PhysRevX.6.041001}, and are of interest in cold atom systems \cite{10.1038/nphys4020,PhysRevA.91.033632}, and in the study of thermalization \cite{PhysRevB.97.245122}. Hence, techniques to handle lower frequencies are needed. 

In this paper we improve on the limitations of previous methods, and provide better access to lower frequencies and higher driving strengths. To achieve this we introduce a formalism to remove the time-dependent part of a Hamiltonian using infinitesimal unitary transformations. This results in flow equations for different couplings, reminiscent of renormalization group calculations \cite{RevModPhys.66.129} and Wegner's flow-equation approach to diagonalizing Hamiltonians \cite{Wegner1994,Kehrein2007}. There has also recently been progress in using the Wegner flow to describe the time-evolution of a many-body localized system \cite{PhysRevB.97.060201}, which however still requires the solution of flow equations for each time-step --- a problem we avoid in our construction. 

We note that while a flow equation method for finding effective Floquet Hamiltonians exists in the literature, it uses an approximate version of the Wegner generator (keeping only terms proportional to $\omega$ in the generator) \cite{PhysRevLett.111.175301} in Sambe space \cite{PhysRevA.7.2203}, where the approximation brings up a question as to the range of validity. 
%We note that while a method exists for finding effective Hamiltonians involving flow equations, they used an approximate version of the Wegner generator (keeping only terms proportional to $\omega$ in the generator) \cite{PhysRevLett.111.175301} in Sambe space \cite{PhysRevA.7.2203}, where the approximation brings up a question as to the range of validity.
Our method differs in that we do not need to introduce Sambe space, and our generator is obtained in a constructive manner and differs completely from the Wegner generator. For our method, we describe both the exact flow equations, and ways to approximate them. We apply our method to the Schwinger-Rabi model of a single spin in a magnetic field, and also to four different spin chain Hamiltonians: (i) an integrable $XY$ model with antisymmetric exchange, (ii, iii) two integrability breaking extensions of a $J_1$-$J_2$-type  XXZ model \cite{Giamarchi2004}, and (iv) the transverse field Ising model.  

The extended $XY$ model is driven by a transverse magnetic field, the first $J_1$-$J_2$-type $XXZ$ model is driven locally by a magnetic field in the $x$-direction, 
%$x$-direction magnetic field, 
and the second by a nearest neighbor Ising exchange interaction, making for a time-dependent $J_1$-$J_2$ model \cite{doi:10.1143/JPSJ.62.1123,PhysRevB.25.4925}. For the transverse field Ising model we consider (i) a harmonic driving case, and (ii) a case where the time evolution operator factorizes into two matrix exponentials, which allows us to find a family of different resummations of the Baker-Campbell-Hausdorff (BCH) identity.  This observation leaves open the question of how to construct the optimal effective Hamiltonian for a given time evolution operator (reverse of the usual situation in which one seeks the optimal time evolution operator approximation for a given Hamiltonian). 

In this paper, we study the time evolution of the exact models, and their effective models obtained in our approach.  We compare our results with those obtained by the Magnus expansion. The integrability breaking models are studied numerically using full exact diagonalization, which provides an unbiased test of the validity of our approach. We find that our flow method generally outperforms the Magnus expansion, with significantly greater accuracy as the resonant regime is approached, as well as in the case when the time-dependent term in the Hamiltonian is large.  Both of these cases are of direct physical relevance and interest.  Our method thus opens new possibilities in the analytical and numerical simulation of time-dependent quantum many-particle systems, and will facilitate the search for novel  prethermal,  and non-equilibrium regimes.

Our paper is organized as follows. In Sec.\ref{sec:formalism} we develop the general flow-equation formalism, and discuss its structure and approximations. In Sec.\ref{sec:approximations} we relate the general results obtained from the flow equation approach to various high-frequency expansions used in the literature. In Sec. \ref{exact_flow_properties} we test the flow equations on an exactly solvable two level system and discuss in detail the properties of the fixed points of the flow equations and their stability.  This discussion is continued in Sec. \ref{sec:exflowtruncansatz} for a many body-system studied via a truncated ansatz where we show it outperforms a high frequency Magnus expansion and rotating wave approximation. In Sec.\ref{sec:models} we introduce four different one-dimensional spin chain Hamiltonians we will use to assess the performance of the approximate method described in Sec.\ref{sec:approximations}.  In Sec.\ref{sec:results} we summarize our results for the different models. In Sec.\ref{sec:BCH} we compare our results to a resummation of the Baker-Campbell-Hausdorff identity that was of recent interest \cite{PhysRevLett.120.200607}. We also show what advantages our approach has over a standard rotating frame approximation---namely that it can be truncated when a rotating frame transformation is not practically possible and that it still performs well under these circumstances. In Sec.\ref{sec:conclusions} we present our main conclusions.  Various technical details and formulas appear in the appendices.

%%%%%%%%%%%%%%%%%%%%%
% General Formalism %
%%%%%%%%%%%%%%%%%%%%%

\section{General Formalism}
\label{sec:formalism}
We take the Schr\"odinger equation of a periodically driven many-particle system as our starting point. Following Ref.~\cite{PhysRevB.95.014112}, the Hamiltonian $H(t)$ is split into a constant part $H_0=\frac{1}{T}\int_0^T dtH(t)$, and a time-periodic term $V(t)=\frac{1}{T}\int_0^T dt_1(H(t)-H(t_1))$ that averages to zero over one period, $\frac{1}{T}\int_0^T dt V(t)=0$.  Thus, the time-dependent Schr\"odinger equation takes the form,
\begin{equation}
i\partial_t \ket{\psi_0}=\left(H_0+V(t)\right)\ket{\psi_0},
\label{eq:SE}
\end{equation}
where we have set Planck's reduced constant $\hbar=1$.

We introduce a unitary transformation, $U=e^{\delta\Omega(t)}$, generated by an as yet undetermined quantity $\delta \Omega$ that will be chosen to reduce the time-dependent term $V(t)$.  The $\delta$ in front of the $\Omega$ indicates we keep the generator infinitesimal, which ensures that the exponential can be safely expanded to lowest order. 

Let us now introduce a new wavefunction $\ket{\phi_{\delta s}}=U^\dag\ket{\psi_0}=[1-\delta\Omega(t)]\ket{\psi_0}$ and act with $U(t)^\dag=1-\delta\Omega(t)$ (to leading order in $\delta \Omega$) from the left on the Schr\"odinger equation. This new wavefunction now fulfills the modified Schr\"odinger equation (keeping lowest order in $\delta \Omega$ only),
\begin{equation}
i\partial_t\ket{\phi_{\delta s}}= \left (H(t)-i\partial_t \delta\Omega(t)-\comm{\delta\Omega(t)}{H(t)}\right)\ket{\phi_{\delta s}}.
\end{equation}

One may read off a new Hamiltonian, which, since  $\delta\Omega$ is infinitesimal, can be written as
\begin{equation}
	\tilde H(t)=H(t)-i\partial_t \delta\Omega(t)-\comm{\delta\Omega(t)}{H(t)}.
	\label{eqHam}
\end{equation}

Up to this point, this treatment coincides with the use of time-dependent generators \cite{0295-5075-93-4-47011}.  We now, however, choose $\delta\Omega$ very different from the Wegner generator. We choose it such that it reduces the time dependent part of the Hamiltonian $V(t)\to (1-\delta s)V(t)$ by some infinitesimal value $\delta s$, 
\begin{equation}
	\delta \Omega=-\frac{i}{T}\delta s\int_0^t dt_1\int_0^T dt_2(H(t_1)-H(t_2)),
	\label{eq:generator}
\end{equation}
where the generator in Eq.\eqref{eq:generator} also has the nice property that it vanishes at stroboscopic times $T$. Therefore, at stroboscopic times, expectation values $\langle \hat O \rangle$ of operators $\hat O$ can be calculated without a change of basis. The behavior at other times can be found by applying the unitary transformation to the operator $\hat O$.

One could now repeat the procedure of splitting the Hamiltonian into a constant and a  time average zero part and then apply this infinitesimal unitary transformation to find the Floquet Hamiltonian after an infinite amount of steps (or an approximation to it by stopping after a finite amount of steps). To simplify the process, we recognize that one can track the progress of the unitary transformations by a single flow parameter, $s$. To do so we extend the functional dependencies of the Hamiltonian to include this parameter, replacing $H(t)\to H(s,t)$ and $\tilde H(t)\to H(s+\delta s,t)$. Note that $H(s,t)$ represents a family of effective Hamiltonians interpolating between a starting Hamiltonian $H(0,t)$, and a  Hamiltonian $H(\infty,t)$. $H(\infty,t)$ is the Floquet Hamiltonian $H_F$ if $V(\infty,t)=0$. It seems plausible that $V(\infty,t)=0$ and we find this to be true in an explicit example and some limiting cases but it remains to be shown rigorously.  We set appropriate boundary conditions by enforcing that $s=0$ corresponds to the initial, non-transformed Hamiltonian. 

With this notation, Eq.~\eqref{eqHam} takes the form
\begin{equation}
\begin{aligned}
	H(s+\delta s,t)&=H(s,t)-\delta s V(s,t)\\
	&+i\delta s\int_0^t dt_1\comm{V(s,t_1)}{H(s,t)},
\end{aligned}
\end{equation}
with $V(s,t)=\frac{1}{T}\int_0^T dt_1(H(s,t)-H(s,t_1))$. One may note that this leaves a residual time-dependence of $\delta s \comm{ V(t)}{H(t)}$ in Eq.\eqref{eqHam},  which is small in magnitude if $\delta s$ is small. 

Taylor expanding the left hand side since $\delta s$ is infinitesimal we find, 
\begin{equation}
	\frac{dH(s,t)}{ds}=-V(s,t)+i\int_0^t dt_1\comm{V(s,t_1)}{H(s,t)},
	\label{exact_flow}
\end{equation}
which is a central result of this work.  We refer to Eq.\eqref{exact_flow} as the exact flow equation. This equation is similar in spirit to the infinitesimal unitary transforms that Wegner \cite{Wegner1994} employs to diagonalize an interacting Hamiltonian in the equilibrium case. 

One can readily see that Eq.\eqref{exact_flow} has a fixed point with the desired property $V(s,t)=0$. This fixed point is guaranteed to be stable for sufficiently large $\omega$ because in this case the commutator term can be neglected. Under these circumstances the time independent parts of $H(s,t)$ remain unchanged. More precisely equation \eqref{exact_flow}  then reduces to
\begin{equation}
		\frac{dV(s,t)}{ds}\approx-V(s,t),
\end{equation}
which trivially has the stable fixed point $V(s,t)=0$ since $V(s,t)$ for this case can be treated like a scalar. 

But what about smaller $\omega$? Because an analytic understanding is difficult to achieve, we will discuss this in the context of an explicit example in Sec.\ref{exact_flow_properties}. While our discussion gives a mechanism by which the fixed points can be stable in general it does not give a rigorous proof.

How should one interpret the flow of $s$ in Eq.\eqref{exact_flow}? Note that $H(s,t)$ is a Hamiltonian and therefore a linear sum of the various energy contributions, and can be expressed as a sum of linear operators with coefficients $c_i(s,t)$, $H(s,t)=\sum_i c_i(s,t)\hat O_i$ (similar in spirit to a Landau-Ginzburg energy functional). The $\hat O_i$ operators are nothing other than kinetic and potential energy terms appearing in a Hamiltonian, such as a hopping term $c_i^\dagger c_j$ in a lattice model, an interaction term $n_{i\uparrow}n_{i\downarrow}$ on a lattice, or a multiple-spin term $(\vec S_i\cdot \vec S_j)(\vec S_k \cdot \vec S_k)$ in a spin model, among many other possibilities.  The coefficients $c_i(s,t)$ describe the coupling constants (strength) of these terms.

This mathematical structure of $H(s,t)=\sum_i c_i(s,t)\hat O_i$ in turn also implies that $-V(s,t)+i\int_0^t dt_1\comm{V(s,t_1)}{H(s,t)}=-\sum_i g_i(t,[c_j(s,t^\prime)])\hat O_i$. Here $g_i$ has a functional dependence on the $c_j(s,t^\prime)$ with $t^\prime\in\left[0,T\right]$, because $V(s,t)$ itself depends on the $c_j(s,t)$ and it appears under an integral.  

One may therefore write Eq.~\eqref{exact_flow} as
\begin{equation}
	\frac{dc_i(s,t)}{ds}=-g_i(t,[c_j(s,t^\prime)]);\quad t^\prime\in\left[0,T\right],
	\label{coupling_eq}
\end{equation}
which is just a flow equation for the coupling parameters $c_i(s,t)$ at different times. Note that the set of operators ${\hat O_i}$ may include both the original operators, and ones generated from the kinetic and potential energy terms of the original Hamiltonian, Eq.\eqref{eq:SE}, as the Hamiltonian flows. 
%The operators $\hat O_i$ are generated from the kinetic and potential energy terms of the original Hamiltonian, Eq.\eqref{eq:SE}, as the Hamiltonian flows.  
In general, new terms are generated such as hopping and interaction terms that involve more and more sites of a lattice as the order of the transformation increases.  These new terms can in principle change the balance of kinetic and potential energy in the effective time-independent Hamiltonian and therefore may lead to new physical regimes for a periodically driven many-particle quantum system. The reason we write the flow equations in this form is to emphasize that Eq.\eqref{exact_flow} actually describes couplings that flow as we reduce out the time dependence and to show how this operator equation corresponds to a numerically tractable scheme to determine couplings.

\section{Approximations to the flow equations}
\label{sec:approximations}
It is important to note that Eq.\eqref{exact_flow} offers a convenient starting point to approximate the Floquet Hamiltonian. In particular, it allows us to improve on the various high frequency expansions of the Floquet Hamiltonian that have appeared in the literature.  As an example, we can find an analytically tractable equation if we set $s=0$ only for the terms $V(s,t)$. This corresponds to removing the original time dependent part $V(t)$ from the Hamiltonian via the rotating frame transformation \cite{doi:10.1080/00018732.2015.1055918} $e^{-i\int_0^t dtV(t)}$, while generating other new time dependences. (This approximation is for convenience. Indeed in the following section we will present an example in which we exactly solve Eq.~\eqref{exact_flow} without taking $s=0$ in $V(s,t)$.)  To ensure that this approximation actually corresponds to the aforementioned unitary transformation we also need to restrict the range of $s$ to $[0,1]$, rather than the previous $[0,\infty)$.

Let us justify this approximation slightly more carefully by using an analogy.  One may notice that Eq.\eqref{exact_flow} is very similar in structure to the classical problem of a first order differential equation,
\begin{equation}
\frac{d f(t)}{dt}=g(t,f(t)),
\end{equation}
where $g(t,f(t))$ would correspond to $-V(s,t)+i\int_0^t dt_1[V(s,t_1),H(s,t)]$ in our case and all the couplings in $H$ and $V$ correspond to $f(t)$.

A standard method of solving this class of problems\cite{birkhoff1989ordinary} is plugging in the initial condition $f(t)\to f_0=f(t=0)$ on the right hand side. Integrating both sides of the equation one finds a first approximation to $f(t)$, which we call $f_1(t)$. One may then repeat the procedure and plug successive approximations $f_n(t)$ into the right-hand side. This procedure is called Picard iteration. In our case, it is the same as replacing $V(s,t)\to V(0,t)$ and $H(s,t)\to H(0,t)$.

A variant of Picard iteration that quite often works better is to only set $f(t)=f_n(t)$ in some places of $g(t,f(t))$ but keep it as $f(t)$ in others. This is a particularly helpful improvement when this is done in such a way that some symmetries are explicitly kept that would otherwise be destroyed\cite{Blanes2009}. For our case, if we only replace the first two $V(s,t)\to V(0,t)$ but keep $H(s,t)$ then we find approximate flow equations,
\begin{equation}
\frac{dH(s,t)}{ds}=-V(0,t)+i\int_0^t dt_1\comm{V(0,t_1)}{H(s,t)},
\label{approx_floweq}
\end{equation}
where $s$ is set to run from zero to one only. As required above, this still implements a unitary transformation, which can be seen explicitly by reconstructing Eq. \eqref{approx_floweq} from a unitary transformation. 

Introducing the dependence on the flow parameter $s$, Eq. \eqref{eqHam} reads
\begin{equation}
H\left( s+\delta s,t\right) =H\left( s,t\right)-i\partial_t \delta\Omega(s,t)-\comm{\delta\Omega(s,t)}{H(s,t)}.
\label{eqHamExplicit}
\end{equation}
One may plug in the manifestly anti-hermitian generator $\delta\Omega (s,t) \equiv \delta\Omega(0,t)=-i\delta s\int_0^t dt V(0,t)$, corresponding to a unitary transformation $U$. The result is Eq.\eqref{approx_floweq}. Therefore, making such an approximation is a particularly convenient improvement on a Picard iteration.

One may ask why $s$  should run from zero to one as claimed above. One reason for this is that in the lowest order improved Picard iteration we neglect terms that are proportional to $s$. Neglecting such terms is only justified if $s\leq1$. Therefore, we let the flow parameter run from zero to 1. If we reach a fixed point in this range of values or come close to it, then it is a good approximation.  Letting $s$ run to higher values would not be justified and may yield a bad result. Another reason we apply this approach is that we know that for infinite frequencies one reaches a stable fixed point for $s=1$. This can be seen easily because Eq.\eqref{approx_floweq} is then approximately given as
\begin{equation}
	\frac{dV(s,t)}{ds}\approx-V(0,t).
\end{equation}
 This procedure also works well in other cases because often at $s=1$ one may be close to an unstable fixed point (see, for example Fig.\ref{fig:flowingcouplings}). We should also mention that the multitude of different possible fixed points (all $V(s,t)=0$) and their corresponding $s$ value makes it difficult to estimate the size of the error from %how much of a mistake is made by 
letting $s$ only run from zero to one.  After all, often $s=1$ is close to a fixed point but there may be more fixed points further out (for larger values of $s$). We will see this explicitly in the next section where we work with the exact flow equations.

Now let us return to discussing Eq.\eqref{approx_floweq}. One finds that this also can be rewritten in terms of coupling constants as,
  \begin{equation}
  \frac{dc_i(s,t)}{ds}=-g_i(t,c_j(s,t)),
  \end{equation}
where  one can write $g_i(t,c_j(s,t))=\sum_j \gamma_{ij}(t)c_j(s,t)$ as a linear combination of couplings $c_j(s,t)$. We are therefore left with a first order linear differential equation that doesn't couple coefficients $c_j$ at different times.  Gone is the more complicated structure of a functional in the $c_i$. The effective time-independent Hamiltonian is then given by
   \begin{equation}
   	H_\mathrm{eff}=\sum_i \bar c_i\hat O_i,
   \end{equation}
 with $\bar c_i=\frac{1}{T}\int_0^Tdt\, c_i(1,t)$, where we have taken an average over one period, which is physically meaningful if one is only looking at stroboscopic times. If one is interested in micromotions, one could in principle retain the time-dependence of $c_i(1,t)$--the important ``flow" having been taken into account in the parameter $s$, which has now been set to unity.
 
The approximation in Eq.\eqref{approx_floweq}, setting $s=0$ in $V(s,t)$, does not make any implicit assumptions,  such as $V(t)$ is small.   By contrast, many other high frequency approximations do make the assumption of smallness.  As a result, our approach like the rotating frame approximation, works especially well in the limit of strong $V(t)$. We will demonstrate this explicitly in later sections of this work. 
 
It is important to pause for a moment and stress the advantages our approximate method,  Eq.~\eqref{approx_floweq}, offers over a rotating frame approximation, if the latter is carried out exactly. Firstly, if the driving is complicated it is often not possible to calculate the matrix exponential needed for a rotating frame approximation or the rotation induced on operators by such a matrix exponential because it will generate infinitely many components of the operator algebra. This is indeed the case with one of our example models namely the square-wave driven Ising model we discuss later. In this case our method allows one to keep all orders in $1/\omega$ with a truncated ansatz for the Hamiltonian. That this method performs well can be seen in the plot shown in Sec.\ref{sec:BCH}. 

It is also important to recognize that, even if a rotating frame approximation can be done exactly, usually most terms in the Hamiltonian become time-dependent. In most cases this makes a second rotating frame approximation not possible. Our method allows one to avoid this issue 
by truncating the ansatz Hamiltonian. Lastly, in some cases one would like to prevent the generation of any new terms and see what happens to the coupling constants of a restricted set of terms.  Thus, our method provides a convenient starting point for many different approximation schemes. 

We would also like to stress that Eq. \eqref{approx_floweq} implements a unitary transformation \emph{exactly}. Its solution therefore still retains the full information of the original Hamiltonian. In this paper we will be content with discussing results from the first order iteration only.  Again, the formalism we present here lays the groundwork for further development of approximation schemes.
    
Let us explicitly relate the first order iteration to the more common high frequency approximations. For the moment, neglect $\int_0^t dt_1\comm{V(0,t_1)}{H(s,t)}$, which assumes that all couplings in the Hamiltonian are negligible compared to the driving frequency. This is an approximation common to many of the high frequency approximations. We then find that
    \begin{equation}
    	H(s,t)\approx H(0,t)-sV(t).
    \end{equation}
   Inserting this back into Eq. \eqref{approx_floweq} and taking a time average we find 
   \begin{equation}
   	 H(1,t)\approx H_0+\frac{i}{T}\int_0^T dt\int_0^t dt_1\comm{V(t_1)}{H_0+\frac{1}{2}V(t)},
    \end{equation}
    which is the lowest order of many common high frequency approximations. Hence, our approximation agrees with other approximations in the high frequency limit.  
    
 One should also note that there are other ways to approximately solve the exact flow equation by directly working with  Eq.\eqref{exact_flow} and a truncated ansatz rather than solving Eq.\eqref{approx_floweq}. We we will do this in one example in Sec.\ref{sec:exflowtruncansatz} and will find that it indeed offers an improvement to the methods above (rotating frame and high frequency expansion) and opens the door to many semi-analytical schemes.
    
 Next, we turn to an application of our method to a number of different Hamiltonians and compare our results with other approaches.  We find the method nearly always provides more accurate evolution than other approximations, and in many cases our method works substantially better, particularly as the strong coupling resonant regime is approached. This is also true if we solve Eq.\eqref{approx_floweq} with a truncated ansatz like on one of the cases in Sec.\ref{sec:BCH}. 
    %Therefore in the high frequency limit our approximation always agrees with the other more common approximations.
   
%%%%%%%%%%%%%%%%%%%%%%%%%%%%%%%%%%%%%%%%%%
% Understanding the exact flow equations %
%%%%%%%%%%%%%%%%%%%%%%%%%%%%%%%%%%%%%%%%%%
   \section{Fixed point stability and the properties of the exact flow equations}
   \label{exact_flow_properties}
Because it is difficult to discuss the stability of the flow equations in Eq.\eqref{approx_floweq} analytically in full generality, we consider a simple example model where the exact flow equations can be written down explicitly. This will allow us to identify a mechanism that makes the fixed point stable. It is conjectured, but we stress not rigorously proven, that this mechanism will persist even for more complicated systems. In Sec.\ref{sec:exflowtruncansatz} this conjecture will be further supported. The current section serves as a means to gain some insight into how the flow equations work. We consider the Schwinger-Rabi model of a spin in a rotating magnetic field,
   \begin{equation}
   	H=B_z\sigma_z+B_p(\sin(\omega t)\sigma_y+\cos(\omega t)\sigma_x).
   	\label{Rabi-Ham}
   \end{equation}
For this model the Floquet Hamiltonian,
   \begin{equation}
   	H_\mathrm{F}=-\frac{\omega}{2}+B_p\sigma_x+(B_z-\frac{\omega}{2})\sigma_z,
   	\label{Heffnew}
   \end{equation} 
 can be found for all frequencies (see for instance \cite{1367-2630-20-9-093022}).
   
 Let us discuss how the flow equations apply to this model. After repeatedly inserting the form of the original Hamiltonian in our exact flow equations in Eq.\eqref{exact_flow} (always including newly generated terms) we find that the Hamiltonian $H(s,t)$ takes the form,
   \begin{equation}
   \begin{aligned}
   H(s,t)=&Z_0(s)\sigma_z+X_0(s)\sigma_x+Y_S(s)\sin(\omega t)\sigma_y\\
   &+X_C(s)\cos(\omega t)\sigma_x+Z_C(s)\cos(\omega t)\sigma_z,
   \label{eq:H_exact_s}
   \end{aligned}
   \end{equation}
   and the flow equations for the couplings $\{Z_0,X_0,Y_S,X_C,Z_C\}$ are given as,
   \begin{equation}
   \begin{aligned}
   Z_0'(s)&=\frac{2 Y_S(s) (X_0(s)-X_C(s))}{\omega },\\
   X_0'(s)&=\frac{2 Y_S(s) (Z_C(s)-Z_0(s))}{\omega },\\
   Y_S'(s)&=\frac{2 (Z_0(s) X_C(s)-Z_C(s) X_0(s))}{\omega }-Y_S(s),\\
   Z_C'(s)&=\frac{2 Y_S(s) (X_C(s)-X_0(s))}{\omega }-Z_C(s),\\
   X_C'(s)&=\frac{2 Y_S(s) (Z_0(s)-Z_C(s))}{\omega }-X_C(s),
   \end{aligned}
   \label{flow_spininmagfield}
   \end{equation}
  (where the $'$ denotes the derivative with respect to $s$) with initial conditions,
   \begin{equation}
   	\begin{aligned}
   &Z_0(0)=B_z,\quad Y_S(0)=X_C(0)=B_p,\\
   	 &Z_C(0)=X_0(0)=0.
   	\end{aligned}
   \end{equation}

As expected from Eq.\eqref{exact_flow}, we find that the fixed point is $Y_S=X_C=Z_C=0$, with arbitrary $Z_0$ and $X_0$. {\em This is the only fixed point.} For this fixed point we may carry out a stability analysis. That is, we expand Eq.~\eqref{flow_spininmagfield} around the fixed point to find linearized equations $\vect C'(s)=J\vect C(s)$, where $\vect C=\{Z_0,X_0,Y_S,X_C,Z_C\}$  is a vector of the couplings. The eigenvalues of the corresponding Jacobian $J$ are given as,
\begin{equation}
	\lambda_{1}=\lambda_2=0;\quad \lambda_{3,4}=-1\pm\frac{2}{\omega}\sqrt{Z_0^2+X_0^2};\quad \lambda_5=-1.
\end{equation}

It would appear that not all eigenvalues are guaranteed to be non-positive. In particular, one of the eigenvalues $\lambda_{3,4}$ could be positive, which would imply that the fixed point is unstable, and that the flow equations break down. If the form of the Hamiltonian at the fixed point reproduces that of Eq.~\eqref{Heffnew} this could indeed be the case, since there $Z_0$ and $X_0$ would be finite for arbitrarily small $\omega$. One might thus expect that flow equations would be unable to reach a stable fixed point for low enough frequencies. However, this outcome is avoided. 

\begin{figure}
	\centering
	\includegraphics[width=1\linewidth]{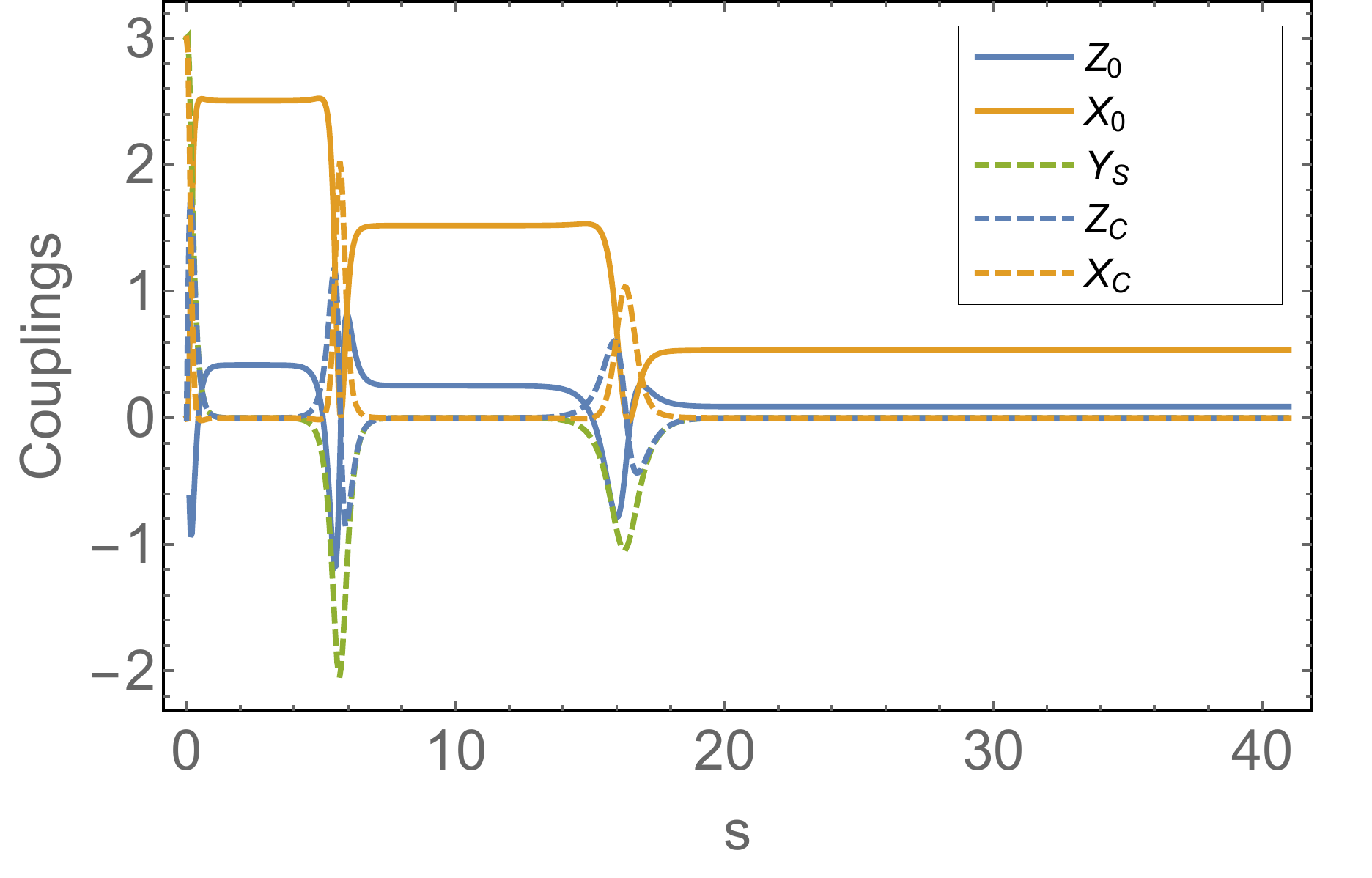}
	\caption{(Color online.) The couplings as a function of flow parameter $s$ for the Hamiltonian in Eq.~\eqref{eq:H_exact_s} with $B_p=3$, $B_z=1$, and $\omega=1$ in Eq.~\eqref{Rabi-Ham}. This corresponds to a low-frequency regime.  Note that while the couplings exhibit a non-trivial dependence on $s$ until sufficiently large $s$, the unitary evolution remains stable down to small frequencies, as seen in the red curve (exact flow) in Fig. \ref{fig:exactflowplot}.  The couplings after the range of the plot do not change within the limits of the line thicknesses.}
	\label{fig:flowingcouplings}
\end{figure}

To see how this works, recall that the Floquet Hamiltonian $H_F$ is determined only up to some phases by 
%It may seem that not all eigenvalues are negative. Particularly the eigenvalue $\lambda_3$ could be positive, which would imply that the fixed point is not stable and the flow equations break down. If our fixed point were to reproduce Eq.\eqref{Heffnew} this would indeed seem the case since  there $Z_0$ and $X_0$ would be finite for arbitrarily small $\omega$. So this would make it appear the flow equations would not be able to reach a stable fixed point for small enough frequencies. However, one should recall that the Floquet Hamiltonian $H_F$ is determined only up to some phases by 
\begin{equation}
	e^{-iH_F\frac{2\pi}{\omega}}=U\left(\frac{2\pi}{\omega}\right),
	\label{eq:H_F_def}
\end{equation}
where $U(t)$ is the time evolution operator. That means there are many different expressions for $H_F$ that would be valid branches of the matrix logarithm of both sides of Eq.~\eqref{eq:H_F_def}. For very small $\omega$ a valid $H_F$ could be chosen very small. Let us see what happens explicitly for our flow equations. Namely let us choose couplings such that Eq.\eqref{Heffnew} would correspond to an unstable fixed point. How these couplings evolve under the flow equations can be seen in Fig. \ref{fig:flowingcouplings}. 

\begin{figure}
	\centering
	\includegraphics[width=1\linewidth]{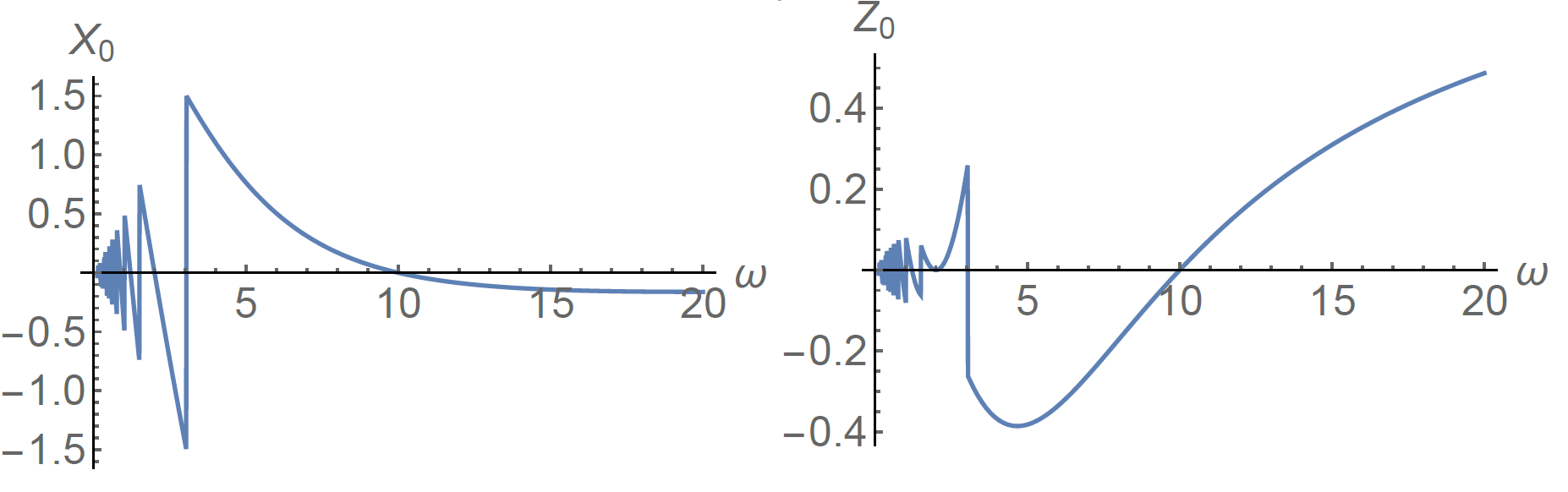}
	\caption{(Color online.) The non-zero couplings as a function of frequency $\omega$ at the end of the flow (large $s$ values in Fig. \ref{fig:flowingcouplings}) for $B_p=3$, $B_z=1$.  Note that in spite of the rapid oscillations for small $\omega$, the resultant unitary evolution remains stable, as seen in the red curve (exact flow) in Fig. \ref{fig:exactflowplot}.}
	\label{fig:rabicouplings}
\end{figure}

In Fig.\ref{fig:flowingcouplings} we see that the couplings made a few approaches to a fixed point $V(s,t)=0$, but it wasn't stable. However, the couplings $Z_0$ and $X_0$ kept shrinking until a stable fixed point was reached.  The matrix logarithm, $\log(U(T))$, has branches with relatively small $H_F$ and the couplings continued flowing until a branch with sufficiently small couplings to have a stable fixed point was reached.  In the language of the exact flow equations,  Eq.\eqref{exact_flow},  there existed a branch of the matrix logarithm $\log(U(T))$ such that $H(s,t)$ became sufficiently small that the commutator $\int_0^t dt_1[V(s,t_1),H(s,t)]$ could be neglected when compared to $V(s,t)$ and therefore a stable fixed point was reached. We were able to observe this effect in all cases we studied and it is plausible that this could be a general mechanism that leads to stable fixed points in our flow equations. This is illustrated in Fig. \ref{fig:rabicouplings}.

From Figs. \ref{fig:flowingcouplings}-\ref{fig:rabicouplings}, one may suspect numerical issues. However, this is not the case. Rather, the oscillations stem from the fact that the flow equations do not consistently stay on one branch of the matrix logarithm for $H_F$. Flowing to a stable fixed point means choosing the branch of the matrix logarithm that corresponds to a stable fixed point. 

Indeed, if we take the time independent couplings in Fig. \ref{fig:rabicouplings} to calculate the time evolution operator at stroboscopic times and compare it to the time evolution operator calculated via the standard method of a Trotter expansion we find them to be identical. More specifically we calculate the $l_2$ distance between two unitary operators,
\begin{equation}
	\frac{1}{2\sqrt{D_{\rm dim}}}\norm{U_1(T)-U_2(T)}_\textnormal{Frob};\quad \norm{A}_\textnormal{Frob}=\sqrt{\mathrm{tr}AA^\dag},	\label{l2distance}
\end{equation}
that was normed such that it takes values between zero and one ($D_{\rm dim}$ is the dimension of the Hilbert space), where one corresponds to the maximum distance between two unitary operators and zero to agreement between the two operators.   A comparison is shown in Fig. \ref{fig:exactflowplot}.  Details of the rotating frame approximation and Magnus expansion are given in appendix \ref{app:spin_in_rotfield}).
\begin{figure}
	\centering
	\includegraphics[width=1\linewidth]{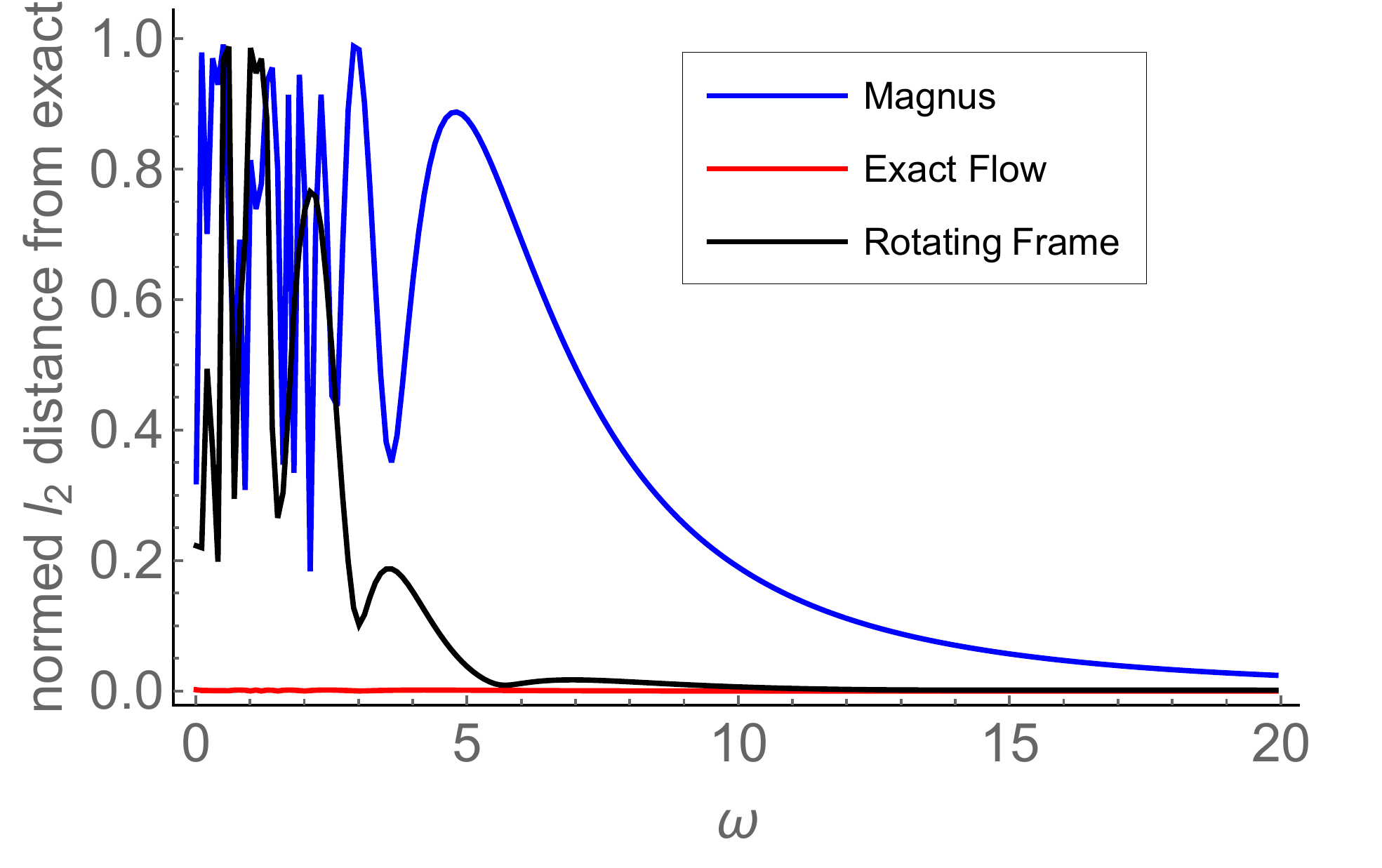}
	\caption{(Color online.)  Plot of the $l_2$ distance between the time evolution operator found by a Trotter expansion and the exact time evolution operator obtained by exactly solving the flow equations in Eq.~\eqref{exact_flow} given by Eq.~\eqref{flow_spininmagfield} (red), the Magnus expansion (blue) and the rotating frame approximation given by Eq.~\eqref{approx_floweq} (black).}
	\label{fig:exactflowplot}
\end{figure}
We find that the exact flow equations--despite the couplings rapidly changing--fully agree with the Trotter expansion as they should. The wildly jumping couplings are therefore not a numerical artifact.

%%%%%%%%%%%%%%%%%%%%%%%%%%%%%%%%%%%%%%%%%%%%%%%%
% Exact Flow equations with a truncated ansatz %
%%%%%%%%%%%%%%%%%%%%%%%%%%%%%%%%%%%%%%%%%%%%%%%%

\section{Exact flow equations with a truncated ansatz}
\label{sec:exflowtruncansatz}
In this section we discuss how the results from the previous section seem to be quite generic by considering a many-body system. We limit ourselves to a specific strongly driven Ising model given by,
\begin{equation}
H(t)=\sum_i [ \sigma_i^z\sigma_{i+1}^z+4\cos(\omega t) \sigma_i^z+4\sin(\omega t)\sigma_i^x].
\label{model}
\end{equation}
We choose this Hamiltonian because: (i) it has a relatively strong external drive, (ii) a time-dependent term that does not commute with itself at different times, and (iii) because the time-dependence is convenient for studying the flow equations.  One may find flow equations by making the truncated ansatz,
\begin{equation}
H(s)=H_0(s)+e^{i\omega t}H_+(s)+e^{-i\omega t}H_-(s),
\end{equation}
with
\begin{equation}
\begin{aligned}
H_{a}&=\sum_i[C_x^a\sigma_i^x+C_y^a\sigma_i^y+C_z^a\sigma_i^z+C_{xx}^a\sigma_i^x\sigma_{i+1}^x\\
&+C_{xy}^a(\sigma_i^x\sigma_{i+1}^y+\sigma_i^x\sigma_{i-1}^y)+C_{yy}^a\sigma_i^y\sigma_{i+1}^y\\
&+C_{xz}^a(\sigma_i^x\sigma_{i+1}^z+\sigma_i^x\sigma_{i-1}^z)+C_{zz}^a\sigma_i^z\sigma_{i+1}^z\\
&+C_{yz}^a(\sigma_i^y\sigma_{i+1}^z+\sigma_i^y\sigma_{i-1}^z)+C_{xzz}^a\sigma_i^x\sigma_{i-1}^z\sigma_{i+1}^z],
\end{aligned}
\end{equation}
where $a\in\{0,+,-\}$ and the $s$ dependence of the coupling constants $C^a$ was dropped for notational simplicity. We do not discuss the specific form of the flow equations here because they are rather complicated and not insightful. Let us rather first have a look at how some of the couplings behave for this system.  Specifically let us first look at one representative coupling as a function of flow parameter $s$.
\begin{figure}[H]
	\centering
	\includegraphics[width=1\linewidth]{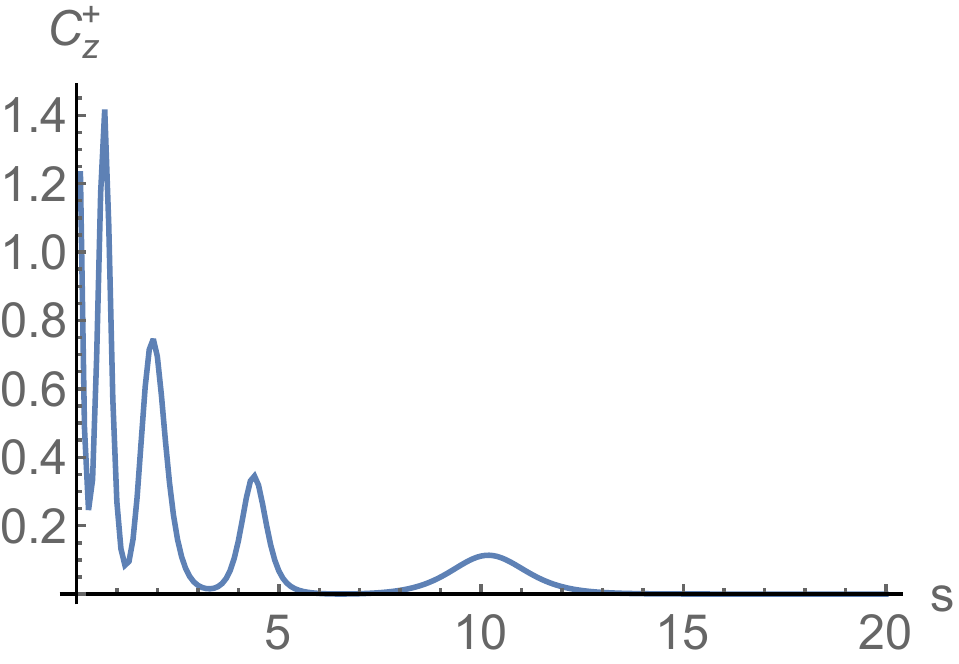}
	\caption{Coupling constant $C_z^+$ as a function of flow parameter $s$ plotted for $\omega=1.2$.}
	\label{fig:isingmodelszasfunctionofs}
\end{figure}
As one may see from Fig.\ref{fig:isingmodelszasfunctionofs}, the coupling constant $C_z^+$ behaves similar to the ones in Fig.\ref{fig:rabicouplings}  for the two level system we solved exactly in the previous section.  In particular, the coupling constant nearly approaches zero for the fixed point multiple times before eventually a stable fixed point is reached. This strengthens our interpretation that our method might be kept stable by the mechanism we provided in Sec.\ref{exact_flow_properties}.  
\begin{figure}[H]
	\centering
	\includegraphics[width=1\linewidth]{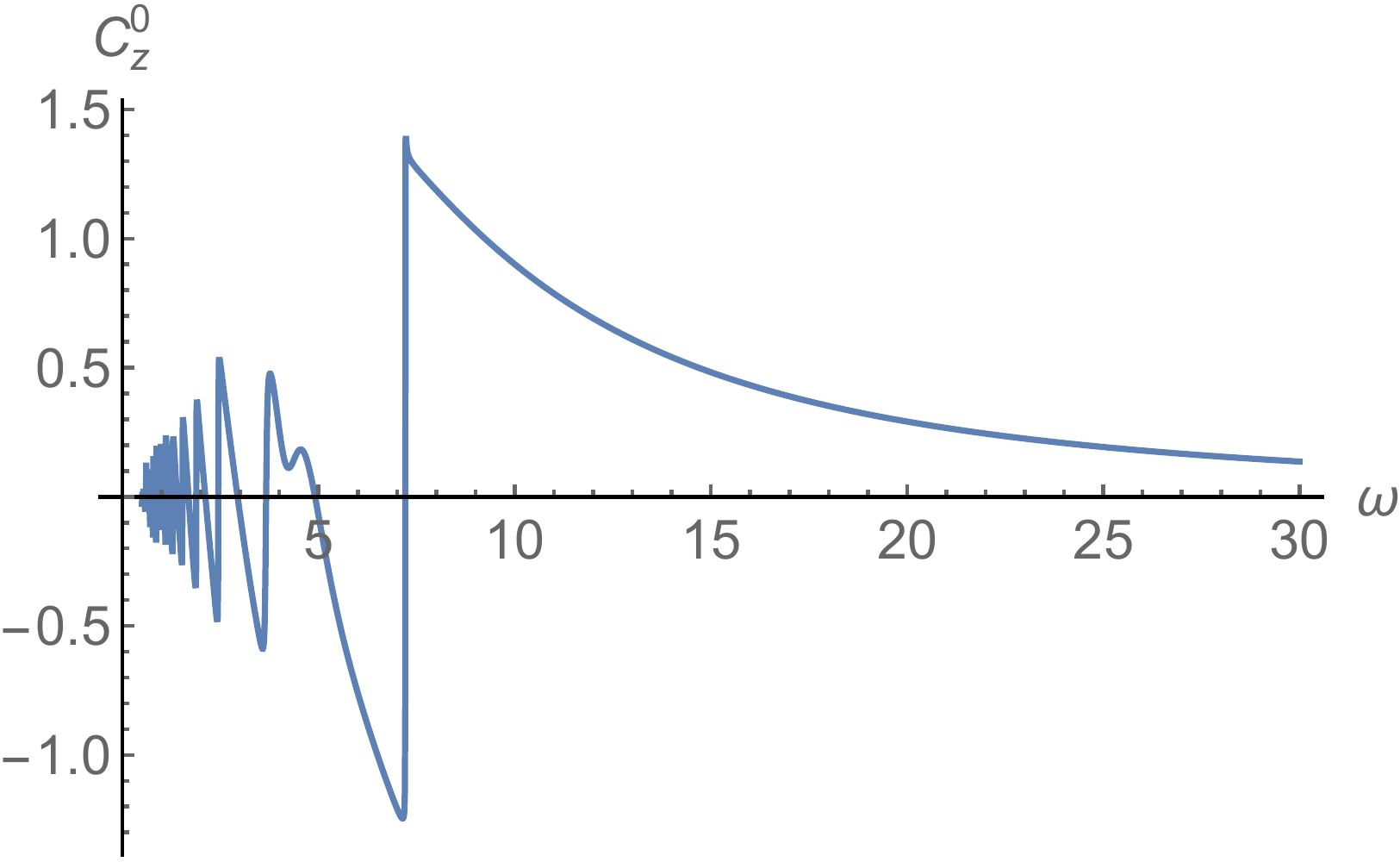}
	\caption{Plot of the coupling constant $C_z^0$ as a function of  $\omega$ for the flow equations solved up to a point $s=2,000,000$, which is long after the fixed point has been reached.}
	\label{fig:isingmodelszasfunctionofomega}
\end{figure}

To get further evidence of this we plot in Fig.\ref{fig:isingmodelszasfunctionofomega} one of the couplings as a function of $\omega$ and find it again to be consistent with the mechanism we proposed in Sec.\ref{exact_flow_properties} and illustrated in Fig.\ref{fig:rabicouplings}.
We stress that this is not a rigorous proof of our understanding of how the flow equations manage to converge, but it is does provide good evidence for the general structure of the convergence.

Let us now discuss these results further. One finds numerically that letting $s\to \infty$ only certain terms survive. Namely, as expected from the fixed point $C_i^{\pm}\to 0$, one is left with
\begin{equation}
\hspace{-0.3cm}
\begin{aligned}
H(s\to\infty)&=\sum_i[C_y\sigma_i^y+C_z\sigma_i^z+C_{xx}\sigma_i^x\sigma_{i+1}^x+C_{yy}\sigma_i^y\sigma_{i+1}^y\\
&\hspace{0.7cm}+C_{zz}\sigma_i^z\sigma_{i+1}^z+C_{yz}(\sigma_i^y\sigma_{i+1}^z+\sigma_i^y\sigma_{i-1}^z)].
\end{aligned}	
\label{trunc_flow}
\end{equation}
The couplings in the range $\omega\in [8,40]$ are well approximated by
\begin{equation}
\begin{aligned}
&C_{xx}^0=\frac{0.99}{\omega }-0.12+0.006 \omega -0.00013 \omega ^2,\\
&C_{yy}^0=\frac{4.87}{\omega }+0.18\,-0.021 \omega  +0.0006 \omega ^2,\\
&C_{zz}^0=-\frac{5.75}{\omega }+0.87+0.022 \omega -0.00067 \omega ^2,\\
&C_{yz}=-\frac{52.07}{\omega ^2}+\frac{7.35}{\omega }+0.1\, -0.0017 \omega, \\
&C_y^0=\frac{78.2}{\omega ^2}-\frac{18.87}{\omega }+0.03\, +0.00013 \omega, \\
&C_z^0=\frac{13.24}{\omega }-0.49+0.0066 \omega  -0.000011 \omega ^2.
\end{aligned}
\label{largeOmegaCoupl}
\end{equation}
Such fitted couplings allow for a semi-analytic understanding in some cases.  One should note that for smaller $\omega$ the expressions become much more complicated because of the non-analytic behavior of the couplings as seen in Fig.\ref{fig:isingmodelszasfunctionofomega}.

\begin{figure}[H]
	\centering
	\includegraphics[width=1\linewidth]{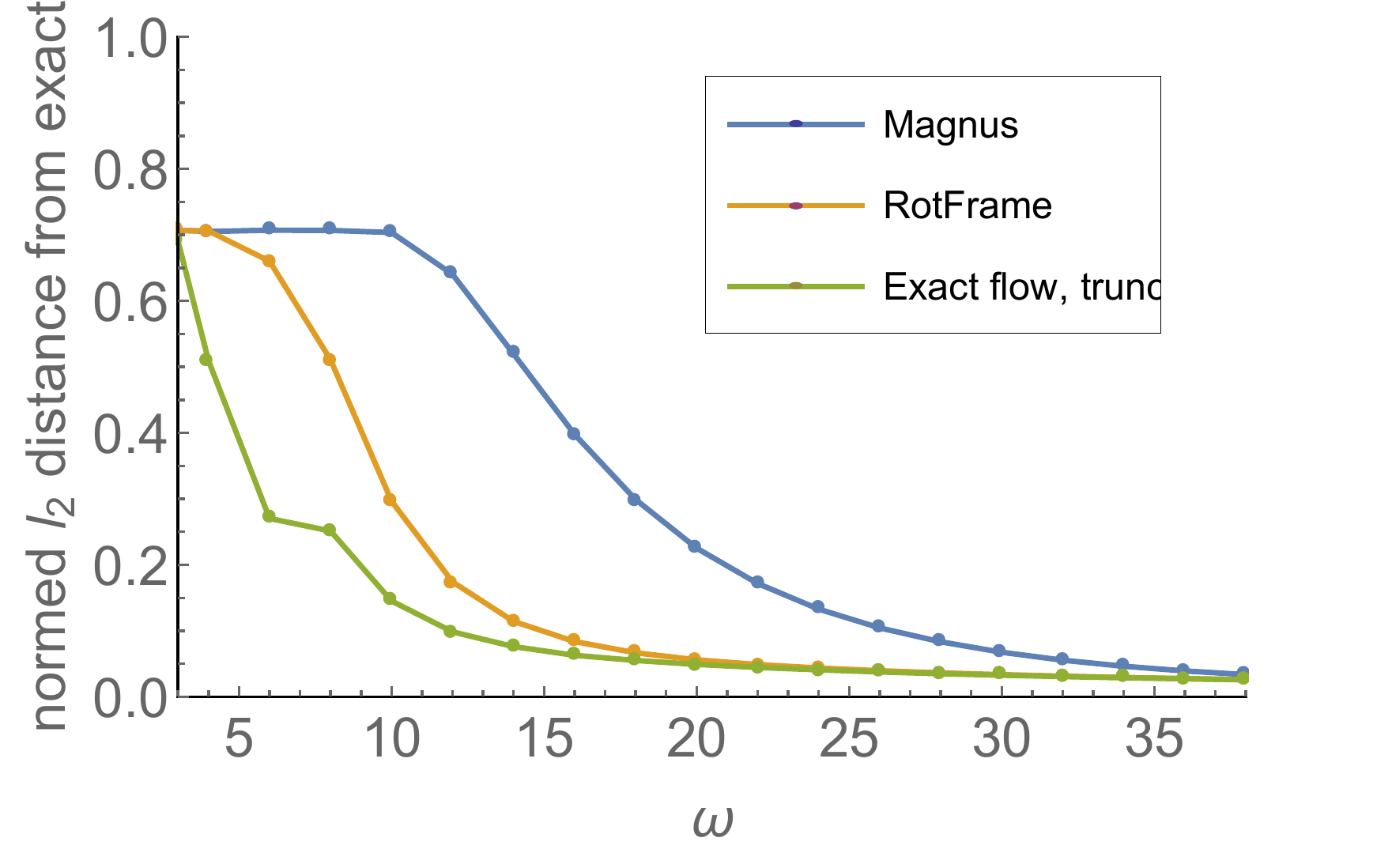}
	\caption{Plot of the $l_2$ distance between the exact time evolution operator and the Magnus expansion (blue), rotating frame approximation (orange) and the solution of the exact flow equations but with a truncated ansatz (green). The system size was $L=14$ sites.}
	\label{fig:exacttruncflowplot}
\end{figure}

Let us show below how well our approximation [also using results for smaller $\omega$ and not just the expression in Eq. \eqref{largeOmegaCoupl}] does when compared to the rotating frame approximation and the Magnus expansion. We do not explicitly give the expressions for the couplings in the Magnus expansion and the rotating frame approximation because they are cumbersome and do not provide much physical insight. Instead, we refer the interested reader Ref.[\onlinecite{2019arXiv190207237V}].  From Fig.\ref{fig:exacttruncflowplot}, one finds that the flow equations (with a truncated ansatz) perform better than both the Magnus expansion and the rotating frame approximation. To stress that the comparison to the rotating frame approximation is a fair one, we note that the operators in Eq.\eqref{trunc_flow} are the same as those appearing within the rotating frame approximation. From this example, one sees that the exact flow equations allowed one to find better coefficients than those afforded by the rotating frame approximation.

   %%%%%%%%%%%%%%%%%%
   % Example Models %
   %%%%%%%%%%%%%%%%%%
   
\section{Example models}
\label{sec:models}
To demonstrate the power and validity of the flow equation approach for a wider range of many-body systems we will next consider a selection of quantum spin chain ($S=\frac{1}{2}$) models. Recall that the spin operators $S_n^{x,y,z}$ fulfill the commutation relations,
\begin{equation}
	\comm{S_{m}^j}{S_{n}^k}=i\epsilon_{jkl}\delta_{mn}S_{m}^l,
\end{equation}
($j,k,l \in \{x,y,z\}$ and $m,n$ label lattice sites) with the special condition for  $S=\frac{1}{2}$ that
\begin{equation}
	(S_n^j)^2=\frac{1}{4}\mathbb{1}_{\mathcal{H}},
\end{equation}
where $\mathbb{1}_{\mathcal{H}}$ is the unit operator in the many-body Hilbert space.  Here $\epsilon_{jkl}$ is the fully antisymmetric tensor and $\delta_{mn}$ is the Kronecker delta function.

In this section we introduce four different spin models that exhibit different functional dependences of the time-dependent term.  The first model (XY spin chain) is integrable, and in particular one-particle reducible.  The next two models are integrability-breaking modifications of the XXZ spin chain, and the final model is a transverse field Ising model which will be discussed independently in Sec.\ref{sec:BCH}.  These models possess a range of different symmetries and form of the driving term. They will illustrate the generality and mathematical structure of the flow equation approach.

\subsection{XY spin chain with antisymmetric exchange in a driven magnetic field}
As a first example model we choose an XY spin-chain with an antisymmetric Dzyaloshinskii-Moriya exchange interaction and a time-periodic magnetic field that both point along the $z$-axis,

\begin{equation}
H(t)=H_0+V(t),
\label{hinit}
\end{equation}
where
\begin{equation}
H_0=\sum_i(J_xS_i^xS_{i+1}^x+J_y S_i^yS_{i+1}^y+D(\vec S_i\times\vec S_{i+1})_z+h_0S_i^z),
\label{hinit10}
\end{equation}
and
\begin{equation}
V(t)=h\sin(\omega t)\sum_i S_i^z.
\end{equation}
Here, $J_{x/y}$ is the strength of the exchange interaction in the $x/y-$direction, $D$ the strength of the antisymmetric exchange, $h_0$ the static magnetic field strength, and $h$ the strength of the magnetic field driving.  This model has the advantage that its instantaneous Hamiltonian can be diagonalized by applying a Jordan-Wigner transformation, followed by a Bogoliubov transformation \cite{Giamarchi2004}. Furthermore, it has multiple coefficients, which can be varied to check the validity of our approximation based on the flow equations in a variety of cases.  Note that the driving term does not generally commute with the static part of the Hamiltonian.
 
 \subsection{\texorpdfstring{$J_1$-$J_2$}{J1-J2}-model with a driven magnetic field in the isotropic plane}
In order to find out if a new approximation scheme is valuable for more realistic interacting systems,
it is important to go beyond non-interacting models.  To this end, we study the $J_1$-$J_2$-model \cite{doi:10.1143/JPSJ.62.1123,PhysRevB.25.4925},
\begin{equation}
H(t)=H_0+V(t),
\label{hinit2}
\end{equation}
where
\begin{equation}
H_0=\sum_{n=1}^2\sum_i (J_n\vect S^\perp_i\vect S^\perp_{i+n}+J_n^z S_i^zS_{i+n}^z),
\end{equation}
and
\begin{equation}
V(t)=\sum_i h(t)S_i^x,
\end{equation}
with a time periodic magnetic field in the $x$-direction 
\begin{equation}
h(t)=B\cdot\begin{cases}
1;\quad  2n\pi<\omega t<2n\pi+\pi\\
-1;\quad  2n\pi+\pi< \omega t<2(n+1)\pi
\end{cases};\quad n\in\mathbb{Z},
\end{equation}
where the time dependence was chosen to simplify the numerical treatment done by exact diagonalization.  None of our physical conclusions---nor our flow equation method---rely on this piecewise constant form of the time-dependence.  It should be noted that $J_n$ is the strength of the $n$-th neighbor exchange interaction in the isotropic plane and  $J_n^z$ is the exchange interaction in $z$-direction. For a more compact notation we defined $\vect S^\perp_i=(S_i^x,S_i^y,0)$. We chose this model because the external magnetic field breaks magnetization conservation and it therefore also allows us to see if the flow equation approach works under circumstances where the driving breaks a symmetry of the static part of the Hamiltonian.

 \subsection{\texorpdfstring{$J_1$-$J_2$}{J1-J2}-model with time-dependent exchange terms}
We also applied the flow equation approach to a model in which one of the spin-spin interaction terms is time-dependent. The model we consider is another $J_1$-$J_2$ model given by,
\begin{equation}
H(t)=H_0+V(t),
\label{hinit3}
\end{equation}
where
\begin{equation}
H_0=\sum_i\left[J_1\vect S_i^\perp \cdot \vect S_{i+1}^\perp+J_1^zS_i^zS_{i+1}^z+J_2\vect S_{i}\cdot \vect S_{i+2}\right],
\end{equation}
and
\begin{equation}
V(t)=J_{1,0}^z\mathrm{sign}\left(\frac{\pi}{\omega}-t\cdot{\rm mod}(2\pi/\omega)\right)\sum_i S_i^zS_{i+1}^z,
\end{equation}
where $\rm mod$ denotes modulo.  In this model, the time-dependence is in an interaction term. 

 In Sec.\ref{sec:BCH}, we will consider one further spin model (Ising model) separately because the structure of the fixed point Hamiltonian is different than the three models introduced in this section.  Together, these four spin models and the example given in Sec.\ref{exact_flow_properties} should provide a compelling picture for the generality and power of our method.

%%%%%%%%%%%
% Results %
%%%%%%%%%%%

\section{Results}
\label{sec:results}
In this section we study how well our flow equation approach performs compared to common high frequency approximations. We compare the approximate time evolution operators obtained through various approximations to the exact time evolution operator (obtained by exact diagonalization) at stroboscopic times. 

We adhere to the following procedure: We first make use of the translational invariance of our models and calculate the exact time evolution operator $U^k_{ex}(T)$ and the approximate time evolution operator $U^k_{approx}(T)$ at different points in $k$-space (momentum space). Then, we calculate the mismatch of the approximate time evolution operator and the exact time evolution operator via, 
\begin{equation}
	E=\frac{1}{2N\sqrt{D_{\rm dim}}}\sum_k\norm{U^k_\textnormal{ex}(T)-U^k_\textnormal{approx}(T)}_\textnormal{Frob},
	\label{rel_error}
\end{equation}
which is a quantity that takes values on the interval $[0,1]$, with zero meaning perfect agreement and one meaning the largest possible disagreement.  Here, $D_{\rm dim}$ is the dimensionality of the Hilbert space for any given $k$-point, $N$ is the number of $k$-points that the sum runs over, and $\norm{A}_\textnormal{Frob}:=\sqrt{\tr{AA^\dag}}$ is the Frobenius norm.  

Let us motivate this quantity: For a given point in $k$-space this is just the $l_2$ distance, Eq.~\eqref{l2distance}, between two unitary operators at this point in $k$-space divided by the maximum $l_2$ distance of two unitary operators. We average this quantity over all points of $k$-space. The Frobenius norm provides us with a basis-independent measure of how accurate unitary evolution of a quantum system will be with various time-independent approximations to the full time-dependent Hamiltonian.  Similar formulas are used in the context of quantum information science.

%%%%%%%%%%%%%%%%%%%%%%%%%%%%%%%%%%%%%%%%%%%%%%
% XY Spin chain with anti-symmetric exchange %
%%%%%%%%%%%%%%%%%%%%%%%%%%%%%%%%%%%%%%%%%%%%%%
\subsection{XY Spin chain with anti-symmetric exchange}
Both the Magnus expansion [see Appendix~\ref{app:Heff}]  and the approximation via flow equations yield an effective Hamiltonian of the form,
\begin{equation}
\label{eq:XY_eff}
\begin{aligned}
H_\mathrm{eff}=\sum_i(&J_x^{(a)}S_i^xS_{i+1}^x+J_y^{(a)} S_i^yS_{i+1}^y+D_+^{(a)}S_i^xS_{i+1}^y\\
&+D_-^{(a)}S_i^yS_{i+1}^x+h_0S_i^z),
\end{aligned}
\end{equation}
where $a$ labels the approximation scheme, with different coupling constants for different approximation schemes. The details of the derivation are given in Appendix \ref{app:Heff}. 

There are newly generated terms in Eq.~\eqref{eq:XY_eff} compared to Eq.~\eqref{hinit10}.  We note that a suitably chosen rotation in spin space gives back the original undriven Hamiltonian with $\Delta J=J_x-J_y$ modified. 

The coupling constants for the leading order Magnus expansion are,
\begin{equation}
J_{x,y}^{(M)}=J_{x,y},\;\;\; D^{(M)}_\pm=\pm D-\frac{(J_x-J_y) h}{\omega},
\end{equation}
and the results for the flow equation approach are,
\begin{eqnarray}
J_x^{(F)}=\frac{J_x+J_y}{2}+\frac{J_x-J_y}{2}  \cos \left(\frac{2 h}{\omega }\right) J_0\left(\frac{2 h}{\omega }\right),\\
J_y^{(F)}=\frac{J_x+J_y}{2}-\frac{J_x-J_y}{2}  \cos \left(\frac{2 h}{\omega }\right) J_0\left(\frac{2 h}{\omega }\right),\\
D_{\pm}^{(F)}=\pm D-\frac{J_x-J_y}{2}\sin \left(\frac{2 h}{\omega }\right) J_0\left(\frac{2 h}{\omega }\right).
\end{eqnarray}

It should be emphasized that both approximations agree in the limit of $\omega\to\infty$ --- a general result mentioned previously at the end of Sec.\ref{sec:formalism}. We also stress that, in the case that $h$ is much larger than all other coefficients, the flow equation approximation works well even when expanded around $2h/\omega\gg1$, which is not what one would normally expect from a high frequency expansion.  The flow equation approach does not make the assumption at any point that $V(t)$ is small, and therefore it handles this regime more accurately.

\begin{figure}[h]
	\centering
	\includegraphics[width=0.4\textwidth]{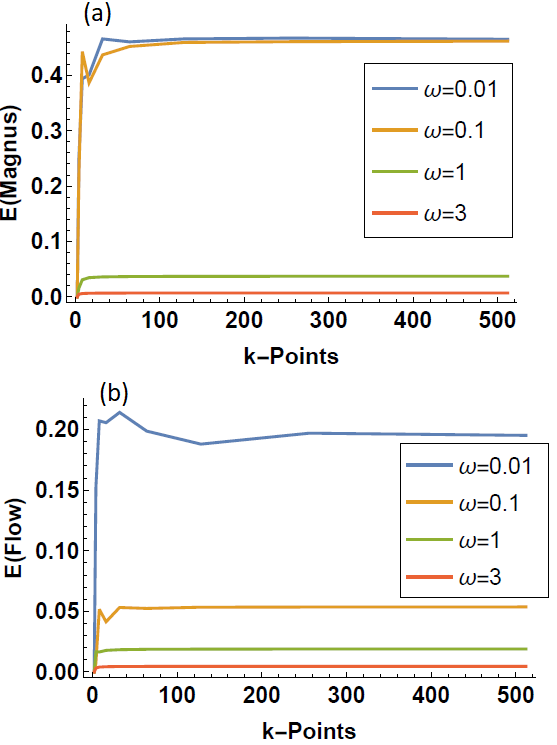}
	\caption{(Color online.) Relative error, Eq.\eqref{rel_error}, of the time evolution operator as a function of sampled $k$-points  for the (a) Magnus expansion and (b) flow equation approach. Different driving frequencies $\omega=0.01$, $\omega=0.1$, $\omega=1$ and $\omega=3$ are considered.  Note that the flow equation error is much smaller than the Magnus expansion error, particularly at the lowest frequencies.  In both approximations, the error decreases as the frequency increases. We consider the case of $D=0.1$, $J_x=1$, $J_y=1.1$, $h_0=1$ and $h=1$}
	\label{fig:Nkpoints}
\end{figure}

We are particularly interested in quantum many-particle systems with a large number of degrees of freedom.  We therefore compute the mismatch
$E$, Eq.\eqref{rel_error}, of the time evolution operators for a long spin chain. We plot the relative error $E$ as a function of the number of $k$-points to find out how many $k$-points are needed for a stable result. (The details on how the time evolution operator was calculated are given in Appendix \ref{time-ev-XY}.) The plot for the Magnus expansion and for the flow equation approach are given in Fig.\ref{fig:Nkpoints}. 
From Fig.~\ref{fig:Nkpoints} one can see that at $256$ $k$-points the value of the relative error $E$ has stabilized. Therefore, for this model all further plots will be done sampling 256 $k$-points.

\begin{figure}[H]
	\centering
	\includegraphics[width=\linewidth]{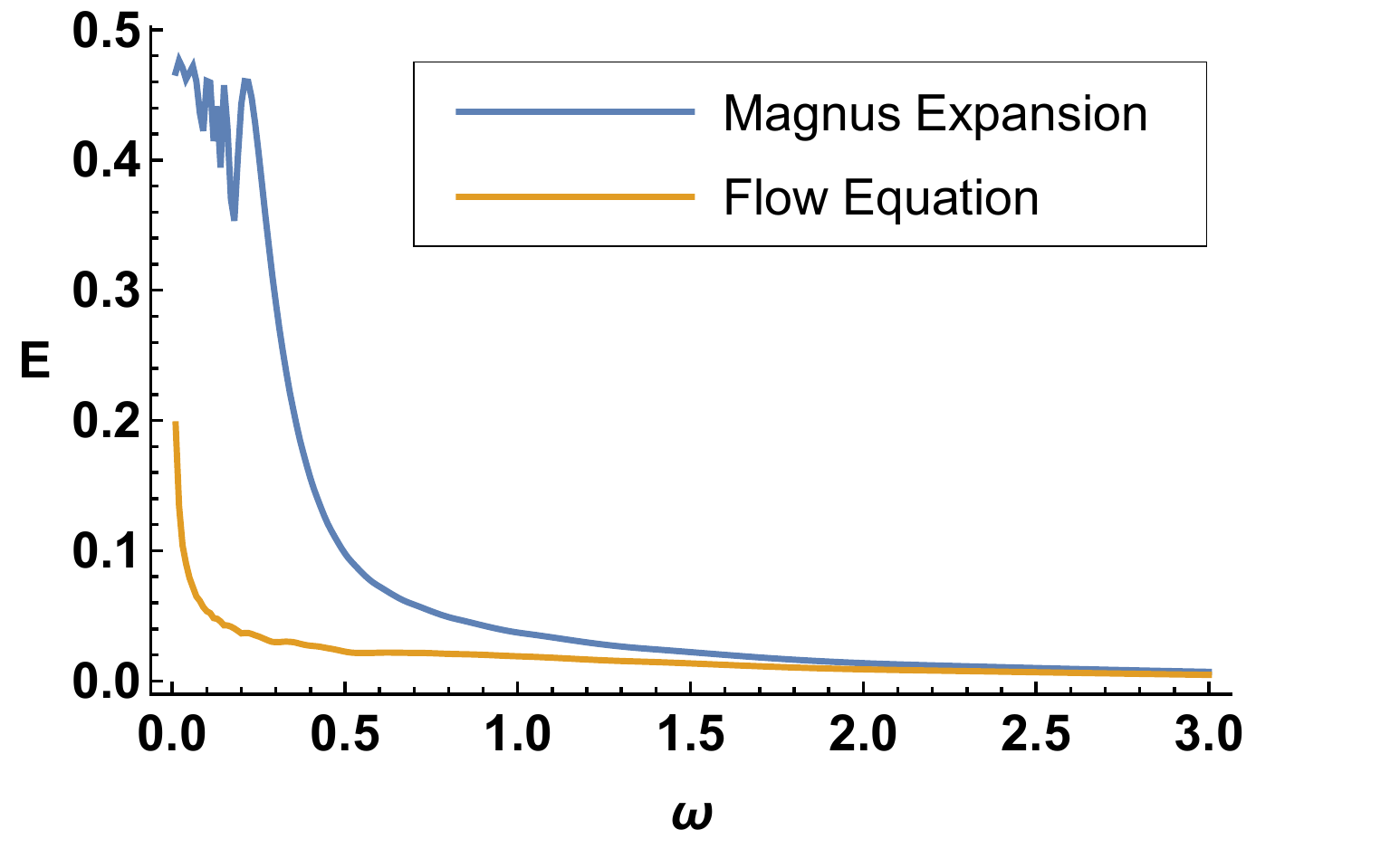}
	\caption{(Color online.) Relative error $E$, Eq.\eqref{rel_error}, for $D=0.1$, $J_x=1$, $J_y=1.1$, $h_0=1$ and $h=1$ as a function of driving frequency $\omega$.  Note how the flow equation approach outperforms the Magnus expansion, particularly at smaller $\omega$.}
	\label{fig:omegaplotXYDzyalosh}
\end{figure}

To study the accuracy of the different approximations as a function of frequency, we choose a set of coefficients $D=0.1$, $J_x=1$, $J_y=1.1$, $h_0=1$, and $h=1$, where $J_x$ was fixed at unity because one may divide the Hamiltonian by $J_x$ to make it dimensionless. The strength of $D$ was chosen to be small since often the anti-symmetric exchange is small when compared to the exchange interactions. The other values were chosen to to be in a similar range. The plot of the relative error $E$ as a function of frequency $\omega$ is given in Fig.~\ref{fig:omegaplotXYDzyalosh}.

From Fig.~\ref{fig:omegaplotXYDzyalosh} one can see that the results from the flow equation approach are valid down to much lower frequencies $\omega$. In fact, one can expect higher order Magnus expansions to become worse at lower frequencies than the first order Magnus expansion we plotted. This is because the optimal cut-off order of the Magnus expansion (and a number of other high frequency expansions) shrinks with decreasing frequencies \cite{PhysRevB.95.014112} unless couplings are small enough to suppress this effect. It should also be noted that the stuttering (wiggles) at low frequencies seen in the plot is an effect that happens because the $U_k$ matrices are relatively small. For larger matrices this averages out as we will see in interacting models to follow.

\begin{figure*}[htb]
\includegraphics[width=\textwidth]{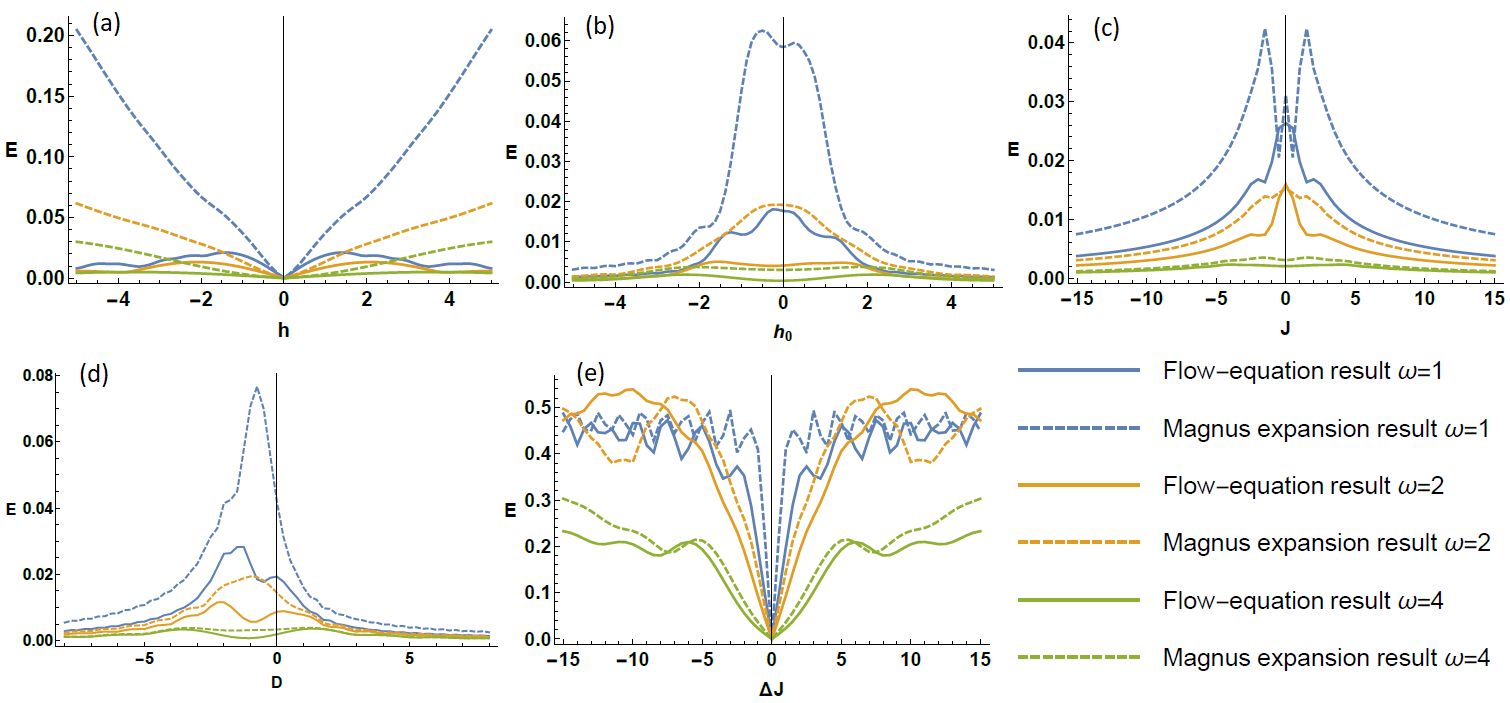}
	\caption{(Color online.) Relative error $E$, Eq.\eqref{rel_error}, as function of the different coupling constants in the Hamiltonian, Eq.\eqref{hinit}. Only one coefficient is varied in each subfigure, while the ones that are not varied are fixed at values of  $D=0.1$ , $\Delta J=J_x-J_y=0.1$, $J=\frac{J_x+J_y}{2}=1.05$,  $h_0=1$ and  $h=1$. We  vary in (a) the driving magnetic field $h$, (b) the static magnetic field $h_0$, (c) the average exchange interaction $J=\frac{J_x+J_y}{2}$, (d) the anti-symmetric exchange strength $D$ and (e) the anisotropy of the exchange interaction $\Delta J=J_x-J_y$.}
	\label{fig:XYErrors}
\end{figure*}
In Fig. \ref{fig:XYErrors} we show how well the approximation does as a function of various couplings.  From the plots it is clear that the results obtained via the flow equation approach are generally more accurate than the results from the Magnus expansion.  As expected from general arguments, we find that the approximation does increasingly well for large values of driving $h$.  We now turn to non-integrable models.

%%%%%%%%%%%%%%%%%%%%%%%%%%%%%%%%%%%%%%%%%%%%%%
% $J_1-J_2$ model with time dependent magnetic field in $x$-direction %
%%%%%%%%%%%%%%%%%%%%%%%%%%%%%%%%%%%%%%%%%%%%%%

\subsection{\texorpdfstring{$J_1$-$J_2$}{J1-J2} model with time-dependent magnetic field in the $x$-direction}
For this model both the Magnus expansion and the flow equation approach yield effective Hamiltonians of the form (for a general model, the terms--quantum operators--appearing in the effective Hamiltonians need not be the same),
\begin{align}
	H_\mathrm{eff}	&=	\sum_{n=1}^2 \sum_i\Bigl( J_nS_i^xS_{i+1}^x+J_n^{y,(a)} S_i^yS_{i+n}^y \nonumber\\
			&+	J_n^{z,(a)}S_i^zS_{i+n}^z +\Gamma_n^{(a)}(S_i^zS_{i+n}^y+S_i^yS_{i+n}^z) \bigr),	\label{heff2}
\end{align}
where $(a)$ labels the approximation scheme (either flow or Magnus). The details of the calculation are given in Appendix \ref{app:Heff}.

It is important to note that one of the new terms, $\Gamma_n$, can be removed by a suitable rotation in spin space, which tells us that we went from an XXZ model to a XYZ model followed by a rotation in spin space.  The effective coefficients for the Magnus expansion are
\begin{equation}
	\begin{aligned}
	&J_n^{y,M}=J_n,\\
	&J_n^{z,M}= J_n^z,\\
	&\Gamma_n^{(M)}=B\pi\frac{J_n^z-J_n}{2\omega},
	\end{aligned}
\end{equation}
and for the flow equation approach
\begin{equation}
\begin{aligned}
&J_n^{y,M}=\frac{1}{2} \left((J_n^z+J_n) -(J_n^z -J_n )  \mathrm{sinc} \left(\frac{2 \pi  B}{\omega }\right) \right),\\
&J_n^{z,M}=\frac{1}{2}\left((J_n^z +J_n )+(J_n^z -J_n )   \mathrm{sinc} \left(\frac{2 \pi  B}{\omega }\right) \right),\\
&\Gamma_n^{(M)}=\frac{(J_n^z -J_n)  \omega  \sin ^2\left(\frac{\pi  B}{\omega }\right)}{2 \pi  B}.
\end{aligned}
\end{equation}

\begin{figure}[!htbp]
	\centering
		\includegraphics[width=0.5\textwidth]{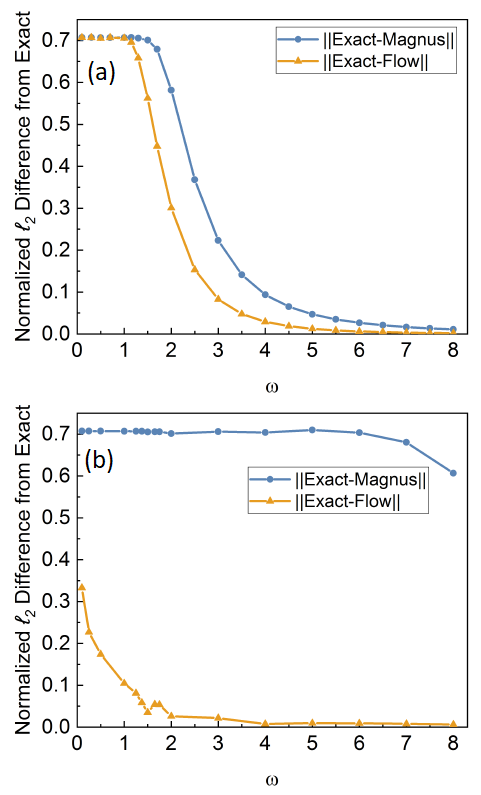}
	\caption{(Color online.) Plot of the normalized $l_2$ distance between exact and approximate time evolution operators for a chain of length $L=16$, and parameters $J_1=-0.5$, $J^z_1=1$ and $J_2=J_2^z=0$ plotted as a function of frequency for driving magnetic field strengths (a) B=0.5 and (b) B=5.}
	\label{fig:J1J2MagFrequ}
\end{figure}

Calculating higher orders in the Magnus expansion for this model yields extremely complicated effective Hamiltonians. The second order Magnus expansion already gives a Hamiltonian that is a sum of 60 different operators with complicated prefactors. One tractable way to improve on the first order Magnus expansion is via the flow equation approach. The plots in Fig.~\ref{fig:J1J2MagFrequ}  illustrate the quality of the approximation for different frequencies. These results are obtained numerically using exact diagonalization for finite size systems, as described in Appendix \ref{app:ED}.

One finds that the flow equations outperform the Magnus expansion for all frequencies. For the plot of strong driving magnetic field $h$ this is especially pronounced. There, the Magnus expansion for a large range of frequencies gives poor results and the flow equations generally give quite precise results.

\begin{figure*}
	\centering
		\includegraphics[width=\textwidth]{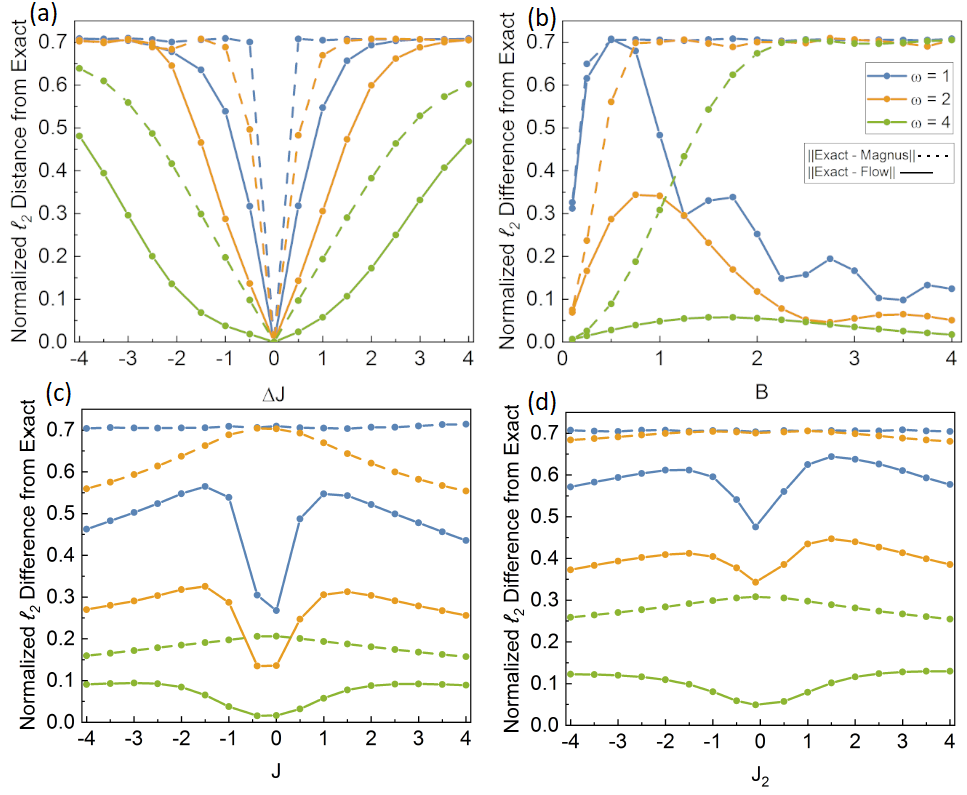}
	\caption{(Color online.) Plots of the normalized $l_2$ distance for a chain of length $L=14$ where we vary the various coupling constants while keeping others fixed. Particularily we are (a) varying nearest neighbor exchange anisotropy $\Delta J=J_1-J_1^z $, while keeping $B=1$, average nearest neighbor exchange $J=\frac{J_1+J_1^z}{2}=1$ and $J_2=J_2^z=0$ fixed, (b) varying $B$, while keeping $J_1=-0.5$, $J_1^z=1$ and $J_2=J_2^z=0$ fixed, (c) varying $J$, while keeping $\Delta J=1$, $B=1$ and $J_2=0$ fixed and (d) varying $J_2$, while keeping $\Delta J=1$, $B=1$ and $J=1$ fixed.}
	\label{fig:J1J2MagPARAMETERS}
\end{figure*}

One may also in this case ask how well the approximation does as a function of all the different coefficients. In Fig.~\ref{fig:J1J2MagPARAMETERS}, we show a plot for different values of the coefficients. These plots were done only including the sector $k=0$ in $k$-space because this is numerically quicker and because other points in $k$-space reproduce the same results.

Similar to the previous integrable model, for this non-integrable model one can see that the flow equation approach outperforms the Magnus expansion for most parameters. The much higher accuracy for large values of $h$ should be emphasized.  The details on how the time evolution operator was obtained are contained in Appendix \ref{app:ED}.
 
 %%%%%%%%%%%%%%%%%%%%%%%%%%%%%%%%%%%%%%%%%%%%%%%%%%%%%%%%%%%
 %$J_1-J_2$ model with time dependent exchange interaction %
 %%%%%%%%%%%%%%%%%%%%%%%%%%%%%%%%%%%%%%%%%%%%%%%%%%%%%%%%%%%

\subsection{\texorpdfstring{$J_1$-$J_2$}{J1-J2} model with time-dependent exchange interaction}
Both the Magnus expansion and the flow equations yield an effective Hamiltonian of the form (some of the terms are zero for the Magnus case),
\begin{eqnarray}
H_\mathrm{eff}=\sum_i\sum_{n=1}^2\Big\{J_n^{(a)}\vect S_i^\perp \vect S_{i+n}^\perp+J_{z,n}^{(a)}S_i^zS_{i+n}^z \nonumber \\
+D_n^{(a)}\left[\vect S_{i+n+1}\times \vect S_{i+1}+\vect S_{i-n-1}\times \vect S_{i-1}\right]_zS_i^z\nonumber \\
+Q_n^{(a)}\left[S_i^xS_{i+n}^x+S_i^yS_{i+n}^y\right]S_{i-1}^zS_{i+n+1}^z\Big\},
\label{heff3}
\end{eqnarray}
where $\vect S_i^\perp=(S_i^x,S_i^y,0)$ and $(a)$ labels the approximation scheme. 

The last two terms of Eq.\eqref{heff3} are newly generated terms in the Hamiltonian. If $S_i^z$ has an approximately uniform orientation the terms proportional to $D_n^{(a)}$ can be interpreted as different range antisymmetric exchange terms - treating $S_i^z$ as a mean-field term. By the same token, in a mean field approximation the term proportional to $Q_n^{(a)}$ can be interpreted as exchange terms.  Beyond the mean-field case it is clear that higher order spin interactions are generated.  Such terms can lead to new physics and can drive new phases.

The coupling constants within the flow equation approach [solving Eq.\eqref{approx_floweq} exactly] are given by,
\begin{equation}
	\begin{aligned}
	&J_n^F=\frac{1}{2} J_n \left[1+\text{sinc}\left(\frac{\pi  J_{1,o}^z}{\omega }\right)\right],\\
	&J_{z,1}^F=J_1^z;\quad J_{z,2}^F=J_2,\\
	&D_n^F=\frac{J_n \omega }{ \pi  J_{1,o}^z}\left[\cos \left(\frac{\pi  J_{1,o}^z}{\omega }\right)-1\right],\\
	&Q_n^F=2 J_n \left[1-\text{sinc}\left(\frac{\pi  J_{1,o}^z}{\omega }\right)\right],
	\end{aligned}
\end{equation}
and within the Magnus expansion,
\begin{equation}
	\begin{aligned}
	&J_n^M=J_n;\quad J_{z,1}^F=J_1^z;\quad J_{z,2}^M=J_2,\\
	&D_n^M=-\frac{\pi}{2\omega}J_nJ_{1,o}^z;\quad Q_n^M=0.
	\end{aligned}
\end{equation}

While the form of the Hamiltonian in Eq.\eqref{heff3} is already complicated (with three and four-spin interactions) it is worth noting that the second order Magnus expansion would become forbiddingly complicated with a sum of over 100 operators, which makes even a numerical implementation impractical. Therefore, the result from the flow equations, while also complicated, is a significant improvement on the first order Magnus expansion.

\begin{figure*}[ht]
	\centering
		\includegraphics[width=\textwidth]{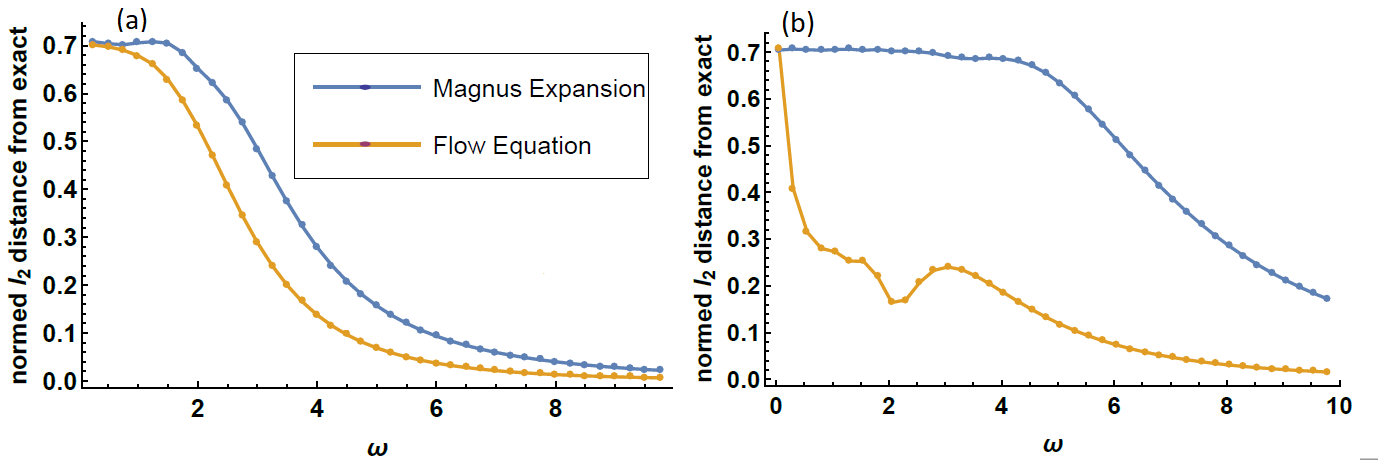}
	\caption{(Color online.) Plot of the normalized $l_2$ distance between exact and approximate time evolution operators for $J_1=1$, $J^z_1=2$ and $J_2=0.2$ plotted as a function of frequency for a spin chain with $L=14$ sites. The driving strength of the nearest neighbor exchange term in z-direction is (a) $J_{1,o}^z=2$ and (b) $J_{1,o}^z=6$. }
	\label{fig:J1J2J1zdrivenMagFrequ}
\end{figure*}

\begin{figure*}[!htbp]
	\centering
		\includegraphics[width=\textwidth]{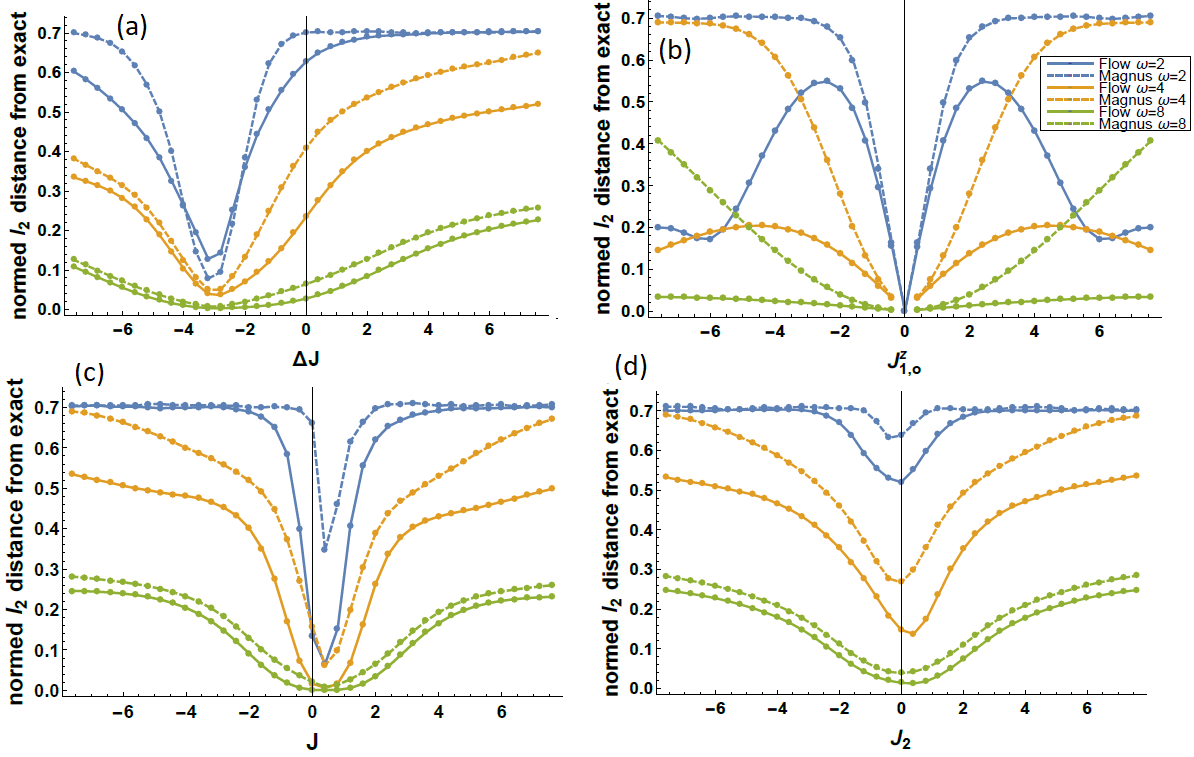}
	\caption{(Color online.)  Shown are plots of the $l_2$ distance as a function of various coupling constants for a chain of length $L=14$. In plot (a) we vary the nearest neighbor exchange anisotropy $\Delta J=J_1-J_1^z$ and keep $J_{1,o}^z=2$, $J=\frac{1}{2}(J_1+J_1^z)=1.5$ and $J_2=0.2$ fixed, in (b) we vary the driving strength of the exchange interaction in z-direction $J_{1,o}^z$ and keep $\Delta J=-1$, $J=1.5$ and $J_2=0.2$ fixed, in (c) we vary the average nearest neighbor exchange interaction $J$ and keep $J_{1,o}^z=2$,  $\Delta J=-1$ and $J_2=0.2$ fixed and in (d) we vary the next nearest neighbor exchange $J_2$ while keeping $J_{1,o}^z=2$,  $\Delta J=-1$ and $J=1.5$ fixed.}
	\label{fig:J1J2J1zPARAMETERS}
\end{figure*}

In Fig.\ref{fig:J1J2J1zdrivenMagFrequ} we plot the frequency dependence of the approximation.  One finds that the flow equation result is much better in the lower frequency regime and outperforms the Magnus approximation significantly when the external drive is relatively strong.  The performance of the two approximations as a function of the different couplings is shown in the plots in Fig.\ref{fig:J1J2J1zPARAMETERS}.  Consistent with the models previously discussed, the flow equation approximation does substantially better across all parameter regimes.  For this case we made use of the QuSpin package \cite{SciPostPhys.2.1.003} to obtain a comparison to the exact result.

%%%%%%%%%%%%%%%%%%%%%%%%%%%%%%%%%%%%%%%%%%%%%%%%%%%%%%
%Comparison to Replica Resummation of BCH id         %
%%%%%%%%%%%%%%%%%%%%%%%%%%%%%%%%%%%%%%%%%%%%%%%%%%%%%%

\section{Comparison with resummations of the Baker-Campbell-Hausdorff identity}
\label{sec:BCH}
In this section, we turn the logic around relative to the conventional Hamiltonian--evolution operator relationship.  Up to this point in the manuscript, we have been asking about computing an effective time-independent Hamiltonian for a time-dependent problem, and we have used this effective Hamiltonian to compute the time evolution of the system.  Now, we turn our attention to a situation in which the time evolution operator is known (in our case it takes a specific product form) and we wish to determine an {\em optimal Hamiltonian} that can be used to produce the desired time evolution.  This may be useful in certain quantum computing applications, for example.

 A second goal of this section is to show that our method has advantages over the rotating frame approximation in that one can capture most of its features by a truncated ansatz even when an exact rotating frame approximation cannot be calculated because the effective Hamiltonian would include infinitely many long range interacting terms.  This highlights an another important dimension to our flow equation approach, beyond the examples illustrating its use in earlier sections of the manuscript. 

There has been a recent surge of interest in resummations of the Baker-Campbell-Haussdorff (BCH) identity \cite{PhysRevLett.120.200607}. An important evolution case where the BCH identity is useful is when the time evolution operator factorizes into a product of matrix exponentials $e^{-iH_1 t}e^{-iH_2 t}$.  This structure corresponds to multiple different Schr\"odinger equations. One possible correspondence is to a delta function time dependence in the Schr\"odinger equation.  For example, the kicked transverse field Ising model that is discussed in Ref.~\cite{PhysRevLett.120.200607} has,
 \begin{equation}
	\begin{aligned}
	&H_1=J\sum_i\sigma_i^z\sigma_{i+1}^z,\\
	&H_2=\sum_i(h_x\sigma_x^i+h_z\sigma_i^z),
	\end{aligned}
\end{equation}
and can be put into the form,
\begin{equation}
	\begin{aligned}
	&H(t)=H_0+V(t),\\
	&H_0=\sum_i\left[J_{z}\sigma_i^z\sigma_{i+1}^z+h_x\sigma_i^x+h_z\sigma_i^z\right],\\
	&V(t)=\sum_i\left[h_x\sigma_i^x+h_z\sigma_i^z\right](\delta(t)-1),
	\end{aligned}
\label{transverseIsingdelta}
\end{equation}
where to stay close to the notation of Ref.~[\onlinecite{PhysRevLett.120.200607}] we use Pauli operators $\sigma_i^{x,y,z}=2S_i^{x,y,z}$ rather than the spin operators we used earlier in our manuscript. Here $\delta(t)$ is the Dirac delta function.  

Another possibility is to rewrite the problem in terms of a Heaviside $\theta$ function as,
\begin{equation}
	\begin{aligned}
	&H(t)=H_0+V(t),\\
	&H_0=\sum_i\left[J_{z}\sigma_i^z\sigma_{i+1}^z+h_x\sigma_i^x+h_z\sigma_i^z\right],\\
	&V(t)=\sum_i\left[h_x\sigma_i^x+h_z\sigma_i^z-J_{z}\sigma_i^z\sigma_{i+1}^z\right]\left(2\theta(t-1/2)-1\right).
	\end{aligned}
		\label{transverseIsingsign}
\end{equation}
Both choices lead to different flow equations and can therefore be interpreted as leading to different resummations of the BCH identity.  Thus, we discuss here these two Hamiltonian choices for a given time evolution operator.  As a matter of fact, there are infinitely many ways to make a choice in the time dependence, and likely one is an ideal choice.  However, we will not discuss this issue of the optimal choice any further. 
An important difference between the two formulations is that 
%An important feature of both formulations is that 
the flow equations in one case can be solved exactly and in the other case require truncation. This allows us to assess how useful our method is in a case where a rotating frame approximation cannot be calculated exactly.  This example helps to illustrate the point that even when the flow equations are not solved exactly, they still give results beyond the Magnus expansion.

One finds that within the lowest order in the BCH expansion, the replica approximation used in Ref.~[\onlinecite{PhysRevLett.120.200607}] and our flow equation approach lead to an approximate Floquet Hamiltonian of the form, 
\begin{align}
H_\mathrm{eff}^{(a)}&=\sum_i\Bigg[ C_x^{(a)}\sigma_i^x+C_z^{(a)}\sigma_i^z +C_{xx}^{(a)}\sigma_i^x\sigma_{i+1}^x\nonumber\\
&+C_{yy}^{(a)}\sigma_i^y\sigma_{i+1}^y+C_{zz}^{(a)}\sigma_i^z\sigma_{i+1}^z\nonumber\\
&+C_{xy}^{(a)}\sigma_i^x(\sigma_{i-1}^y+\sigma_{i+1}^y)+C_{xz}^{(a)}\sigma_i^x(\sigma_{i-1}^z+\sigma_{i+1}^z)\nonumber\\
&+C_{yz}^{(a)}\sigma_i^y(\sigma_{i-1}^z+\sigma_{i+1}^z)+C_{xzz}^{(a)}\sigma_{i}^x\sigma_{i-1}^z\sigma_{i+1}^z\Bigg],\label{transverseIsingeffham}
\end{align}
where $a$ labels the approximation scheme. The different approximations only differ in their coefficients (and some coefficients may be zero). The coefficients themselves offer little to illuminate our discussion.  Therefore, their derivation is given in Appendix \ref{appBCHresum}. 

In Fig.~\ref{fig:delta} we show a comparative plot for the $\delta$-type and the Heaviside-type resummations. The plots are done for spin chains of length $L=14$ to get a smooth plot. There are only small numerical differences for longer spin chains. 

\begin{figure*}
	\centering
	\includegraphics[width=\textwidth]{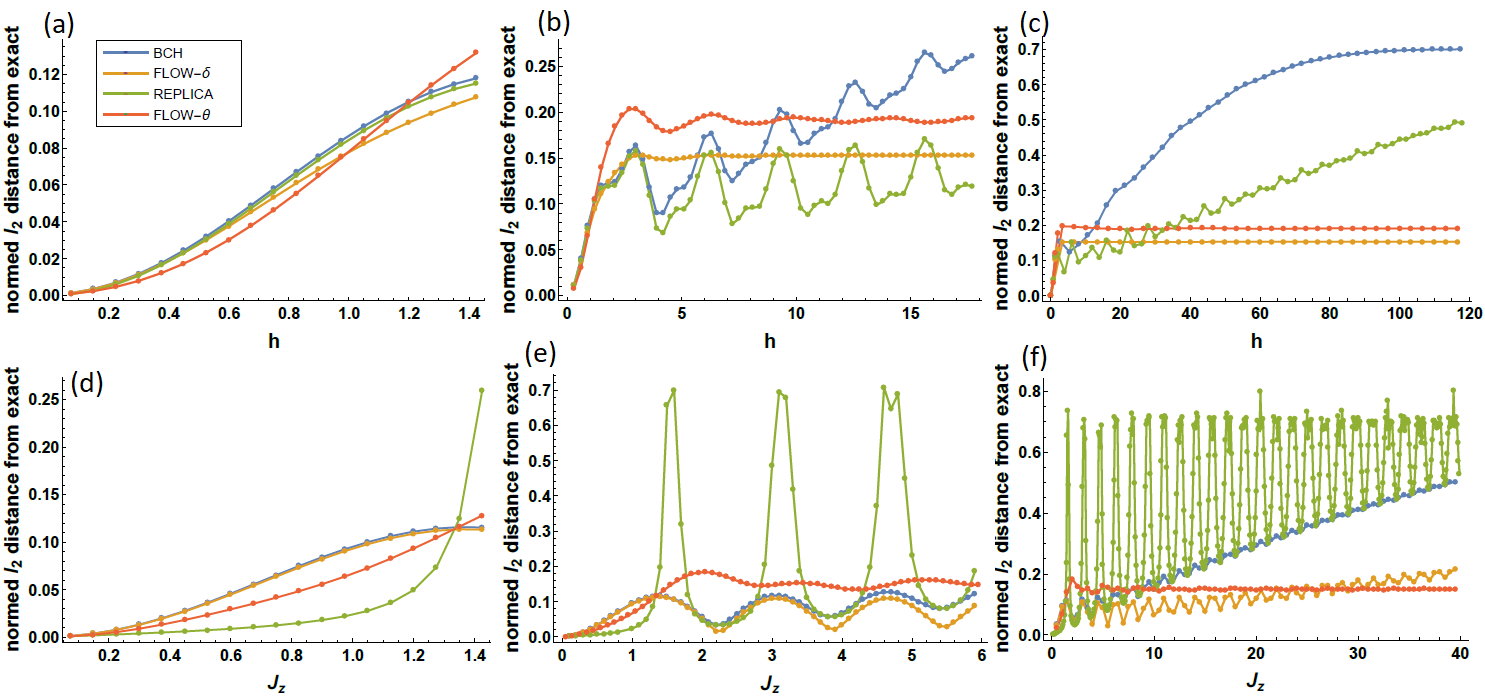}
	\caption{(Color online.)  Plot of the normalized $l_2$ distance between the exact time evolution operator and the approximate time evolution operator for $h_x=h\cos\theta$, $h_z=h\sin\theta $, and $\theta=0.643501$ using the ``$\delta$-formulation" of the flow equations in orange and the ``Heaviside $\theta$-formulation" in red. In plots (a-c) we keep $J_z=0.1$ fixed and plot different ranges of $h$ values, which are (a) short range $h$, (b) medium range $h$ and (c) long range $h$. Similarily in plots (d-f) we keep $h=0.1$ fixed and plot different ranges of $J_z$ values, which are (d) short range $J_z$, (e) medium range $J_z$ and (f) long range $J_z$}
	\label{fig:delta}
\end{figure*}

In the plots one can see that the flow equation approach Eq.\eqref{approx_floweq} does better for small values of coupling strength than the Magnus expansion --- in some cases also better than the replica expansion. For large couplings, it outperforms both.  

From Fig.~\ref{fig:delta}, one can see that the flow equation approach is the most reliable approximation with the mismatch in some cases plateauing at values of around 0.1. For those values one is still able to capture at least qualitative features of the time-evolution. Thus, the flow equation approach offers a useful numerical strategy for finding a Hamiltonian describing a given time-evolution.  This may be of practical importance in a wide variety of applications where %the underlying Hamiltonian may be hard 
it is difficult to determine the underlying Hamiltonian 
%to determine 
from microscopic considerations, such as may be the case in various types of quantum information scenarios. 

We would also like to stress that for the step-wise drive the exact rotating frame transformation was not possible to calculate and therefore a truncated ansatz for the Hamiltonian had to be employed to solve Eq.\eqref{approx_floweq}. One can see that this truncated ansatz performs well (red curve). It should be stressed that the truncated ansatz performed similar to the case where an exact rotating frame approximation was possible. Our method therefore allows one to capture properties of a rotating frame approximation even when calculating a rotating frame approximation exactly is not possible.

%One should also note that while the two different flow equation results are comparable for most parameter values,  there should be an optimal choice and a procedure of determining that that choice maybe.  However, it is beyond the scope of the paper to determine such a protocol.

%%%%%%%%%%%%%
%Conclusion %
%%%%%%%%%%%%%
\section{Conclusions}
\label{sec:conclusions}
In conclusion, we have introduced an accurate ``flow equation" approach to compute effective time-independent Hamiltonians, valid for finite times (which may be exponentially long) for periodically driven quantum many-particle systems.  We have demonstrated the power of the flow equation approach by illustrating how one can reach into perturbatively inaccessible frequency regimes, and shown that the approximation generally yields an improvement over the Magnus expansion, and that it can also outperform the rotating frame approximation. Furthermore, in many instances the results from the flow equation approach also yield a practically accessible improvement on the first order Magnus expansion where no other method appears to be available.  A straightforward application of the Magnus expansion leads to an explosion in the number of different operators that contribute to the effective Hamiltonian with coefficients that are tedious to evaluate.  In our approach, one is able to truncate the number of operators contributing to the flow equations in a controlled way, which allows one to keep fewer terms but find highly accurate coefficients.  We have also demonstrated that our method compares favorably to resummations of the Baker-Campbell-Hausdorff identity, illustrating it shows its strength even in niche applications, where more powerful methods are to be expected. Our approach also has a wider range of applicability than standard rotating frame approximations because, even if a rotating frame approximation is impractical or not possible because the matrix exponential or the rotation of operators induced by it cannot be calculated, our method allows for a truncated ansatz that may still capture the important features of the transformation.

In summary, we hope that the demonstration of the validity of our approximate method illustrates its power and potential impact on time-dependent quantum many-body systems. The method is completely general and applicable to any form of time-dependent terms in the Hamiltonian--be it through the potential energy, kinetic energy, or both.  With the accurate effective time-independent Hamiltonians that one obtains, new access is granted to potential prethermal regimes with properties not present in the equilibrium phase diagram of the original Hamiltonian.  Our results also open the door to new opportunities for quantum control through Hamiltonian engineering to create desired properties out-of-equilibrium.   The effective Hamiltonian can be used to compute any observable over finite times through the standard formulas of statistical mechanics, in addition to accurately governing the evolution of the quantum states themselves.  We hope our approach will inspire new studies that exploit its flexibility and expand the range of approximation schemes that can be employed within it.  With it, new regimes of cold atom, condensed matter, and other systems will likely be uncovered and manipulated in new ways.

\acknowledgments
We gratefully acknowledge funding from Army Research Office Grant No. W911NF-14-1-0579, NSF Grant No. DMR-1507621, and NSF Materials Research Science and Engineering Center Grant No. DMR-1720595.   We acknowledge the Texas Advanced Computing Center (TACC) at The University of Texas at Austin for providing computing resources that have contributed to the research results reported within this paper. \url{www.tacc.utexas.edu} GAF acknowledges support from a Simons Fellowship.

%\bibliography{Floquet,literature}

\begin{thebibliography}{100}%
\makeatletter
\providecommand \@ifxundefined [1]{%
 \@ifx{#1\undefined}
}%
\providecommand \@ifnum [1]{%
 \ifnum #1\expandafter \@firstoftwo
 \else \expandafter \@secondoftwo
 \fi
}%
\providecommand \@ifx [1]{%
 \ifx #1\expandafter \@firstoftwo
 \else \expandafter \@secondoftwo
 \fi
}%
\providecommand \natexlab [1]{#1}%
\providecommand \enquote  [1]{``#1''}%
\providecommand \bibnamefont  [1]{#1}%
\providecommand \bibfnamefont [1]{#1}%
\providecommand \citenamefont [1]{#1}%
\providecommand \href@noop [0]{\@secondoftwo}%
\providecommand \href [0]{\begingroup \@sanitize@url \@href}%
\providecommand \@href[1]{\@@startlink{#1}\@@href}%
\providecommand \@@href[1]{\endgroup#1\@@endlink}%
\providecommand \@sanitize@url [0]{\catcode `\\12\catcode `\$12\catcode
  `\&12\catcode `\#12\catcode `\^12\catcode `\_12\catcode `\%12\relax}%
\providecommand \@@startlink[1]{}%
\providecommand \@@endlink[0]{}%
\providecommand \url  [0]{\begingroup\@sanitize@url \@url }%
\providecommand \@url [1]{\endgroup\@href {#1}{\urlprefix }}%
\providecommand \urlprefix  [0]{URL }%
\providecommand \Eprint [0]{\href }%
\providecommand \doibase [0]{http://dx.doi.org/}%
\providecommand \selectlanguage [0]{\@gobble}%
\providecommand \bibinfo  [0]{\@secondoftwo}%
\providecommand \bibfield  [0]{\@secondoftwo}%
\providecommand \translation [1]{[#1]}%
\providecommand \BibitemOpen [0]{}%
\providecommand \bibitemStop [0]{}%
\providecommand \bibitemNoStop [0]{.\EOS\space}%
\providecommand \EOS [0]{\spacefactor3000\relax}%
\providecommand \BibitemShut  [1]{\csname bibitem#1\endcsname}%
\let\auto@bib@innerbib\@empty
%</preamble>
\bibitem [{\citenamefont {Polkovnikov}\ \emph {et~al.}(2011)\citenamefont
  {Polkovnikov}, \citenamefont {Sengupta}, \citenamefont {Silva},\ and\
  \citenamefont {Vengalattore}}]{RevModPhys.83.863}%
  \BibitemOpen
  \bibfield  {author} {\bibinfo {author} {\bibfnamefont {Anatoli}\ \bibnamefont
  {Polkovnikov}}, \bibinfo {author} {\bibfnamefont {Krishnendu}\ \bibnamefont
  {Sengupta}}, \bibinfo {author} {\bibfnamefont {Alessandro}\ \bibnamefont
  {Silva}}, \ and\ \bibinfo {author} {\bibfnamefont {Mukund}\ \bibnamefont
  {Vengalattore}},\ }\bibfield  {title} {\enquote {\bibinfo {title}
  {Colloquium: Nonequilibrium dynamics of closed interacting quantum
  systems},}\ }\href {\doibase 10.1103/RevModPhys.83.863} {\bibfield  {journal}
  {\bibinfo  {journal} {Rev. Mod. Phys.}\ }\textbf {\bibinfo {volume} {83}},\
  \bibinfo {pages} {863--883} (\bibinfo {year} {2011})}\BibitemShut {NoStop}%
\bibitem [{\citenamefont {Eckardt}(2017)}]{RevModPhys.89.011004}%
  \BibitemOpen
  \bibfield  {author} {\bibinfo {author} {\bibfnamefont {Andr\'e}\ \bibnamefont
  {Eckardt}},\ }\bibfield  {title} {\enquote {\bibinfo {title} {Colloquium:
  Atomic quantum gases in periodically driven optical lattices},}\ }\href
  {\doibase 10.1103/RevModPhys.89.011004} {\bibfield  {journal} {\bibinfo
  {journal} {Rev. Mod. Phys.}\ }\textbf {\bibinfo {volume} {89}},\ \bibinfo
  {pages} {011004} (\bibinfo {year} {2017})}\BibitemShut {NoStop}%
\bibitem [{\citenamefont {Bloch}\ \emph {et~al.}(2008)\citenamefont {Bloch},
  \citenamefont {Dalibard},\ and\ \citenamefont {Zwerger}}]{RevModPhys.80.885}%
  \BibitemOpen
  \bibfield  {author} {\bibinfo {author} {\bibfnamefont {Immanuel}\
  \bibnamefont {Bloch}}, \bibinfo {author} {\bibfnamefont {Jean}\ \bibnamefont
  {Dalibard}}, \ and\ \bibinfo {author} {\bibfnamefont {Wilhelm}\ \bibnamefont
  {Zwerger}},\ }\bibfield  {title} {\enquote {\bibinfo {title} {Many-body
  physics with ultracold gases},}\ }\href {\doibase 10.1103/RevModPhys.80.885}
  {\bibfield  {journal} {\bibinfo  {journal} {Rev. Mod. Phys.}\ }\textbf
  {\bibinfo {volume} {80}},\ \bibinfo {pages} {885--964} (\bibinfo {year}
  {2008})}\BibitemShut {NoStop}%
\bibitem [{\citenamefont {Dalibard}\ \emph {et~al.}(2011)\citenamefont
  {Dalibard}, \citenamefont {Gerbier}, \citenamefont
  {Juzeli\ifmmode~\bar{u}\else \={u}\fi{}nas},\ and\ \citenamefont
  {\"Ohberg}}]{RevModPhys.83.1523}%
  \BibitemOpen
  \bibfield  {author} {\bibinfo {author} {\bibfnamefont {Jean}\ \bibnamefont
  {Dalibard}}, \bibinfo {author} {\bibfnamefont {Fabrice}\ \bibnamefont
  {Gerbier}}, \bibinfo {author} {\bibfnamefont {Gediminas}\ \bibnamefont
  {Juzeli\ifmmode~\bar{u}\else \={u}\fi{}nas}}, \ and\ \bibinfo {author}
  {\bibfnamefont {Patrik}\ \bibnamefont {\"Ohberg}},\ }\bibfield  {title}
  {\enquote {\bibinfo {title} {Colloquium: Artificial gauge potentials for
  neutral atoms},}\ }\href {\doibase 10.1103/RevModPhys.83.1523} {\bibfield
  {journal} {\bibinfo  {journal} {Rev. Mod. Phys.}\ }\textbf {\bibinfo {volume}
  {83}},\ \bibinfo {pages} {1523--1543} (\bibinfo {year} {2011})}\BibitemShut
  {NoStop}%
\bibitem [{\citenamefont {Basov}\ \emph {et~al.}(2017)\citenamefont {Basov},
  \citenamefont {Averitt},\ and\ \citenamefont {Hsieh}}]{Basov2017}%
  \BibitemOpen
  \bibfield  {author} {\bibinfo {author} {\bibfnamefont {D.~N.}\ \bibnamefont
  {Basov}}, \bibinfo {author} {\bibfnamefont {R.~D.}\ \bibnamefont {Averitt}},
  \ and\ \bibinfo {author} {\bibfnamefont {D.}~\bibnamefont {Hsieh}},\
  }\bibfield  {title} {\enquote {\bibinfo {title} {Towards properties on demand
  in quantum materials},}\ }\href {\doibase 10.1038/nmat5017} {\bibfield
  {journal} {\bibinfo  {journal} {Nat. Mat.}\ }\textbf {\bibinfo {volume}
  {16}},\ \bibinfo {pages} {1077} (\bibinfo {year} {2017})}\BibitemShut
  {NoStop}%
\bibitem [{\citenamefont {Zhang}\ and\ \citenamefont
  {Averitt}(2014)}]{doi:10.1146/annurev-matsci-070813-113258}%
  \BibitemOpen
  \bibfield  {author} {\bibinfo {author} {\bibfnamefont {J.}~\bibnamefont
  {Zhang}}\ and\ \bibinfo {author} {\bibfnamefont {R.D.}\ \bibnamefont
  {Averitt}},\ }\bibfield  {title} {\enquote {\bibinfo {title} {Dynamics and
  control in complex transition metal oxides},}\ }\href {\doibase
  10.1146/annurev-matsci-070813-113258} {\bibfield  {journal} {\bibinfo
  {journal} {Annu. Rev. Mater. Res.}\ }\textbf {\bibinfo {volume} {44}},\
  \bibinfo {pages} {19--43} (\bibinfo {year} {2014})}\BibitemShut {NoStop}%
\bibitem [{\citenamefont {Basov}\ \emph {et~al.}(2011)\citenamefont {Basov},
  \citenamefont {Averitt}, \citenamefont {van~der Marel}, \citenamefont
  {Dressel},\ and\ \citenamefont {Haule}}]{RevModPhys.83.471}%
  \BibitemOpen
  \bibfield  {author} {\bibinfo {author} {\bibfnamefont {D.~N.}\ \bibnamefont
  {Basov}}, \bibinfo {author} {\bibfnamefont {Richard~D.}\ \bibnamefont
  {Averitt}}, \bibinfo {author} {\bibfnamefont {Dirk}\ \bibnamefont {van~der
  Marel}}, \bibinfo {author} {\bibfnamefont {Martin}\ \bibnamefont {Dressel}},
  \ and\ \bibinfo {author} {\bibfnamefont {Kristjan}\ \bibnamefont {Haule}},\
  }\bibfield  {title} {\enquote {\bibinfo {title} {Electrodynamics of
  correlated electron materials},}\ }\href {\doibase 10.1103/RevModPhys.83.471}
  {\bibfield  {journal} {\bibinfo  {journal} {Rev. Mod. Phys.}\ }\textbf
  {\bibinfo {volume} {83}},\ \bibinfo {pages} {471--541} (\bibinfo {year}
  {2011})}\BibitemShut {NoStop}%
\bibitem [{\citenamefont {Giannetti}\ \emph {et~al.}(2016)\citenamefont
  {Giannetti}, \citenamefont {Capone}, \citenamefont {Fausti}, \citenamefont
  {Fabrizio}, \citenamefont {Parmigiani},\ and\ \citenamefont
  {Mihailovic}}]{doi:10.1080/00018732.2016.1194044}%
  \BibitemOpen
  \bibfield  {author} {\bibinfo {author} {\bibfnamefont {Claudio}\ \bibnamefont
  {Giannetti}}, \bibinfo {author} {\bibfnamefont {Massimo}\ \bibnamefont
  {Capone}}, \bibinfo {author} {\bibfnamefont {Daniele}\ \bibnamefont
  {Fausti}}, \bibinfo {author} {\bibfnamefont {Michele}\ \bibnamefont
  {Fabrizio}}, \bibinfo {author} {\bibfnamefont {Fulvio}\ \bibnamefont
  {Parmigiani}}, \ and\ \bibinfo {author} {\bibfnamefont {Dragan}\ \bibnamefont
  {Mihailovic}},\ }\bibfield  {title} {\enquote {\bibinfo {title} {Ultrafast
  optical spectroscopy of strongly correlated materials and high-temperature
  superconductors: a non-equilibrium approach},}\ }\href {\doibase
  10.1080/00018732.2016.1194044} {\bibfield  {journal} {\bibinfo  {journal}
  {Adv. Phys.}\ }\textbf {\bibinfo {volume} {65}},\ \bibinfo {pages} {58--238}
  (\bibinfo {year} {2016})}\BibitemShut {NoStop}%
\bibitem [{\citenamefont {Gandolfi}\ \emph {et~al.}(2017)\citenamefont
  {Gandolfi}, \citenamefont {Celardo}, \citenamefont {Borgonovi}, \citenamefont
  {Ferrini}, \citenamefont {Avella}, \citenamefont {Banfi},\ and\ \citenamefont
  {Giannetti}}]{1402-4896-92-3-034004}%
  \BibitemOpen
  \bibfield  {author} {\bibinfo {author} {\bibfnamefont {M.}~\bibnamefont
  {Gandolfi}}, \bibinfo {author} {\bibfnamefont {G.~L.}\ \bibnamefont
  {Celardo}}, \bibinfo {author} {\bibfnamefont {F.}~\bibnamefont {Borgonovi}},
  \bibinfo {author} {\bibfnamefont {G.}~\bibnamefont {Ferrini}}, \bibinfo
  {author} {\bibfnamefont {A.}~\bibnamefont {Avella}}, \bibinfo {author}
  {\bibfnamefont {F.}~\bibnamefont {Banfi}}, \ and\ \bibinfo {author}
  {\bibfnamefont {C.}~\bibnamefont {Giannetti}},\ }\bibfield  {title} {\enquote
  {\bibinfo {title} {Emergent ultrafast phenomena in correlated oxides and
  heterostructures},}\ }\href {http://stacks.iop.org/1402-4896/92/i=3/a=034004}
  {\bibfield  {journal} {\bibinfo  {journal} {Phys. Scr.}\ }\textbf {\bibinfo
  {volume} {92}},\ \bibinfo {pages} {034004} (\bibinfo {year}
  {2017})}\BibitemShut {NoStop}%
\bibitem [{\citenamefont {Mentink}\ and\ \citenamefont
  {Eckstein}(2014)}]{PhysRevLett.113.057201}%
  \BibitemOpen
  \bibfield  {author} {\bibinfo {author} {\bibfnamefont {J.~H.}\ \bibnamefont
  {Mentink}}\ and\ \bibinfo {author} {\bibfnamefont {M.}~\bibnamefont
  {Eckstein}},\ }\bibfield  {title} {\enquote {\bibinfo {title} {Ultrafast
  quenching of the exchange interaction in a {Mott} insulator},}\ }\href
  {\doibase 10.1103/PhysRevLett.113.057201} {\bibfield  {journal} {\bibinfo
  {journal} {Phys. Rev. Lett.}\ }\textbf {\bibinfo {volume} {113}},\ \bibinfo
  {pages} {057201} (\bibinfo {year} {2014})}\BibitemShut {NoStop}%
\bibitem [{\citenamefont {Mentink}\ \emph {et~al.}(2015)\citenamefont
  {Mentink}, \citenamefont {Balzer},\ and\ \citenamefont
  {Eckstein}}]{Mentink2015}%
  \BibitemOpen
  \bibfield  {author} {\bibinfo {author} {\bibfnamefont {J.H.}\ \bibnamefont
  {Mentink}}, \bibinfo {author} {\bibfnamefont {K.}~\bibnamefont {Balzer}}, \
  and\ \bibinfo {author} {\bibfnamefont {M.}~\bibnamefont {Eckstein}},\
  }\bibfield  {title} {\enquote {\bibinfo {title} {Ultrafast and reversible
  control of the exchange interaction in {Mott} insulators},}\ }\href {\doibase
  10.1038/ncomms7708} {\bibfield  {journal} {\bibinfo  {journal} {Nat. Comms.}\
  }\textbf {\bibinfo {volume} {6}},\ \bibinfo {pages} {6708} (\bibinfo {year}
  {2015})}\BibitemShut {NoStop}%
\bibitem [{\citenamefont {Fausti}\ \emph {et~al.}(2011)\citenamefont {Fausti},
  \citenamefont {Tobey}, \citenamefont {Dean}, \citenamefont {Kaiser},
  \citenamefont {Dienst}, \citenamefont {Hoffmann}, \citenamefont {Pyon},
  \citenamefont {Takayama}, \citenamefont {Takagi},\ and\ \citenamefont
  {Cavalleri}}]{Fausti2011}%
  \BibitemOpen
  \bibfield  {author} {\bibinfo {author} {\bibfnamefont {D.}~\bibnamefont
  {Fausti}}, \bibinfo {author} {\bibfnamefont {R.~I.}\ \bibnamefont {Tobey}},
  \bibinfo {author} {\bibfnamefont {N.}~\bibnamefont {Dean}}, \bibinfo {author}
  {\bibfnamefont {S.}~\bibnamefont {Kaiser}}, \bibinfo {author} {\bibfnamefont
  {A.}~\bibnamefont {Dienst}}, \bibinfo {author} {\bibfnamefont {M.~C.}\
  \bibnamefont {Hoffmann}}, \bibinfo {author} {\bibfnamefont {S.}~\bibnamefont
  {Pyon}}, \bibinfo {author} {\bibfnamefont {T.}~\bibnamefont {Takayama}},
  \bibinfo {author} {\bibfnamefont {H.}~\bibnamefont {Takagi}}, \ and\ \bibinfo
  {author} {\bibfnamefont {A.}~\bibnamefont {Cavalleri}},\ }\bibfield  {title}
  {\enquote {\bibinfo {title} {Light-induced superconductivity in a
  stripe-ordered cuprate},}\ }\href {\doibase 10.1126/science.1197294}
  {\bibfield  {journal} {\bibinfo  {journal} {Science}\ }\textbf {\bibinfo
  {volume} {331}},\ \bibinfo {pages} {189--191} (\bibinfo {year}
  {2011})}\BibitemShut {NoStop}%
\bibitem [{\citenamefont
  {Cavalleri}(2018)}]{doi:10.1080/00107514.2017.1406623}%
  \BibitemOpen
  \bibfield  {author} {\bibinfo {author} {\bibfnamefont {Andrea}\ \bibnamefont
  {Cavalleri}},\ }\bibfield  {title} {\enquote {\bibinfo {title} {Photo-induced
  superconductivity},}\ }\href {\doibase 10.1080/00107514.2017.1406623}
  {\bibfield  {journal} {\bibinfo  {journal} {Contemp. Phys.}\ }\textbf
  {\bibinfo {volume} {59}},\ \bibinfo {pages} {31--46} (\bibinfo {year}
  {2018})}\BibitemShut {NoStop}%
\bibitem [{\citenamefont {Stojchevska}\ \emph {et~al.}(2014)\citenamefont
  {Stojchevska}, \citenamefont {Vaskivskyi}, \citenamefont {Mertelj},
  \citenamefont {Kusar}, \citenamefont {Svetin}, \citenamefont {Brazovskii},\
  and\ \citenamefont {Mihailovic}}]{Stojchevska2014}%
  \BibitemOpen
  \bibfield  {author} {\bibinfo {author} {\bibfnamefont {L.}~\bibnamefont
  {Stojchevska}}, \bibinfo {author} {\bibfnamefont {I.}~\bibnamefont
  {Vaskivskyi}}, \bibinfo {author} {\bibfnamefont {T.}~\bibnamefont {Mertelj}},
  \bibinfo {author} {\bibfnamefont {P.}~\bibnamefont {Kusar}}, \bibinfo
  {author} {\bibfnamefont {D.}~\bibnamefont {Svetin}}, \bibinfo {author}
  {\bibfnamefont {S.}~\bibnamefont {Brazovskii}}, \ and\ \bibinfo {author}
  {\bibfnamefont {D.}~\bibnamefont {Mihailovic}},\ }\bibfield  {title}
  {\enquote {\bibinfo {title} {Ultrafast switching to a stable hidden quantum
  state in an electronic crystal},}\ }\href {\doibase 10.1126/science.1241591}
  {\bibfield  {journal} {\bibinfo  {journal} {Science}\ }\textbf {\bibinfo
  {volume} {344}},\ \bibinfo {pages} {177--180} (\bibinfo {year}
  {2014})}\BibitemShut {NoStop}%
\bibitem [{\citenamefont {{Gerasimenko}}\ \emph {et~al.}(2017)\citenamefont
  {{Gerasimenko}}, \citenamefont {{Vaskivskyi}},\ and\ \citenamefont
  {{Mihailovic}}}]{Gerasimenko2017}%
  \BibitemOpen
  \bibfield  {author} {\bibinfo {author} {\bibfnamefont {Y.~A.}\ \bibnamefont
  {{Gerasimenko}}}, \bibinfo {author} {\bibfnamefont {I.}~\bibnamefont
  {{Vaskivskyi}}}, \ and\ \bibinfo {author} {\bibfnamefont {D.}~\bibnamefont
  {{Mihailovic}}},\ }\href {https://arxiv.org/abs/1704.08149} {\enquote
  {\bibinfo {title} {Long range electronic order in a metastable state created
  by ultrafast topological transformation},}\ } (\bibinfo {year} {2017}),\
  \bibinfo {note} {arXiv:1704.08149}\BibitemShut {NoStop}%
\bibitem [{\citenamefont {Zhang}\ \emph
  {et~al.}(2017{\natexlab{a}})\citenamefont {Zhang}, \citenamefont {Hess},
  \citenamefont {Kyprianidis}, \citenamefont {Becker}, \citenamefont {Lee},
  \citenamefont {Smith}, \citenamefont {Pagano}, \citenamefont {Potirniche},
  \citenamefont {Potter}, \citenamefont {Vishwanath}, \citenamefont {Yao},\
  and\ \citenamefont {Monroe}}]{10.1038/nature21413}%
  \BibitemOpen
  \bibfield  {author} {\bibinfo {author} {\bibfnamefont {J.}~\bibnamefont
  {Zhang}}, \bibinfo {author} {\bibfnamefont {P.~W.}\ \bibnamefont {Hess}},
  \bibinfo {author} {\bibfnamefont {A.}~\bibnamefont {Kyprianidis}}, \bibinfo
  {author} {\bibfnamefont {P.}~\bibnamefont {Becker}}, \bibinfo {author}
  {\bibfnamefont {A.}~\bibnamefont {Lee}}, \bibinfo {author} {\bibfnamefont
  {J.}~\bibnamefont {Smith}}, \bibinfo {author} {\bibfnamefont
  {G.}~\bibnamefont {Pagano}}, \bibinfo {author} {\bibfnamefont {I.-D.}\
  \bibnamefont {Potirniche}}, \bibinfo {author} {\bibfnamefont {A.~C.}\
  \bibnamefont {Potter}}, \bibinfo {author} {\bibfnamefont {A.}~\bibnamefont
  {Vishwanath}}, \bibinfo {author} {\bibfnamefont {N.~Y.}\ \bibnamefont {Yao}},
  \ and\ \bibinfo {author} {\bibfnamefont {C.}~\bibnamefont {Monroe}},\
  }\bibfield  {title} {\enquote {\bibinfo {title} {Observation of a discrete
  time crystal},}\ }\href {\doibase 10.1038/nature21413} {\bibfield  {journal}
  {\bibinfo  {journal} {Nature}\ }\textbf {\bibinfo {volume} {543}},\ \bibinfo
  {pages} {217--220} (\bibinfo {year} {2017}{\natexlab{a}})}\BibitemShut
  {NoStop}%
\bibitem [{\citenamefont {Choi}\ \emph {et~al.}(2017)\citenamefont {Choi},
  \citenamefont {Choi}, \citenamefont {Landig}, \citenamefont {Kucsko},
  \citenamefont {Zhou}, \citenamefont {Isoya}, \citenamefont {Jelezko},
  \citenamefont {Onoda}, \citenamefont {Sumiya}, \citenamefont {Khemani},
  \citenamefont {von Keyserlingk}, \citenamefont {Yao}, \citenamefont
  {Demler},\ and\ \citenamefont {Lukin}}]{10.1038/nature21426}%
  \BibitemOpen
  \bibfield  {author} {\bibinfo {author} {\bibfnamefont {Soonwon}\ \bibnamefont
  {Choi}}, \bibinfo {author} {\bibfnamefont {Joonhee}\ \bibnamefont {Choi}},
  \bibinfo {author} {\bibfnamefont {Renate}\ \bibnamefont {Landig}}, \bibinfo
  {author} {\bibfnamefont {Georg}\ \bibnamefont {Kucsko}}, \bibinfo {author}
  {\bibfnamefont {Hengyun}\ \bibnamefont {Zhou}}, \bibinfo {author}
  {\bibfnamefont {Junichi}\ \bibnamefont {Isoya}}, \bibinfo {author}
  {\bibfnamefont {Fedor}\ \bibnamefont {Jelezko}}, \bibinfo {author}
  {\bibfnamefont {Shinobu}\ \bibnamefont {Onoda}}, \bibinfo {author}
  {\bibfnamefont {Hitoshi}\ \bibnamefont {Sumiya}}, \bibinfo {author}
  {\bibfnamefont {Vedika}\ \bibnamefont {Khemani}}, \bibinfo {author}
  {\bibfnamefont {Curt}\ \bibnamefont {von Keyserlingk}}, \bibinfo {author}
  {\bibfnamefont {Norman~Y.}\ \bibnamefont {Yao}}, \bibinfo {author}
  {\bibfnamefont {Eugene}\ \bibnamefont {Demler}}, \ and\ \bibinfo {author}
  {\bibfnamefont {Mikhail~D.}\ \bibnamefont {Lukin}},\ }\bibfield  {title}
  {\enquote {\bibinfo {title} {Observation of discrete time-crystalline order
  in a disordered dipolar many-body system},}\ }\href {\doibase
  10.1038/nature21426} {\bibfield  {journal} {\bibinfo  {journal} {Nature}\
  }\textbf {\bibinfo {volume} {543}},\ \bibinfo {pages} {221--225} (\bibinfo
  {year} {2017})}\BibitemShut {NoStop}%
\bibitem [{\citenamefont {Potter}\ \emph {et~al.}(2016)\citenamefont {Potter},
  \citenamefont {Morimoto},\ and\ \citenamefont
  {Vishwanath}}]{PhysRevX.6.041001}%
  \BibitemOpen
  \bibfield  {author} {\bibinfo {author} {\bibfnamefont {Andrew~C.}\
  \bibnamefont {Potter}}, \bibinfo {author} {\bibfnamefont {Takahiro}\
  \bibnamefont {Morimoto}}, \ and\ \bibinfo {author} {\bibfnamefont {Ashvin}\
  \bibnamefont {Vishwanath}},\ }\bibfield  {title} {\enquote {\bibinfo {title}
  {Classification of interacting topological {Floquet} phases in one
  dimension},}\ }\href {\doibase 10.1103/PhysRevX.6.041001} {\bibfield
  {journal} {\bibinfo  {journal} {Phys. Rev. X}\ }\textbf {\bibinfo {volume}
  {6}},\ \bibinfo {pages} {041001} (\bibinfo {year} {2016})}\BibitemShut
  {NoStop}%
\bibitem [{\citenamefont {Khemani}\ \emph {et~al.}(2016)\citenamefont
  {Khemani}, \citenamefont {Lazarides}, \citenamefont {Moessner},\ and\
  \citenamefont {Sondhi}}]{PhysRevLett.116.250401}%
  \BibitemOpen
  \bibfield  {author} {\bibinfo {author} {\bibfnamefont {Vedika}\ \bibnamefont
  {Khemani}}, \bibinfo {author} {\bibfnamefont {Achilleas}\ \bibnamefont
  {Lazarides}}, \bibinfo {author} {\bibfnamefont {Roderich}\ \bibnamefont
  {Moessner}}, \ and\ \bibinfo {author} {\bibfnamefont {S.~L.}\ \bibnamefont
  {Sondhi}},\ }\bibfield  {title} {\enquote {\bibinfo {title} {Phase structure
  of driven quantum systems},}\ }\href {\doibase
  10.1103/PhysRevLett.116.250401} {\bibfield  {journal} {\bibinfo  {journal}
  {Phys. Rev. Lett.}\ }\textbf {\bibinfo {volume} {116}},\ \bibinfo {pages}
  {250401} (\bibinfo {year} {2016})}\BibitemShut {NoStop}%
\bibitem [{\citenamefont {Bukov}\ \emph
  {et~al.}(2015{\natexlab{a}})\citenamefont {Bukov}, \citenamefont
  {D'Alessio},\ and\ \citenamefont {Polkovnikov}}]{Bukov2015}%
  \BibitemOpen
  \bibfield  {author} {\bibinfo {author} {\bibfnamefont {Marin}\ \bibnamefont
  {Bukov}}, \bibinfo {author} {\bibfnamefont {Luca}\ \bibnamefont {D'Alessio}},
  \ and\ \bibinfo {author} {\bibfnamefont {Anatoli}\ \bibnamefont
  {Polkovnikov}},\ }\bibfield  {title} {\enquote {\bibinfo {title} {Universal
  high-frequency behavior of periodically driven systems: from dynamical
  stabilization to {Floquet} engineering},}\ }\href {\doibase
  10.1080/00018732.2015.1055918} {\bibfield  {journal} {\bibinfo  {journal}
  {Adv. Phys.}\ }\textbf {\bibinfo {volume} {64}},\ \bibinfo {pages} {139--226}
  (\bibinfo {year} {2015}{\natexlab{a}})}\BibitemShut {NoStop}%
\bibitem [{\citenamefont {Shirley}(1965)}]{PhysRev.138.B979}%
  \BibitemOpen
  \bibfield  {author} {\bibinfo {author} {\bibfnamefont {Jon~H.}\ \bibnamefont
  {Shirley}},\ }\bibfield  {title} {\enquote {\bibinfo {title} {Solution of the
  {Schr\"odinger} equation with a {Hamiltonian} periodic in time},}\ }\href
  {\doibase 10.1103/PhysRev.138.B979} {\bibfield  {journal} {\bibinfo
  {journal} {Phys. Rev.}\ }\textbf {\bibinfo {volume} {138}},\ \bibinfo {pages}
  {B979--B987} (\bibinfo {year} {1965})}\BibitemShut {NoStop}%
\bibitem [{\citenamefont {{Moessner}}\ and\ \citenamefont
  {{Sondhi}}(2017)}]{10.1038/nphys4106}%
  \BibitemOpen
  \bibfield  {author} {\bibinfo {author} {\bibfnamefont {R.}~\bibnamefont
  {{Moessner}}}\ and\ \bibinfo {author} {\bibfnamefont {S.~L.}\ \bibnamefont
  {{Sondhi}}},\ }\bibfield  {title} {\enquote {\bibinfo {title} {Equilibration
  and order in quantum {Floquet} matter},}\ }\href {\doibase 10.1038/nphys4106}
  {\bibfield  {journal} {\bibinfo  {journal} {Nat. Phys.}\ }\textbf {\bibinfo
  {volume} {13}},\ \bibinfo {pages} {424} (\bibinfo {year} {2017})}\BibitemShut
  {NoStop}%
\bibitem [{\citenamefont {Lazarides}\ \emph
  {et~al.}(2014{\natexlab{a}})\citenamefont {Lazarides}, \citenamefont {Das},\
  and\ \citenamefont {Moessner}}]{PhysRevLett.112.150401}%
  \BibitemOpen
  \bibfield  {author} {\bibinfo {author} {\bibfnamefont {Achilleas}\
  \bibnamefont {Lazarides}}, \bibinfo {author} {\bibfnamefont {Arnab}\
  \bibnamefont {Das}}, \ and\ \bibinfo {author} {\bibfnamefont {Roderich}\
  \bibnamefont {Moessner}},\ }\bibfield  {title} {\enquote {\bibinfo {title}
  {Periodic thermodynamics of isolated quantum systems},}\ }\href {\doibase
  10.1103/PhysRevLett.112.150401} {\bibfield  {journal} {\bibinfo  {journal}
  {Phys. Rev. Lett.}\ }\textbf {\bibinfo {volume} {112}},\ \bibinfo {pages}
  {150401} (\bibinfo {year} {2014}{\natexlab{a}})}\BibitemShut {NoStop}%
\bibitem [{\citenamefont {Berdanier}\ \emph {et~al.}(2017)\citenamefont
  {Berdanier}, \citenamefont {Kolodrubetz}, \citenamefont {Vasseur},\ and\
  \citenamefont {Moore}}]{PhysRevLett.118.260602}%
  \BibitemOpen
  \bibfield  {author} {\bibinfo {author} {\bibfnamefont {William}\ \bibnamefont
  {Berdanier}}, \bibinfo {author} {\bibfnamefont {Michael}\ \bibnamefont
  {Kolodrubetz}}, \bibinfo {author} {\bibfnamefont {Romain}\ \bibnamefont
  {Vasseur}}, \ and\ \bibinfo {author} {\bibfnamefont {Joel~E.}\ \bibnamefont
  {Moore}},\ }\bibfield  {title} {\enquote {\bibinfo {title} {Floquet dynamics
  of boundary-driven systems at criticality},}\ }\href {\doibase
  10.1103/PhysRevLett.118.260602} {\bibfield  {journal} {\bibinfo  {journal}
  {Phys. Rev. Lett.}\ }\textbf {\bibinfo {volume} {118}},\ \bibinfo {pages}
  {260602} (\bibinfo {year} {2017})}\BibitemShut {NoStop}%
\bibitem [{\citenamefont {{Wen}}\ and\ \citenamefont
  {{Wu}}(2018)}]{2018arXiv180500031W}%
  \BibitemOpen
  \bibfield  {author} {\bibinfo {author} {\bibfnamefont {X.}~\bibnamefont
  {{Wen}}}\ and\ \bibinfo {author} {\bibfnamefont {J.-Q.}\ \bibnamefont
  {{Wu}}},\ }\href {https://arxiv.org/abs/1805.00031} {\enquote {\bibinfo
  {title} {{Floquet conformal field theory}},}\ } (\bibinfo {year} {2018}),\
  \bibinfo {note} {arXiv:1805.00031}\BibitemShut {NoStop}%
\bibitem [{\citenamefont {Ponte}\ \emph
  {et~al.}(2015{\natexlab{a}})\citenamefont {Ponte}, \citenamefont {Chandran},
  \citenamefont {Papić},\ and\ \citenamefont {Abanin}}]{Ponte2015}%
  \BibitemOpen
  \bibfield  {author} {\bibinfo {author} {\bibfnamefont {Pedro}\ \bibnamefont
  {Ponte}}, \bibinfo {author} {\bibfnamefont {Anushya}\ \bibnamefont
  {Chandran}}, \bibinfo {author} {\bibfnamefont {Z.}~\bibnamefont {Papić}}, \
  and\ \bibinfo {author} {\bibfnamefont {Dmitry~A.}\ \bibnamefont {Abanin}},\
  }\bibfield  {title} {\enquote {\bibinfo {title} {Periodically driven ergodic
  and many-body localized quantum systems},}\ }\href {\doibase
  https://doi.org/10.1016/j.aop.2014.11.008} {\bibfield  {journal} {\bibinfo
  {journal} {Ann. Phys. (N.Y.)}\ }\textbf {\bibinfo {volume} {353}},\ \bibinfo
  {pages} {196 -- 204} (\bibinfo {year} {2015}{\natexlab{a}})}\BibitemShut
  {NoStop}%
\bibitem [{\citenamefont {Ponte}\ \emph
  {et~al.}(2015{\natexlab{b}})\citenamefont {Ponte}, \citenamefont
  {Papi\ifmmode~\acute{c}\else \'{c}\fi{}}, \citenamefont {Huveneers},\ and\
  \citenamefont {Abanin}}]{PhysRevLett.114.140401}%
  \BibitemOpen
  \bibfield  {author} {\bibinfo {author} {\bibfnamefont {Pedro}\ \bibnamefont
  {Ponte}}, \bibinfo {author} {\bibfnamefont {Z.}~\bibnamefont
  {Papi\ifmmode~\acute{c}\else \'{c}\fi{}}}, \bibinfo {author} {\bibfnamefont
  {Fran\ifmmode \mbox{\c{c}}\else{}\c{c}\fi{}ois}\ \bibnamefont {Huveneers}}, \
  and\ \bibinfo {author} {\bibfnamefont {Dmitry~A.}\ \bibnamefont {Abanin}},\
  }\bibfield  {title} {\enquote {\bibinfo {title} {Many-body localization in
  periodically driven systems},}\ }\href {\doibase
  10.1103/PhysRevLett.114.140401} {\bibfield  {journal} {\bibinfo  {journal}
  {Phys. Rev. Lett.}\ }\textbf {\bibinfo {volume} {114}},\ \bibinfo {pages}
  {140401} (\bibinfo {year} {2015}{\natexlab{b}})}\BibitemShut {NoStop}%
\bibitem [{\citenamefont {Lazarides}\ \emph
  {et~al.}(2014{\natexlab{b}})\citenamefont {Lazarides}, \citenamefont {Das},\
  and\ \citenamefont {Moessner}}]{PhysRevE.90.012110}%
  \BibitemOpen
  \bibfield  {author} {\bibinfo {author} {\bibfnamefont {Achilleas}\
  \bibnamefont {Lazarides}}, \bibinfo {author} {\bibfnamefont {Arnab}\
  \bibnamefont {Das}}, \ and\ \bibinfo {author} {\bibfnamefont {Roderich}\
  \bibnamefont {Moessner}},\ }\bibfield  {title} {\enquote {\bibinfo {title}
  {Equilibrium states of generic quantum systems subject to periodic
  driving},}\ }\href {\doibase 10.1103/PhysRevE.90.012110} {\bibfield
  {journal} {\bibinfo  {journal} {Phys. Rev. E}\ }\textbf {\bibinfo {volume}
  {90}},\ \bibinfo {pages} {012110} (\bibinfo {year}
  {2014}{\natexlab{b}})}\BibitemShut {NoStop}%
\bibitem [{\citenamefont {Wang}\ \emph {et~al.}(2013)\citenamefont {Wang},
  \citenamefont {Steinberg}, \citenamefont {Jarillo-Herrero},\ and\
  \citenamefont {Gedik}}]{Wang2013a}%
  \BibitemOpen
  \bibfield  {author} {\bibinfo {author} {\bibfnamefont {Y.~H.}\ \bibnamefont
  {Wang}}, \bibinfo {author} {\bibfnamefont {H.}~\bibnamefont {Steinberg}},
  \bibinfo {author} {\bibfnamefont {P.}~\bibnamefont {Jarillo-Herrero}}, \ and\
  \bibinfo {author} {\bibfnamefont {N.}~\bibnamefont {Gedik}},\ }\bibfield
  {title} {\enquote {\bibinfo {title} {Observation of {Floquet-Bloch} states on
  the surface of a topological insulator},}\ }\href {\doibase
  10.1126/science.1239834} {\bibfield  {journal} {\bibinfo  {journal}
  {Science}\ }\textbf {\bibinfo {volume} {342}},\ \bibinfo {pages} {453--457}
  (\bibinfo {year} {2013})}\BibitemShut {NoStop}%
\bibitem [{\citenamefont {Jiang}\ \emph {et~al.}(2011)\citenamefont {Jiang},
  \citenamefont {Kitagawa}, \citenamefont {Alicea}, \citenamefont {Akhmerov},
  \citenamefont {Pekker}, \citenamefont {Refael}, \citenamefont {Cirac},
  \citenamefont {Demler}, \citenamefont {Lukin},\ and\ \citenamefont
  {Zoller}}]{PhysRevLett.106.220402}%
  \BibitemOpen
  \bibfield  {author} {\bibinfo {author} {\bibfnamefont {Liang}\ \bibnamefont
  {Jiang}}, \bibinfo {author} {\bibfnamefont {Takuya}\ \bibnamefont
  {Kitagawa}}, \bibinfo {author} {\bibfnamefont {Jason}\ \bibnamefont
  {Alicea}}, \bibinfo {author} {\bibfnamefont {A.~R.}\ \bibnamefont
  {Akhmerov}}, \bibinfo {author} {\bibfnamefont {David}\ \bibnamefont
  {Pekker}}, \bibinfo {author} {\bibfnamefont {Gil}\ \bibnamefont {Refael}},
  \bibinfo {author} {\bibfnamefont {J.~Ignacio}\ \bibnamefont {Cirac}},
  \bibinfo {author} {\bibfnamefont {Eugene}\ \bibnamefont {Demler}}, \bibinfo
  {author} {\bibfnamefont {Mikhail~D.}\ \bibnamefont {Lukin}}, \ and\ \bibinfo
  {author} {\bibfnamefont {Peter}\ \bibnamefont {Zoller}},\ }\bibfield  {title}
  {\enquote {\bibinfo {title} {Majorana fermions in equilibrium and in driven
  cold-atom quantum wires},}\ }\href {\doibase 10.1103/PhysRevLett.106.220402}
  {\bibfield  {journal} {\bibinfo  {journal} {Phys. Rev. Lett.}\ }\textbf
  {\bibinfo {volume} {106}},\ \bibinfo {pages} {220402} (\bibinfo {year}
  {2011})}\BibitemShut {NoStop}%
\bibitem [{\citenamefont {Molignini}\ \emph {et~al.}(2017)\citenamefont
  {Molignini}, \citenamefont {van Nieuwenburg},\ and\ \citenamefont
  {Chitra}}]{PhysRevB.96.125144}%
  \BibitemOpen
  \bibfield  {author} {\bibinfo {author} {\bibfnamefont {Paolo}\ \bibnamefont
  {Molignini}}, \bibinfo {author} {\bibfnamefont {Evert}\ \bibnamefont {van
  Nieuwenburg}}, \ and\ \bibinfo {author} {\bibfnamefont {R.}~\bibnamefont
  {Chitra}},\ }\bibfield  {title} {\enquote {\bibinfo {title} {Sensing
  {Floquet-Majorana} fermions via heat transfer},}\ }\href {\doibase
  10.1103/PhysRevB.96.125144} {\bibfield  {journal} {\bibinfo  {journal} {Phys.
  Rev. B}\ }\textbf {\bibinfo {volume} {96}},\ \bibinfo {pages} {125144}
  (\bibinfo {year} {2017})}\BibitemShut {NoStop}%
\bibitem [{\citenamefont {Jim\'enez-Garc\'{\i}a}\ \emph
  {et~al.}(2015)\citenamefont {Jim\'enez-Garc\'{\i}a}, \citenamefont {LeBlanc},
  \citenamefont {Williams}, \citenamefont {Beeler}, \citenamefont {Qu},
  \citenamefont {Gong}, \citenamefont {Zhang},\ and\ \citenamefont
  {Spielman}}]{PhysRevLett.114.125301}%
  \BibitemOpen
  \bibfield  {author} {\bibinfo {author} {\bibfnamefont {K.}~\bibnamefont
  {Jim\'enez-Garc\'{\i}a}}, \bibinfo {author} {\bibfnamefont {L.~J.}\
  \bibnamefont {LeBlanc}}, \bibinfo {author} {\bibfnamefont {R.~A.}\
  \bibnamefont {Williams}}, \bibinfo {author} {\bibfnamefont {M.~C.}\
  \bibnamefont {Beeler}}, \bibinfo {author} {\bibfnamefont {C.}~\bibnamefont
  {Qu}}, \bibinfo {author} {\bibfnamefont {M.}~\bibnamefont {Gong}}, \bibinfo
  {author} {\bibfnamefont {C.}~\bibnamefont {Zhang}}, \ and\ \bibinfo {author}
  {\bibfnamefont {I.~B.}\ \bibnamefont {Spielman}},\ }\bibfield  {title}
  {\enquote {\bibinfo {title} {Tunable spin-orbit coupling via strong driving
  in ultracold-atom systems},}\ }\href {\doibase
  10.1103/PhysRevLett.114.125301} {\bibfield  {journal} {\bibinfo  {journal}
  {Phys. Rev. Lett.}\ }\textbf {\bibinfo {volume} {114}},\ \bibinfo {pages}
  {125301} (\bibinfo {year} {2015})}\BibitemShut {NoStop}%
\bibitem [{\citenamefont {Klinovaja}\ \emph {et~al.}(2016)\citenamefont
  {Klinovaja}, \citenamefont {Stano},\ and\ \citenamefont
  {Loss}}]{PhysRevLett.116.176401}%
  \BibitemOpen
  \bibfield  {author} {\bibinfo {author} {\bibfnamefont {Jelena}\ \bibnamefont
  {Klinovaja}}, \bibinfo {author} {\bibfnamefont {Peter}\ \bibnamefont
  {Stano}}, \ and\ \bibinfo {author} {\bibfnamefont {Daniel}\ \bibnamefont
  {Loss}},\ }\bibfield  {title} {\enquote {\bibinfo {title} {Topological
  {Floquet} phases in driven coupled {Rashba} nanowires},}\ }\href {\doibase
  10.1103/PhysRevLett.116.176401} {\bibfield  {journal} {\bibinfo  {journal}
  {Phys. Rev. Lett.}\ }\textbf {\bibinfo {volume} {116}},\ \bibinfo {pages}
  {176401} (\bibinfo {year} {2016})}\BibitemShut {NoStop}%
\bibitem [{\citenamefont {Du}\ \emph {et~al.}(2017)\citenamefont {Du},
  \citenamefont {Zhou},\ and\ \citenamefont {Fiete}}]{PhysRevB.95.035136}%
  \BibitemOpen
  \bibfield  {author} {\bibinfo {author} {\bibfnamefont {Liang}\ \bibnamefont
  {Du}}, \bibinfo {author} {\bibfnamefont {Xiaoting}\ \bibnamefont {Zhou}}, \
  and\ \bibinfo {author} {\bibfnamefont {Gregory~A.}\ \bibnamefont {Fiete}},\
  }\bibfield  {title} {\enquote {\bibinfo {title} {Quadratic band touching
  points and flat bands in two-dimensional topological {Floquet} systems},}\
  }\href {\doibase 10.1103/PhysRevB.95.035136} {\bibfield  {journal} {\bibinfo
  {journal} {Phys. Rev. B}\ }\textbf {\bibinfo {volume} {95}},\ \bibinfo
  {pages} {035136} (\bibinfo {year} {2017})}\BibitemShut {NoStop}%
\bibitem [{\citenamefont {Chen}\ \emph {et~al.}(2018)\citenamefont {Chen},
  \citenamefont {Du},\ and\ \citenamefont {Fiete}}]{PhysRevB.97.035422}%
  \BibitemOpen
  \bibfield  {author} {\bibinfo {author} {\bibfnamefont {Qi}~\bibnamefont
  {Chen}}, \bibinfo {author} {\bibfnamefont {Liang}\ \bibnamefont {Du}}, \ and\
  \bibinfo {author} {\bibfnamefont {Gregory~A.}\ \bibnamefont {Fiete}},\
  }\bibfield  {title} {\enquote {\bibinfo {title} {Floquet band structure of a
  semi-{Dirac} system},}\ }\href {\doibase 10.1103/PhysRevB.97.035422}
  {\bibfield  {journal} {\bibinfo  {journal} {Phys. Rev. B}\ }\textbf {\bibinfo
  {volume} {97}},\ \bibinfo {pages} {035422} (\bibinfo {year}
  {2018})}\BibitemShut {NoStop}%
\bibitem [{\citenamefont {Titum}\ \emph {et~al.}(2016)\citenamefont {Titum},
  \citenamefont {Berg}, \citenamefont {Rudner}, \citenamefont {Refael},\ and\
  \citenamefont {Lindner}}]{PhysRevX.6.021013}%
  \BibitemOpen
  \bibfield  {author} {\bibinfo {author} {\bibfnamefont {Paraj}\ \bibnamefont
  {Titum}}, \bibinfo {author} {\bibfnamefont {Erez}\ \bibnamefont {Berg}},
  \bibinfo {author} {\bibfnamefont {Mark~S.}\ \bibnamefont {Rudner}}, \bibinfo
  {author} {\bibfnamefont {Gil}\ \bibnamefont {Refael}}, \ and\ \bibinfo
  {author} {\bibfnamefont {Netanel~H.}\ \bibnamefont {Lindner}},\ }\bibfield
  {title} {\enquote {\bibinfo {title} {Anomalous {Floquet-Anderson} insulator
  as a nonadiabatic quantized charge pump},}\ }\href {\doibase
  10.1103/PhysRevX.6.021013} {\bibfield  {journal} {\bibinfo  {journal} {Phys.
  Rev. X}\ }\textbf {\bibinfo {volume} {6}},\ \bibinfo {pages} {021013}
  (\bibinfo {year} {2016})}\BibitemShut {NoStop}%
\bibitem [{\citenamefont {D'Alessio}\ and\ \citenamefont
  {Rigol}(2015)}]{DAlessio2015}%
  \BibitemOpen
  \bibfield  {author} {\bibinfo {author} {\bibfnamefont {Luca}\ \bibnamefont
  {D'Alessio}}\ and\ \bibinfo {author} {\bibfnamefont {Marcos}\ \bibnamefont
  {Rigol}},\ }\bibfield  {title} {\enquote {\bibinfo {title} {Dynamical
  preparation of {Floquet Chern} insulators},}\ }\href {\doibase
  10.1038/ncomms9336} {\bibfield  {journal} {\bibinfo  {journal} {Nat. Comms.}\
  }\textbf {\bibinfo {volume} {6}},\ \bibinfo {pages} {8836} (\bibinfo {year}
  {2015})}\BibitemShut {NoStop}%
\bibitem [{\citenamefont {Martin}\ \emph {et~al.}(2017)\citenamefont {Martin},
  \citenamefont {Refael},\ and\ \citenamefont {Halperin}}]{PhysRevX.7.041008}%
  \BibitemOpen
  \bibfield  {author} {\bibinfo {author} {\bibfnamefont {Ivar}\ \bibnamefont
  {Martin}}, \bibinfo {author} {\bibfnamefont {Gil}\ \bibnamefont {Refael}}, \
  and\ \bibinfo {author} {\bibfnamefont {Bertrand}\ \bibnamefont {Halperin}},\
  }\bibfield  {title} {\enquote {\bibinfo {title} {Topological frequency
  conversion in strongly driven quantum systems},}\ }\href {\doibase
  10.1103/PhysRevX.7.041008} {\bibfield  {journal} {\bibinfo  {journal} {Phys.
  Rev. X}\ }\textbf {\bibinfo {volume} {7}},\ \bibinfo {pages} {041008}
  (\bibinfo {year} {2017})}\BibitemShut {NoStop}%
\bibitem [{\citenamefont {Oka}\ and\ \citenamefont
  {Aoki}(2009)}]{PhysRevB.79.081406}%
  \BibitemOpen
  \bibfield  {author} {\bibinfo {author} {\bibfnamefont {Takashi}\ \bibnamefont
  {Oka}}\ and\ \bibinfo {author} {\bibfnamefont {Hideo}\ \bibnamefont {Aoki}},\
  }\bibfield  {title} {\enquote {\bibinfo {title} {Photovoltaic {Hall} effect
  in graphene},}\ }\href {\doibase 10.1103/PhysRevB.79.081406} {\bibfield
  {journal} {\bibinfo  {journal} {Phys. Rev. B}\ }\textbf {\bibinfo {volume}
  {79}},\ \bibinfo {pages} {081406} (\bibinfo {year} {2009})}\BibitemShut
  {NoStop}%
\bibitem [{\citenamefont {Lindner}\ \emph {et~al.}(2011)\citenamefont
  {Lindner}, \citenamefont {Refael},\ and\ \citenamefont
  {Galitski}}]{Lindner2011}%
  \BibitemOpen
  \bibfield  {author} {\bibinfo {author} {\bibfnamefont {Netanel~H.}\
  \bibnamefont {Lindner}}, \bibinfo {author} {\bibfnamefont {Gil}\ \bibnamefont
  {Refael}}, \ and\ \bibinfo {author} {\bibfnamefont {Victor}\ \bibnamefont
  {Galitski}},\ }\bibfield  {title} {\enquote {\bibinfo {title} {Floquet
  topological insulator in semiconductor quantum wells},}\ }\href {\doibase
  10.1038/nphys1926} {\bibfield  {journal} {\bibinfo  {journal} {Nat. Phys.}\
  }\textbf {\bibinfo {volume} {7}},\ \bibinfo {pages} {490} (\bibinfo {year}
  {2011})}\BibitemShut {NoStop}%
\bibitem [{\citenamefont {Kitagawa}\ \emph {et~al.}(2011)\citenamefont
  {Kitagawa}, \citenamefont {Oka}, \citenamefont {Brataas}, \citenamefont
  {Fu},\ and\ \citenamefont {Demler}}]{PhysRevB.84.235108}%
  \BibitemOpen
  \bibfield  {author} {\bibinfo {author} {\bibfnamefont {Takuya}\ \bibnamefont
  {Kitagawa}}, \bibinfo {author} {\bibfnamefont {Takashi}\ \bibnamefont {Oka}},
  \bibinfo {author} {\bibfnamefont {Arne}\ \bibnamefont {Brataas}}, \bibinfo
  {author} {\bibfnamefont {Liang}\ \bibnamefont {Fu}}, \ and\ \bibinfo {author}
  {\bibfnamefont {Eugene}\ \bibnamefont {Demler}},\ }\bibfield  {title}
  {\enquote {\bibinfo {title} {Transport properties of nonequilibrium systems
  under the application of light: Photoinduced quantum {Hall} insulators
  without {Landau} levels},}\ }\href {\doibase 10.1103/PhysRevB.84.235108}
  {\bibfield  {journal} {\bibinfo  {journal} {Phys. Rev. B}\ }\textbf {\bibinfo
  {volume} {84}},\ \bibinfo {pages} {235108} (\bibinfo {year}
  {2011})}\BibitemShut {NoStop}%
\bibitem [{\citenamefont {Gu}\ \emph {et~al.}(2011)\citenamefont {Gu},
  \citenamefont {Fertig}, \citenamefont {Arovas},\ and\ \citenamefont
  {Auerbach}}]{PhysRevLett.107.216601}%
  \BibitemOpen
  \bibfield  {author} {\bibinfo {author} {\bibfnamefont {Zhenghao}\
  \bibnamefont {Gu}}, \bibinfo {author} {\bibfnamefont {H.~A.}\ \bibnamefont
  {Fertig}}, \bibinfo {author} {\bibfnamefont {Daniel~P.}\ \bibnamefont
  {Arovas}}, \ and\ \bibinfo {author} {\bibfnamefont {Assa}\ \bibnamefont
  {Auerbach}},\ }\bibfield  {title} {\enquote {\bibinfo {title} {Floquet
  spectrum and transport through an irradiated graphene ribbon},}\ }\href
  {\doibase 10.1103/PhysRevLett.107.216601} {\bibfield  {journal} {\bibinfo
  {journal} {Phys. Rev. Lett.}\ }\textbf {\bibinfo {volume} {107}},\ \bibinfo
  {pages} {216601} (\bibinfo {year} {2011})}\BibitemShut {NoStop}%
\bibitem [{\citenamefont {Rudner}\ \emph {et~al.}(2013)\citenamefont {Rudner},
  \citenamefont {Lindner}, \citenamefont {Berg},\ and\ \citenamefont
  {Levin}}]{PhysRevX.3.031005}%
  \BibitemOpen
  \bibfield  {author} {\bibinfo {author} {\bibfnamefont {Mark~S.}\ \bibnamefont
  {Rudner}}, \bibinfo {author} {\bibfnamefont {Netanel~H.}\ \bibnamefont
  {Lindner}}, \bibinfo {author} {\bibfnamefont {Erez}\ \bibnamefont {Berg}}, \
  and\ \bibinfo {author} {\bibfnamefont {Michael}\ \bibnamefont {Levin}},\
  }\bibfield  {title} {\enquote {\bibinfo {title} {Anomalous edge states and
  the bulk-edge correspondence for periodically driven two-dimensional
  systems},}\ }\href {\doibase 10.1103/PhysRevX.3.031005} {\bibfield  {journal}
  {\bibinfo  {journal} {Phys. Rev. X}\ }\textbf {\bibinfo {volume} {3}},\
  \bibinfo {pages} {031005} (\bibinfo {year} {2013})}\BibitemShut {NoStop}%
\bibitem [{\citenamefont {Usaj}\ \emph {et~al.}(2014)\citenamefont {Usaj},
  \citenamefont {Perez-Piskunow}, \citenamefont {Foa~Torres},\ and\
  \citenamefont {Balseiro}}]{PhysRevB.90.115423}%
  \BibitemOpen
  \bibfield  {author} {\bibinfo {author} {\bibfnamefont {Gonzalo}\ \bibnamefont
  {Usaj}}, \bibinfo {author} {\bibfnamefont {P.~M.}\ \bibnamefont
  {Perez-Piskunow}}, \bibinfo {author} {\bibfnamefont {L.~E.~F.}\ \bibnamefont
  {Foa~Torres}}, \ and\ \bibinfo {author} {\bibfnamefont {C.~A.}\ \bibnamefont
  {Balseiro}},\ }\bibfield  {title} {\enquote {\bibinfo {title} {Irradiated
  graphene as a tunable {Floquet} topological insulator},}\ }\href {\doibase
  10.1103/PhysRevB.90.115423} {\bibfield  {journal} {\bibinfo  {journal} {Phys.
  Rev. B}\ }\textbf {\bibinfo {volume} {90}},\ \bibinfo {pages} {115423}
  (\bibinfo {year} {2014})}\BibitemShut {NoStop}%
\bibitem [{\citenamefont {Zenesini}\ \emph {et~al.}(2009)\citenamefont
  {Zenesini}, \citenamefont {Lignier}, \citenamefont {Ciampini}, \citenamefont
  {Morsch},\ and\ \citenamefont {Arimondo}}]{PhysRevLett.102.100403}%
  \BibitemOpen
  \bibfield  {author} {\bibinfo {author} {\bibfnamefont {Alessandro}\
  \bibnamefont {Zenesini}}, \bibinfo {author} {\bibfnamefont {Hans}\
  \bibnamefont {Lignier}}, \bibinfo {author} {\bibfnamefont {Donatella}\
  \bibnamefont {Ciampini}}, \bibinfo {author} {\bibfnamefont {Oliver}\
  \bibnamefont {Morsch}}, \ and\ \bibinfo {author} {\bibfnamefont {Ennio}\
  \bibnamefont {Arimondo}},\ }\bibfield  {title} {\enquote {\bibinfo {title}
  {Coherent control of dressed matter waves},}\ }\href {\doibase
  10.1103/PhysRevLett.102.100403} {\bibfield  {journal} {\bibinfo  {journal}
  {Phys. Rev. Lett.}\ }\textbf {\bibinfo {volume} {102}},\ \bibinfo {pages}
  {100403} (\bibinfo {year} {2009})}\BibitemShut {NoStop}%
\bibitem [{\citenamefont {Eckardt}\ \emph {et~al.}(2005)\citenamefont
  {Eckardt}, \citenamefont {Weiss},\ and\ \citenamefont
  {Holthaus}}]{PhysRevLett.95.260404}%
  \BibitemOpen
  \bibfield  {author} {\bibinfo {author} {\bibfnamefont {Andr\'e}\ \bibnamefont
  {Eckardt}}, \bibinfo {author} {\bibfnamefont {Christoph}\ \bibnamefont
  {Weiss}}, \ and\ \bibinfo {author} {\bibfnamefont {Martin}\ \bibnamefont
  {Holthaus}},\ }\bibfield  {title} {\enquote {\bibinfo {title}
  {Superfluid-insulator transition in a periodically driven optical lattice},}\
  }\href {\doibase 10.1103/PhysRevLett.95.260404} {\bibfield  {journal}
  {\bibinfo  {journal} {Phys. Rev. Lett.}\ }\textbf {\bibinfo {volume} {95}},\
  \bibinfo {pages} {260404} (\bibinfo {year} {2005})}\BibitemShut {NoStop}%
\bibitem [{\citenamefont {Yao}\ \emph {et~al.}(2017)\citenamefont {Yao},
  \citenamefont {Potter}, \citenamefont {Potirniche},\ and\ \citenamefont
  {Vishwanath}}]{PhysRevLett.118.030401}%
  \BibitemOpen
  \bibfield  {author} {\bibinfo {author} {\bibfnamefont {N.~Y.}\ \bibnamefont
  {Yao}}, \bibinfo {author} {\bibfnamefont {A.~C.}\ \bibnamefont {Potter}},
  \bibinfo {author} {\bibfnamefont {I.-D.}\ \bibnamefont {Potirniche}}, \ and\
  \bibinfo {author} {\bibfnamefont {A.}~\bibnamefont {Vishwanath}},\ }\bibfield
   {title} {\enquote {\bibinfo {title} {Discrete time crystals: Rigidity,
  criticality, and realizations},}\ }\href {\doibase
  10.1103/PhysRevLett.118.030401} {\bibfield  {journal} {\bibinfo  {journal}
  {Phys. Rev. Lett.}\ }\textbf {\bibinfo {volume} {118}},\ \bibinfo {pages}
  {030401} (\bibinfo {year} {2017})}\BibitemShut {NoStop}%
\bibitem [{\citenamefont {Potirniche}\ \emph {et~al.}(2017)\citenamefont
  {Potirniche}, \citenamefont {Potter}, \citenamefont {Schleier-Smith},
  \citenamefont {Vishwanath},\ and\ \citenamefont
  {Yao}}]{PhysRevLett.119.123601}%
  \BibitemOpen
  \bibfield  {author} {\bibinfo {author} {\bibfnamefont {I.-D.}\ \bibnamefont
  {Potirniche}}, \bibinfo {author} {\bibfnamefont {A.~C.}\ \bibnamefont
  {Potter}}, \bibinfo {author} {\bibfnamefont {M.}~\bibnamefont
  {Schleier-Smith}}, \bibinfo {author} {\bibfnamefont {A.}~\bibnamefont
  {Vishwanath}}, \ and\ \bibinfo {author} {\bibfnamefont {N.~Y.}\ \bibnamefont
  {Yao}},\ }\bibfield  {title} {\enquote {\bibinfo {title} {Floquet
  symmetry-protected topological phases in cold-atom systems},}\ }\href
  {\doibase 10.1103/PhysRevLett.119.123601} {\bibfield  {journal} {\bibinfo
  {journal} {Phys. Rev. Lett.}\ }\textbf {\bibinfo {volume} {119}},\ \bibinfo
  {pages} {123601} (\bibinfo {year} {2017})}\BibitemShut {NoStop}%
\bibitem [{\citenamefont {{Lerose}}\ \emph {et~al.}(2018)\citenamefont
  {{Lerose}}, \citenamefont {{Marino}}, \citenamefont {{Gambassi}},\ and\
  \citenamefont {{Silva}}}]{2018arXiv180304490L}%
  \BibitemOpen
  \bibfield  {author} {\bibinfo {author} {\bibfnamefont {A.}~\bibnamefont
  {{Lerose}}}, \bibinfo {author} {\bibfnamefont {J.}~\bibnamefont {{Marino}}},
  \bibinfo {author} {\bibfnamefont {A.}~\bibnamefont {{Gambassi}}}, \ and\
  \bibinfo {author} {\bibfnamefont {A.}~\bibnamefont {{Silva}}},\ }\href
  {https://arxiv.org/abs/1803.04490} {\enquote {\bibinfo {title} {{Quantum
  many-body Kapitza phases of periodically driven spin systems}},}\ } (\bibinfo
  {year} {2018}),\ \bibinfo {note} {arXiv:1803.04490}\BibitemShut {NoStop}%
\bibitem [{\citenamefont {Rehn}\ \emph {et~al.}(2016)\citenamefont {Rehn},
  \citenamefont {Lazarides}, \citenamefont {Pollmann},\ and\ \citenamefont
  {Moessner}}]{PhysRevB.94.020201}%
  \BibitemOpen
  \bibfield  {author} {\bibinfo {author} {\bibfnamefont {Jorge}\ \bibnamefont
  {Rehn}}, \bibinfo {author} {\bibfnamefont {Achilleas}\ \bibnamefont
  {Lazarides}}, \bibinfo {author} {\bibfnamefont {Frank}\ \bibnamefont
  {Pollmann}}, \ and\ \bibinfo {author} {\bibfnamefont {Roderich}\ \bibnamefont
  {Moessner}},\ }\bibfield  {title} {\enquote {\bibinfo {title} {How periodic
  driving heats a disordered quantum spin chain},}\ }\href {\doibase
  10.1103/PhysRevB.94.020201} {\bibfield  {journal} {\bibinfo  {journal} {Phys.
  Rev. B}\ }\textbf {\bibinfo {volume} {94}},\ \bibinfo {pages} {020201}
  (\bibinfo {year} {2016})}\BibitemShut {NoStop}%
\bibitem [{\citenamefont {D'Alessio}\ and\ \citenamefont
  {Rigol}(2014)}]{PhysRevX.4.041048}%
  \BibitemOpen
  \bibfield  {author} {\bibinfo {author} {\bibfnamefont {Luca}\ \bibnamefont
  {D'Alessio}}\ and\ \bibinfo {author} {\bibfnamefont {Marcos}\ \bibnamefont
  {Rigol}},\ }\bibfield  {title} {\enquote {\bibinfo {title} {Long-time
  behavior of isolated periodically driven interacting lattice systems},}\
  }\href {\doibase 10.1103/PhysRevX.4.041048} {\bibfield  {journal} {\bibinfo
  {journal} {Phys. Rev. X}\ }\textbf {\bibinfo {volume} {4}},\ \bibinfo {pages}
  {041048} (\bibinfo {year} {2014})}\BibitemShut {NoStop}%
\bibitem [{\citenamefont {Abanin}\ \emph {et~al.}(2015)\citenamefont {Abanin},
  \citenamefont {De~Roeck},\ and\ \citenamefont
  {Huveneers}}]{PhysRevLett.115.256803}%
  \BibitemOpen
  \bibfield  {author} {\bibinfo {author} {\bibfnamefont {Dmitry~A.}\
  \bibnamefont {Abanin}}, \bibinfo {author} {\bibfnamefont {Wojciech}\
  \bibnamefont {De~Roeck}}, \ and\ \bibinfo {author} {\bibfnamefont
  {Fran\ifmmode \mbox{\c{c}}\else{}\c{c}\fi{}ois}\ \bibnamefont {Huveneers}},\
  }\bibfield  {title} {\enquote {\bibinfo {title} {Exponentially slow heating
  in periodically driven many-body systems},}\ }\href {\doibase
  10.1103/PhysRevLett.115.256803} {\bibfield  {journal} {\bibinfo  {journal}
  {Phys. Rev. Lett.}\ }\textbf {\bibinfo {volume} {115}},\ \bibinfo {pages}
  {256803} (\bibinfo {year} {2015})}\BibitemShut {NoStop}%
\bibitem [{\citenamefont {Abanin}\ \emph {et~al.}(2017)\citenamefont {Abanin},
  \citenamefont {De~Roeck}, \citenamefont {Ho},\ and\ \citenamefont
  {Huveneers}}]{PhysRevB.95.014112}%
  \BibitemOpen
  \bibfield  {author} {\bibinfo {author} {\bibfnamefont {Dmitry~A.}\
  \bibnamefont {Abanin}}, \bibinfo {author} {\bibfnamefont {Wojciech}\
  \bibnamefont {De~Roeck}}, \bibinfo {author} {\bibfnamefont {Wen~Wei}\
  \bibnamefont {Ho}}, \ and\ \bibinfo {author} {\bibfnamefont {Fran\ifmmode
  \mbox{\c{c}}\else{}\c{c}\fi{}ois}\ \bibnamefont {Huveneers}},\ }\bibfield
  {title} {\enquote {\bibinfo {title} {Effective {Hamiltonians},
  prethermalization, and slow energy absorption in periodically driven
  many-body systems},}\ }\href {\doibase 10.1103/PhysRevB.95.014112} {\bibfield
   {journal} {\bibinfo  {journal} {Phys. Rev. B}\ }\textbf {\bibinfo {volume}
  {95}},\ \bibinfo {pages} {014112} (\bibinfo {year} {2017})}\BibitemShut
  {NoStop}%
\bibitem [{\citenamefont {Mori}\ \emph {et~al.}(2016)\citenamefont {Mori},
  \citenamefont {Kuwahara},\ and\ \citenamefont
  {Saito}}]{PhysRevLett.116.120401}%
  \BibitemOpen
  \bibfield  {author} {\bibinfo {author} {\bibfnamefont {Takashi}\ \bibnamefont
  {Mori}}, \bibinfo {author} {\bibfnamefont {Tomotaka}\ \bibnamefont
  {Kuwahara}}, \ and\ \bibinfo {author} {\bibfnamefont {Keiji}\ \bibnamefont
  {Saito}},\ }\bibfield  {title} {\enquote {\bibinfo {title} {Rigorous bound on
  energy absorption and generic relaxation in periodically driven quantum
  systems},}\ }\href {\doibase 10.1103/PhysRevLett.116.120401} {\bibfield
  {journal} {\bibinfo  {journal} {Phys. Rev. Lett.}\ }\textbf {\bibinfo
  {volume} {116}},\ \bibinfo {pages} {120401} (\bibinfo {year}
  {2016})}\BibitemShut {NoStop}%
\bibitem [{\citenamefont {Else}\ \emph {et~al.}(2017)\citenamefont {Else},
  \citenamefont {Bauer},\ and\ \citenamefont {Nayak}}]{PhysRevX.7.011026}%
  \BibitemOpen
  \bibfield  {author} {\bibinfo {author} {\bibfnamefont {Dominic~V.}\
  \bibnamefont {Else}}, \bibinfo {author} {\bibfnamefont {Bela}\ \bibnamefont
  {Bauer}}, \ and\ \bibinfo {author} {\bibfnamefont {Chetan}\ \bibnamefont
  {Nayak}},\ }\bibfield  {title} {\enquote {\bibinfo {title} {Prethermal phases
  of matter protected by time-translation symmetry},}\ }\href {\doibase
  10.1103/PhysRevX.7.011026} {\bibfield  {journal} {\bibinfo  {journal} {Phys.
  Rev. X}\ }\textbf {\bibinfo {volume} {7}},\ \bibinfo {pages} {011026}
  (\bibinfo {year} {2017})}\BibitemShut {NoStop}%
\bibitem [{\citenamefont {Canovi}\ \emph {et~al.}(2016)\citenamefont {Canovi},
  \citenamefont {Kollar},\ and\ \citenamefont {Eckstein}}]{PhysRevE.93.012130}%
  \BibitemOpen
  \bibfield  {author} {\bibinfo {author} {\bibfnamefont {Elena}\ \bibnamefont
  {Canovi}}, \bibinfo {author} {\bibfnamefont {Marcus}\ \bibnamefont {Kollar}},
  \ and\ \bibinfo {author} {\bibfnamefont {Martin}\ \bibnamefont {Eckstein}},\
  }\bibfield  {title} {\enquote {\bibinfo {title} {Stroboscopic
  prethermalization in weakly interacting periodically driven systems},}\
  }\href {\doibase 10.1103/PhysRevE.93.012130} {\bibfield  {journal} {\bibinfo
  {journal} {Phys. Rev. E}\ }\textbf {\bibinfo {volume} {93}},\ \bibinfo
  {pages} {012130} (\bibinfo {year} {2016})}\BibitemShut {NoStop}%
\bibitem [{\citenamefont {{Machado}}\ \emph {et~al.}(2017)\citenamefont
  {{Machado}}, \citenamefont {{Meyer}}, \citenamefont {{Else}}, \citenamefont
  {{Nayak}},\ and\ \citenamefont {{Yao}}}]{Machado2017}%
  \BibitemOpen
  \bibfield  {author} {\bibinfo {author} {\bibfnamefont {F.}~\bibnamefont
  {{Machado}}}, \bibinfo {author} {\bibfnamefont {G.~D.}\ \bibnamefont
  {{Meyer}}}, \bibinfo {author} {\bibfnamefont {D.~V.}\ \bibnamefont {{Else}}},
  \bibinfo {author} {\bibfnamefont {C.}~\bibnamefont {{Nayak}}}, \ and\
  \bibinfo {author} {\bibfnamefont {N.~Y.}\ \bibnamefont {{Yao}}},\ }\href
  {https://arxiv.org/abs/1708.01620} {\enquote {\bibinfo {title} {Exponentially
  slow heating in short and long-range interacting {Floquet} systems},}\ }
  (\bibinfo {year} {2017}),\ \bibinfo {note} {arXiv:1708.01620}\BibitemShut
  {NoStop}%
\bibitem [{\citenamefont {{Weidinger}}\ and\ \citenamefont
  {{Knap}}(2017)}]{Weidinger2016}%
  \BibitemOpen
  \bibfield  {author} {\bibinfo {author} {\bibfnamefont {S.~A.}\ \bibnamefont
  {{Weidinger}}}\ and\ \bibinfo {author} {\bibfnamefont {M.}~\bibnamefont
  {{Knap}}},\ }\bibfield  {title} {\enquote {\bibinfo {title} {Floquet
  prethermalization and regimes of heating in a periodically driven,
  interacting quantum system},}\ }\href {\doibase 10.1038/srep45382} {\bibfield
   {journal} {\bibinfo  {journal} {Sci. Rep.}\ }\textbf {\bibinfo {volume}
  {7}},\ \bibinfo {pages} {45382} (\bibinfo {year} {2017})}\BibitemShut
  {NoStop}%
\bibitem [{\citenamefont {Kuwahara}\ \emph {et~al.}(2016)\citenamefont
  {Kuwahara}, \citenamefont {Mori},\ and\ \citenamefont
  {Saito}}]{Kuwahara2016}%
  \BibitemOpen
  \bibfield  {author} {\bibinfo {author} {\bibfnamefont {Tomotaka}\
  \bibnamefont {Kuwahara}}, \bibinfo {author} {\bibfnamefont {Takashi}\
  \bibnamefont {Mori}}, \ and\ \bibinfo {author} {\bibfnamefont {Keiji}\
  \bibnamefont {Saito}},\ }\bibfield  {title} {\enquote {\bibinfo {title}
  {{Floquet-Magnus} theory and generic transient dynamics in periodically
  driven many-body quantum systems},}\ }\href {\doibase
  https://doi.org/10.1016/j.aop.2016.01.012} {\bibfield  {journal} {\bibinfo
  {journal} {Ann. Phys. (N.Y.)}\ }\textbf {\bibinfo {volume} {367}},\ \bibinfo
  {pages} {96 -- 124} (\bibinfo {year} {2016})}\BibitemShut {NoStop}%
\bibitem [{\citenamefont {Heyl}\ \emph {et~al.}(2013)\citenamefont {Heyl},
  \citenamefont {Polkovnikov},\ and\ \citenamefont
  {Kehrein}}]{PhysRevLett.110.135704}%
  \BibitemOpen
  \bibfield  {author} {\bibinfo {author} {\bibfnamefont {M.}~\bibnamefont
  {Heyl}}, \bibinfo {author} {\bibfnamefont {A.}~\bibnamefont {Polkovnikov}}, \
  and\ \bibinfo {author} {\bibfnamefont {S.}~\bibnamefont {Kehrein}},\
  }\bibfield  {title} {\enquote {\bibinfo {title} {Dynamical quantum phase
  transitions in the transverse-field {Ising} model},}\ }\href {\doibase
  10.1103/PhysRevLett.110.135704} {\bibfield  {journal} {\bibinfo  {journal}
  {Phys. Rev. Lett.}\ }\textbf {\bibinfo {volume} {110}},\ \bibinfo {pages}
  {135704} (\bibinfo {year} {2013})}\BibitemShut {NoStop}%
\bibitem [{\citenamefont {Jurcevic}\ \emph {et~al.}(2017)\citenamefont
  {Jurcevic}, \citenamefont {Shen}, \citenamefont {Hauke}, \citenamefont
  {Maier}, \citenamefont {Brydges}, \citenamefont {Hempel}, \citenamefont
  {Lanyon}, \citenamefont {Heyl}, \citenamefont {Blatt},\ and\ \citenamefont
  {Roos}}]{PhysRevLett.119.080501}%
  \BibitemOpen
  \bibfield  {author} {\bibinfo {author} {\bibfnamefont {P.}~\bibnamefont
  {Jurcevic}}, \bibinfo {author} {\bibfnamefont {H.}~\bibnamefont {Shen}},
  \bibinfo {author} {\bibfnamefont {P.}~\bibnamefont {Hauke}}, \bibinfo
  {author} {\bibfnamefont {C.}~\bibnamefont {Maier}}, \bibinfo {author}
  {\bibfnamefont {T.}~\bibnamefont {Brydges}}, \bibinfo {author} {\bibfnamefont
  {C.}~\bibnamefont {Hempel}}, \bibinfo {author} {\bibfnamefont {B.~P.}\
  \bibnamefont {Lanyon}}, \bibinfo {author} {\bibfnamefont {M.}~\bibnamefont
  {Heyl}}, \bibinfo {author} {\bibfnamefont {R.}~\bibnamefont {Blatt}}, \ and\
  \bibinfo {author} {\bibfnamefont {C.~F.}\ \bibnamefont {Roos}},\ }\bibfield
  {title} {\enquote {\bibinfo {title} {Direct observation of dynamical quantum
  phase transitions in an interacting many-body system},}\ }\href {\doibase
  10.1103/PhysRevLett.119.080501} {\bibfield  {journal} {\bibinfo  {journal}
  {Phys. Rev. Lett.}\ }\textbf {\bibinfo {volume} {119}},\ \bibinfo {pages}
  {080501} (\bibinfo {year} {2017})}\BibitemShut {NoStop}%
\bibitem [{\citenamefont {{Fl{\"a}schner}}\ \emph {et~al.}(2018)\citenamefont
  {{Fl{\"a}schner}}, \citenamefont {{Vogel}}, \citenamefont {{Tarnowski}},
  \citenamefont {{Rem}}, \citenamefont {{L{\"u}hmann}}, \citenamefont {{Heyl}},
  \citenamefont {{Budich}}, \citenamefont {{Mathey}}, \citenamefont
  {{Sengstock}},\ and\ \citenamefont {{Weitenberg}}}]{Flaeschner2016}%
  \BibitemOpen
  \bibfield  {author} {\bibinfo {author} {\bibfnamefont {N.}~\bibnamefont
  {{Fl{\"a}schner}}}, \bibinfo {author} {\bibfnamefont {D.}~\bibnamefont
  {{Vogel}}}, \bibinfo {author} {\bibfnamefont {M.}~\bibnamefont
  {{Tarnowski}}}, \bibinfo {author} {\bibfnamefont {B.~S.}\ \bibnamefont
  {{Rem}}}, \bibinfo {author} {\bibfnamefont {D.-S.}\ \bibnamefont
  {{L{\"u}hmann}}}, \bibinfo {author} {\bibfnamefont {M.}~\bibnamefont
  {{Heyl}}}, \bibinfo {author} {\bibfnamefont {J.~C.}\ \bibnamefont
  {{Budich}}}, \bibinfo {author} {\bibfnamefont {L.}~\bibnamefont {{Mathey}}},
  \bibinfo {author} {\bibfnamefont {K.}~\bibnamefont {{Sengstock}}}, \ and\
  \bibinfo {author} {\bibfnamefont {C.}~\bibnamefont {{Weitenberg}}},\
  }\bibfield  {title} {\enquote {\bibinfo {title} {Observation of a dynamical
  topological phase transition},}\ }\href {\doibase 10.1038/s41567-017-0013-8}
  {\bibfield  {journal} {\bibinfo  {journal} {Nat. Phys.}\ }\textbf {\bibinfo
  {volume} {14}},\ \bibinfo {pages} {265--268} (\bibinfo {year}
  {2018})}\BibitemShut {NoStop}%
\bibitem [{\citenamefont {Du}\ and\ \citenamefont
  {Fiete}(2018)}]{PhysRevB.97.085152}%
  \BibitemOpen
  \bibfield  {author} {\bibinfo {author} {\bibfnamefont {Liang}\ \bibnamefont
  {Du}}\ and\ \bibinfo {author} {\bibfnamefont {Gregory~A.}\ \bibnamefont
  {Fiete}},\ }\bibfield  {title} {\enquote {\bibinfo {title} {Dynamical
  recovery of {SU(2)} symmetry in the mass-quenched {Hubbard} model},}\ }\href
  {\doibase 10.1103/PhysRevB.97.085152} {\bibfield  {journal} {\bibinfo
  {journal} {Phys. Rev. B}\ }\textbf {\bibinfo {volume} {97}},\ \bibinfo
  {pages} {085152} (\bibinfo {year} {2018})}\BibitemShut {NoStop}%
\bibitem [{\citenamefont {Chen}\ \emph {et~al.}(2011)\citenamefont {Chen},
  \citenamefont {Nascimb\`ene}, \citenamefont {Aidelsburger}, \citenamefont
  {Atala}, \citenamefont {Trotzky},\ and\ \citenamefont
  {Bloch}}]{PhysRevLett.107.210405}%
  \BibitemOpen
  \bibfield  {author} {\bibinfo {author} {\bibfnamefont {Yu-Ao}\ \bibnamefont
  {Chen}}, \bibinfo {author} {\bibfnamefont {Sylvain}\ \bibnamefont
  {Nascimb\`ene}}, \bibinfo {author} {\bibfnamefont {Monika}\ \bibnamefont
  {Aidelsburger}}, \bibinfo {author} {\bibfnamefont {Marcos}\ \bibnamefont
  {Atala}}, \bibinfo {author} {\bibfnamefont {Stefan}\ \bibnamefont {Trotzky}},
  \ and\ \bibinfo {author} {\bibfnamefont {Immanuel}\ \bibnamefont {Bloch}},\
  }\bibfield  {title} {\enquote {\bibinfo {title} {Controlling correlated
  tunneling and superexchange interactions with ac-driven optical lattices},}\
  }\href {\doibase 10.1103/PhysRevLett.107.210405} {\bibfield  {journal}
  {\bibinfo  {journal} {Phys. Rev. Lett.}\ }\textbf {\bibinfo {volume} {107}},\
  \bibinfo {pages} {210405} (\bibinfo {year} {2011})}\BibitemShut {NoStop}%
\bibitem [{\citenamefont {Dutreix}\ and\ \citenamefont
  {Katsnelson}(2017)}]{PhysRevB.95.024306}%
  \BibitemOpen
  \bibfield  {author} {\bibinfo {author} {\bibfnamefont {C.}~\bibnamefont
  {Dutreix}}\ and\ \bibinfo {author} {\bibfnamefont {M.~I.}\ \bibnamefont
  {Katsnelson}},\ }\bibfield  {title} {\enquote {\bibinfo {title} {Dynamical
  control of electron-phonon interactions with high-frequency light},}\ }\href
  {\doibase 10.1103/PhysRevB.95.024306} {\bibfield  {journal} {\bibinfo
  {journal} {Phys. Rev. B}\ }\textbf {\bibinfo {volume} {95}},\ \bibinfo
  {pages} {024306} (\bibinfo {year} {2017})}\BibitemShut {NoStop}%
\bibitem [{\citenamefont {Bukov}\ \emph
  {et~al.}(2015{\natexlab{b}})\citenamefont {Bukov}, \citenamefont
  {Gopalakrishnan}, \citenamefont {Knap},\ and\ \citenamefont
  {Demler}}]{PhysRevLett.115.205301}%
  \BibitemOpen
  \bibfield  {author} {\bibinfo {author} {\bibfnamefont {Marin}\ \bibnamefont
  {Bukov}}, \bibinfo {author} {\bibfnamefont {Sarang}\ \bibnamefont
  {Gopalakrishnan}}, \bibinfo {author} {\bibfnamefont {Michael}\ \bibnamefont
  {Knap}}, \ and\ \bibinfo {author} {\bibfnamefont {Eugene}\ \bibnamefont
  {Demler}},\ }\bibfield  {title} {\enquote {\bibinfo {title} {Prethermal
  {Floquet} steady states and instabilities in the periodically driven, weakly
  interacting {Bose-Hubbard} model},}\ }\href {\doibase
  10.1103/PhysRevLett.115.205301} {\bibfield  {journal} {\bibinfo  {journal}
  {Phys. Rev. Lett.}\ }\textbf {\bibinfo {volume} {115}},\ \bibinfo {pages}
  {205301} (\bibinfo {year} {2015}{\natexlab{b}})}\BibitemShut {NoStop}%
\bibitem [{\citenamefont {Bukov}\ \emph {et~al.}(2016)\citenamefont {Bukov},
  \citenamefont {Kolodrubetz},\ and\ \citenamefont
  {Polkovnikov}}]{PhysRevLett.116.125301}%
  \BibitemOpen
  \bibfield  {author} {\bibinfo {author} {\bibfnamefont {Marin}\ \bibnamefont
  {Bukov}}, \bibinfo {author} {\bibfnamefont {Michael}\ \bibnamefont
  {Kolodrubetz}}, \ and\ \bibinfo {author} {\bibfnamefont {Anatoli}\
  \bibnamefont {Polkovnikov}},\ }\bibfield  {title} {\enquote {\bibinfo {title}
  {{Schrieffer-Wolff} transformation for periodically driven systems: Strongly
  correlated systems with artificial gauge fields},}\ }\href {\doibase
  10.1103/PhysRevLett.116.125301} {\bibfield  {journal} {\bibinfo  {journal}
  {Phys. Rev. Lett.}\ }\textbf {\bibinfo {volume} {116}},\ \bibinfo {pages}
  {125301} (\bibinfo {year} {2016})}\BibitemShut {NoStop}%
\bibitem [{\citenamefont {Peronaci}\ \emph {et~al.}(2018)\citenamefont
  {Peronaci}, \citenamefont {Schir\'o},\ and\ \citenamefont
  {Parcollet}}]{Peronaci2017}%
  \BibitemOpen
  \bibfield  {author} {\bibinfo {author} {\bibfnamefont {Francesco}\
  \bibnamefont {Peronaci}}, \bibinfo {author} {\bibfnamefont {Marco}\
  \bibnamefont {Schir\'o}}, \ and\ \bibinfo {author} {\bibfnamefont {Olivier}\
  \bibnamefont {Parcollet}},\ }\bibfield  {title} {\enquote {\bibinfo {title}
  {Resonant thermalization of periodically driven strongly correlated
  electrons},}\ }\href {\doibase 10.1103/PhysRevLett.120.197601} {\bibfield
  {journal} {\bibinfo  {journal} {Phys. Rev. Lett.}\ }\textbf {\bibinfo
  {volume} {120}},\ \bibinfo {pages} {197601} (\bibinfo {year}
  {2018})}\BibitemShut {NoStop}%
\bibitem [{\citenamefont {Lindner}\ \emph {et~al.}(2017)\citenamefont
  {Lindner}, \citenamefont {Berg},\ and\ \citenamefont
  {Rudner}}]{PhysRevX.7.011018}%
  \BibitemOpen
  \bibfield  {author} {\bibinfo {author} {\bibfnamefont {Netanel~H.}\
  \bibnamefont {Lindner}}, \bibinfo {author} {\bibfnamefont {Erez}\
  \bibnamefont {Berg}}, \ and\ \bibinfo {author} {\bibfnamefont {Mark~S.}\
  \bibnamefont {Rudner}},\ }\bibfield  {title} {\enquote {\bibinfo {title}
  {Universal chiral quasisteady states in periodically driven many-body
  systems},}\ }\href {\doibase 10.1103/PhysRevX.7.011018} {\bibfield  {journal}
  {\bibinfo  {journal} {Phys. Rev. X}\ }\textbf {\bibinfo {volume} {7}},\
  \bibinfo {pages} {011018} (\bibinfo {year} {2017})}\BibitemShut {NoStop}%
\bibitem [{\citenamefont {Plekhanov}\ \emph {et~al.}(2017)\citenamefont
  {Plekhanov}, \citenamefont {Roux},\ and\ \citenamefont
  {Le~Hur}}]{PhysRevB.95.045102}%
  \BibitemOpen
  \bibfield  {author} {\bibinfo {author} {\bibfnamefont {Kirill}\ \bibnamefont
  {Plekhanov}}, \bibinfo {author} {\bibfnamefont {Guillaume}\ \bibnamefont
  {Roux}}, \ and\ \bibinfo {author} {\bibfnamefont {Karyn}\ \bibnamefont
  {Le~Hur}},\ }\bibfield  {title} {\enquote {\bibinfo {title} {Floquet
  engineering of {Haldane Chern} insulators and chiral bosonic phase
  transitions},}\ }\href {\doibase 10.1103/PhysRevB.95.045102} {\bibfield
  {journal} {\bibinfo  {journal} {Phys. Rev. B}\ }\textbf {\bibinfo {volume}
  {95}},\ \bibinfo {pages} {045102} (\bibinfo {year} {2017})}\BibitemShut
  {NoStop}%
\bibitem [{\citenamefont {Blanes}\ \emph {et~al.}(2009)\citenamefont {Blanes},
  \citenamefont {Casas}, \citenamefont {Oteo},\ and\ \citenamefont
  {Ros}}]{Blanes2009}%
  \BibitemOpen
  \bibfield  {author} {\bibinfo {author} {\bibfnamefont {S.}~\bibnamefont
  {Blanes}}, \bibinfo {author} {\bibfnamefont {F.}~\bibnamefont {Casas}},
  \bibinfo {author} {\bibfnamefont {J.A.}\ \bibnamefont {Oteo}}, \ and\
  \bibinfo {author} {\bibfnamefont {J.}~\bibnamefont {Ros}},\ }\bibfield
  {title} {\enquote {\bibinfo {title} {The {Magnus} expansion and some of its
  applications},}\ }\href {\doibase
  http://dx.doi.org/10.1016/j.physrep.2008.11.001} {\bibfield  {journal}
  {\bibinfo  {journal} {Phys. Rep.}\ }\textbf {\bibinfo {volume} {470}},\
  \bibinfo {pages} {151 -- 238} (\bibinfo {year} {2009})}\BibitemShut {NoStop}%
\bibitem [{\citenamefont {Fel'dman}(1984)}]{Feldm1984}%
  \BibitemOpen
  \bibfield  {author} {\bibinfo {author} {\bibfnamefont {E.B.}\ \bibnamefont
  {Fel'dman}},\ }\bibfield  {title} {\enquote {\bibinfo {title} {On the
  convergence of the {Magnus} expansion for spin systems in periodic magnetic
  fields},}\ }\href {\doibase http://dx.doi.org/10.1016/0375-9601(84)90027-6}
  {\bibfield  {journal} {\bibinfo  {journal} {Phys. Lett. A}\ }\textbf
  {\bibinfo {volume} {104}},\ \bibinfo {pages} {479 -- 481} (\bibinfo {year}
  {1984})}\BibitemShut {NoStop}%
\bibitem [{\citenamefont {Magnus}(1954)}]{Magnus1954}%
  \BibitemOpen
  \bibfield  {author} {\bibinfo {author} {\bibfnamefont {Wilhelm}\ \bibnamefont
  {Magnus}},\ }\bibfield  {title} {\enquote {\bibinfo {title} {On the
  exponential solution of differential equations for a linear operator},}\
  }\href {\doibase 10.1002/cpa.3160070404} {\bibfield  {journal} {\bibinfo
  {journal} {Commun. Pure Appl. Math.}\ }\textbf {\bibinfo {volume} {7}},\
  \bibinfo {pages} {649--673} (\bibinfo {year} {1954})}\BibitemShut {NoStop}%
\bibitem [{\citenamefont {Rahav}\ \emph {et~al.}(2003)\citenamefont {Rahav},
  \citenamefont {Gilary},\ and\ \citenamefont {Fishman}}]{PhysRevA.68.013820}%
  \BibitemOpen
  \bibfield  {author} {\bibinfo {author} {\bibfnamefont {Saar}\ \bibnamefont
  {Rahav}}, \bibinfo {author} {\bibfnamefont {Ido}\ \bibnamefont {Gilary}}, \
  and\ \bibinfo {author} {\bibfnamefont {Shmuel}\ \bibnamefont {Fishman}},\
  }\bibfield  {title} {\enquote {\bibinfo {title} {Effective {Hamiltonians} for
  periodically driven systems},}\ }\href {\doibase 10.1103/PhysRevA.68.013820}
  {\bibfield  {journal} {\bibinfo  {journal} {Phys. Rev. A}\ }\textbf {\bibinfo
  {volume} {68}},\ \bibinfo {pages} {013820} (\bibinfo {year}
  {2003})}\BibitemShut {NoStop}%
\bibitem [{\citenamefont {Goldman}\ and\ \citenamefont
  {Dalibard}(2014)}]{PhysRevX.4.031027}%
  \BibitemOpen
  \bibfield  {author} {\bibinfo {author} {\bibfnamefont {N.}~\bibnamefont
  {Goldman}}\ and\ \bibinfo {author} {\bibfnamefont {J.}~\bibnamefont
  {Dalibard}},\ }\bibfield  {title} {\enquote {\bibinfo {title} {Periodically
  driven quantum systems: Effective {Hamiltonians} and engineered gauge
  fields},}\ }\href {\doibase 10.1103/PhysRevX.4.031027} {\bibfield  {journal}
  {\bibinfo  {journal} {Phys. Rev. X}\ }\textbf {\bibinfo {volume} {4}},\
  \bibinfo {pages} {031027} (\bibinfo {year} {2014})}\BibitemShut {NoStop}%
\bibitem [{\citenamefont {Itin}\ and\ \citenamefont
  {Katsnelson}(2015)}]{PhysRevLett.115.075301}%
  \BibitemOpen
  \bibfield  {author} {\bibinfo {author} {\bibfnamefont {A.~P.}\ \bibnamefont
  {Itin}}\ and\ \bibinfo {author} {\bibfnamefont {M.~I.}\ \bibnamefont
  {Katsnelson}},\ }\bibfield  {title} {\enquote {\bibinfo {title} {Effective
  {Hamiltonians} for rapidly driven many-body lattice systems: Induced exchange
  interactions and density-dependent hoppings},}\ }\href {\doibase
  10.1103/PhysRevLett.115.075301} {\bibfield  {journal} {\bibinfo  {journal}
  {Phys. Rev. Lett.}\ }\textbf {\bibinfo {volume} {115}},\ \bibinfo {pages}
  {075301} (\bibinfo {year} {2015})}\BibitemShut {NoStop}%
\bibitem [{\citenamefont {Eckardt}\ and\ \citenamefont
  {Anisimovas}(2015)}]{Eckardt2015}%
  \BibitemOpen
  \bibfield  {author} {\bibinfo {author} {\bibfnamefont {Andr\'e}\ \bibnamefont
  {Eckardt}}\ and\ \bibinfo {author} {\bibfnamefont {Egidijus}\ \bibnamefont
  {Anisimovas}},\ }\bibfield  {title} {\enquote {\bibinfo {title}
  {High-frequency approximation for periodically driven quantum systems from a
  {Floquet}-space perspective},}\ }\href
  {http://stacks.iop.org/1367-2630/17/i=9/a=093039} {\bibfield  {journal}
  {\bibinfo  {journal} {New J. of Phys.}\ }\textbf {\bibinfo {volume} {17}},\
  \bibinfo {pages} {093039} (\bibinfo {year} {2015})}\BibitemShut {NoStop}%
\bibitem [{\citenamefont {Mikami}\ \emph {et~al.}(2016)\citenamefont {Mikami},
  \citenamefont {Kitamura}, \citenamefont {Yasuda}, \citenamefont {Tsuji},
  \citenamefont {Oka},\ and\ \citenamefont {Aoki}}]{PhysRevB.93.144307}%
  \BibitemOpen
  \bibfield  {author} {\bibinfo {author} {\bibfnamefont {Takahiro}\
  \bibnamefont {Mikami}}, \bibinfo {author} {\bibfnamefont {Sota}\ \bibnamefont
  {Kitamura}}, \bibinfo {author} {\bibfnamefont {Kenji}\ \bibnamefont
  {Yasuda}}, \bibinfo {author} {\bibfnamefont {Naoto}\ \bibnamefont {Tsuji}},
  \bibinfo {author} {\bibfnamefont {Takashi}\ \bibnamefont {Oka}}, \ and\
  \bibinfo {author} {\bibfnamefont {Hideo}\ \bibnamefont {Aoki}},\ }\bibfield
  {title} {\enquote {\bibinfo {title} {{Brillouin-Wigner} theory for
  high-frequency expansion in periodically driven systems: Application to
  {Floquet} topological insulators},}\ }\href {\doibase
  10.1103/PhysRevB.93.144307} {\bibfield  {journal} {\bibinfo  {journal} {Phys.
  Rev. B}\ }\textbf {\bibinfo {volume} {93}},\ \bibinfo {pages} {144307}
  (\bibinfo {year} {2016})}\BibitemShut {NoStop}%
\bibitem [{\citenamefont {Mohan}\ \emph {et~al.}(2016)\citenamefont {Mohan},
  \citenamefont {Saxena}, \citenamefont {Kundu},\ and\ \citenamefont
  {Rao}}]{PhysRevB.94.235419}%
  \BibitemOpen
  \bibfield  {author} {\bibinfo {author} {\bibfnamefont {Priyanka}\
  \bibnamefont {Mohan}}, \bibinfo {author} {\bibfnamefont {Ruchi}\ \bibnamefont
  {Saxena}}, \bibinfo {author} {\bibfnamefont {Arijit}\ \bibnamefont {Kundu}},
  \ and\ \bibinfo {author} {\bibfnamefont {Sumathi}\ \bibnamefont {Rao}},\
  }\bibfield  {title} {\enquote {\bibinfo {title} {{Brillouin-Wigner} theory
  for {Floquet} topological phase transitions in spin-orbit-coupled
  materials},}\ }\href {\doibase 10.1103/PhysRevB.94.235419} {\bibfield
  {journal} {\bibinfo  {journal} {Phys. Rev. B}\ }\textbf {\bibinfo {volume}
  {94}},\ \bibinfo {pages} {235419} (\bibinfo {year} {2016})}\BibitemShut
  {NoStop}%
\bibitem [{\citenamefont {Maricq}(1982)}]{PhysRevB.25.6622}%
  \BibitemOpen
  \bibfield  {author} {\bibinfo {author} {\bibfnamefont {M.~Matti}\
  \bibnamefont {Maricq}},\ }\bibfield  {title} {\enquote {\bibinfo {title}
  {Application of average {Hamiltonian} theory to the {NMR} of solids},}\
  }\href {\doibase 10.1103/PhysRevB.25.6622} {\bibfield  {journal} {\bibinfo
  {journal} {Phys. Rev. B}\ }\textbf {\bibinfo {volume} {25}},\ \bibinfo
  {pages} {6622--6632} (\bibinfo {year} {1982})}\BibitemShut {NoStop}%
\bibitem [{\citenamefont {Vajna}\ \emph {et~al.}(2018)\citenamefont {Vajna},
  \citenamefont {Klobas}, \citenamefont {Prosen},\ and\ \citenamefont
  {Polkovnikov}}]{PhysRevLett.120.200607}%
  \BibitemOpen
  \bibfield  {author} {\bibinfo {author} {\bibfnamefont {Szabolcs}\
  \bibnamefont {Vajna}}, \bibinfo {author} {\bibfnamefont {Katja}\ \bibnamefont
  {Klobas}}, \bibinfo {author} {\bibfnamefont
  {Toma\ifmmode{}\check{z}\else{}\v{z}\fi{}}\ \bibnamefont {Prosen}}, \ and\
  \bibinfo {author} {\bibfnamefont {Anatoli}\ \bibnamefont {Polkovnikov}},\
  }\bibfield  {title} {\enquote {\bibinfo {title} {Replica resummation of the
  {Baker-Campbell-Hausdorff} series},}\ }\href {\doibase
  10.1103/PhysRevLett.120.200607} {\bibfield  {journal} {\bibinfo  {journal}
  {Phys. Rev. Lett.}\ }\textbf {\bibinfo {volume} {120}},\ \bibinfo {pages}
  {200607} (\bibinfo {year} {2018})}\BibitemShut {NoStop}%
\bibitem [{\citenamefont {Bordia}\ \emph {et~al.}(2017)\citenamefont {Bordia},
  \citenamefont {L\"uschen}, \citenamefont {Schneider}, \citenamefont {Knap},\
  and\ \citenamefont {Bloch}}]{10.1038/nphys4020}%
  \BibitemOpen
  \bibfield  {author} {\bibinfo {author} {\bibfnamefont {Pranjal}\ \bibnamefont
  {Bordia}}, \bibinfo {author} {\bibfnamefont {Henrik}\ \bibnamefont
  {L\"uschen}}, \bibinfo {author} {\bibfnamefont {Ulrich}\ \bibnamefont
  {Schneider}}, \bibinfo {author} {\bibfnamefont {Michael}\ \bibnamefont
  {Knap}}, \ and\ \bibinfo {author} {\bibfnamefont {Immanuel}\ \bibnamefont
  {Bloch}},\ }\bibfield  {title} {\enquote {\bibinfo {title} {Periodically
  driving a many-body localized quantum system},}\ }\href {\doibase
  10.1038/nphys4020} {\bibfield  {journal} {\bibinfo  {journal} {Nat. Phys.}\
  }\textbf {\bibinfo {volume} {13}},\ \bibinfo {pages} {460--464} (\bibinfo
  {year} {2017})}\BibitemShut {NoStop}%
\bibitem [{\citenamefont {Goldman}\ \emph {et~al.}(2015)\citenamefont
  {Goldman}, \citenamefont {Dalibard}, \citenamefont {Aidelsburger},\ and\
  \citenamefont {Cooper}}]{PhysRevA.91.033632}%
  \BibitemOpen
  \bibfield  {author} {\bibinfo {author} {\bibfnamefont {N.}~\bibnamefont
  {Goldman}}, \bibinfo {author} {\bibfnamefont {J.}~\bibnamefont {Dalibard}},
  \bibinfo {author} {\bibfnamefont {M.}~\bibnamefont {Aidelsburger}}, \ and\
  \bibinfo {author} {\bibfnamefont {N.~R.}\ \bibnamefont {Cooper}},\ }\bibfield
   {title} {\enquote {\bibinfo {title} {Periodically driven quantum matter: The
  case of resonant modulations},}\ }\href {\doibase 10.1103/PhysRevA.91.033632}
  {\bibfield  {journal} {\bibinfo  {journal} {Phys. Rev. A}\ }\textbf {\bibinfo
  {volume} {91}},\ \bibinfo {pages} {033632} (\bibinfo {year}
  {2015})}\BibitemShut {NoStop}%
\bibitem [{\citenamefont {Haldar}\ \emph {et~al.}(2018)\citenamefont {Haldar},
  \citenamefont {Moessner},\ and\ \citenamefont {Das}}]{PhysRevB.97.245122}%
  \BibitemOpen
  \bibfield  {author} {\bibinfo {author} {\bibfnamefont {Asmi}\ \bibnamefont
  {Haldar}}, \bibinfo {author} {\bibfnamefont {Roderich}\ \bibnamefont
  {Moessner}}, \ and\ \bibinfo {author} {\bibfnamefont {Arnab}\ \bibnamefont
  {Das}},\ }\bibfield  {title} {\enquote {\bibinfo {title} {Onset of {Floquet}
  thermalization},}\ }\href {\doibase 10.1103/PhysRevB.97.245122} {\bibfield
  {journal} {\bibinfo  {journal} {Phys. Rev. B}\ }\textbf {\bibinfo {volume}
  {97}},\ \bibinfo {pages} {245122} (\bibinfo {year} {2018})}\BibitemShut
  {NoStop}%
\bibitem [{\citenamefont {Shankar}(1994)}]{RevModPhys.66.129}%
  \BibitemOpen
  \bibfield  {author} {\bibinfo {author} {\bibfnamefont {R.}~\bibnamefont
  {Shankar}},\ }\bibfield  {title} {\enquote {\bibinfo {title}
  {Renormalization-group approach to interacting fermions},}\ }\href {\doibase
  10.1103/RevModPhys.66.129} {\bibfield  {journal} {\bibinfo  {journal} {Rev.
  Mod. Phys.}\ }\textbf {\bibinfo {volume} {66}},\ \bibinfo {pages} {129--192}
  (\bibinfo {year} {1994})}\BibitemShut {NoStop}%
\bibitem [{\citenamefont {Wegner}(1994)}]{Wegner1994}%
  \BibitemOpen
  \bibfield  {author} {\bibinfo {author} {\bibfnamefont {Franz}\ \bibnamefont
  {Wegner}},\ }\bibfield  {title} {\enquote {\bibinfo {title} {Flow-equations
  for {Hamiltonians}},}\ }\href {\doibase 10.1002/andp.19945060203} {\bibfield
  {journal} {\bibinfo  {journal} {Ann. der Phys. (Leipzig)}\ }\textbf {\bibinfo
  {volume} {3}},\ \bibinfo {pages} {77} (\bibinfo {year} {1994})}\BibitemShut
  {NoStop}%
\bibitem [{\citenamefont {Kehrein}(2007)}]{Kehrein2007}%
  \BibitemOpen
  \bibfield  {author} {\bibinfo {author} {\bibfnamefont {S.}~\bibnamefont
  {Kehrein}},\ }\href {https://books.google.com/books?id=haH1BwAAQBAJ} {\emph
  {\bibinfo {title} {The Flow Equation Approach to Many-Particle Systems}}},\
  Springer Tracts in Modern Physics\ (\bibinfo  {publisher} {Springer Berlin
  Heidelberg},\ \bibinfo {year} {2007})\BibitemShut {NoStop}%
\bibitem [{\citenamefont {Thomson}\ and\ \citenamefont
  {Schir\'o}(2018)}]{PhysRevB.97.060201}%
  \BibitemOpen
  \bibfield  {author} {\bibinfo {author} {\bibfnamefont {S.~J.}\ \bibnamefont
  {Thomson}}\ and\ \bibinfo {author} {\bibfnamefont {M.}~\bibnamefont
  {Schir\'o}},\ }\bibfield  {title} {\enquote {\bibinfo {title} {Time evolution
  of many-body localized systems with the flow equation approach},}\ }\href
  {\doibase 10.1103/PhysRevB.97.060201} {\bibfield  {journal} {\bibinfo
  {journal} {Phys. Rev. B}\ }\textbf {\bibinfo {volume} {97}},\ \bibinfo
  {pages} {060201} (\bibinfo {year} {2018})}\BibitemShut {NoStop}%
\bibitem [{\citenamefont {Verdeny}\ \emph {et~al.}(2013)\citenamefont
  {Verdeny}, \citenamefont {Mielke},\ and\ \citenamefont
  {Mintert}}]{PhysRevLett.111.175301}%
  \BibitemOpen
  \bibfield  {author} {\bibinfo {author} {\bibfnamefont {Albert}\ \bibnamefont
  {Verdeny}}, \bibinfo {author} {\bibfnamefont {Andreas}\ \bibnamefont
  {Mielke}}, \ and\ \bibinfo {author} {\bibfnamefont {Florian}\ \bibnamefont
  {Mintert}},\ }\bibfield  {title} {\enquote {\bibinfo {title} {Accurate
  effective {Hamiltonians} via unitary flow in {Floquet} space},}\ }\href
  {\doibase 10.1103/PhysRevLett.111.175301} {\bibfield  {journal} {\bibinfo
  {journal} {Phys. Rev. Lett.}\ }\textbf {\bibinfo {volume} {111}},\ \bibinfo
  {pages} {175301} (\bibinfo {year} {2013})}\BibitemShut {NoStop}%
\bibitem [{\citenamefont {Sambe}(1973)}]{PhysRevA.7.2203}%
  \BibitemOpen
  \bibfield  {author} {\bibinfo {author} {\bibfnamefont {Hideo}\ \bibnamefont
  {Sambe}},\ }\bibfield  {title} {\enquote {\bibinfo {title} {Steady states and
  quasienergies of a quantum-mechanical system in an oscillating field},}\
  }\href {\doibase 10.1103/PhysRevA.7.2203} {\bibfield  {journal} {\bibinfo
  {journal} {Phys. Rev. A}\ }\textbf {\bibinfo {volume} {7}},\ \bibinfo {pages}
  {2203--2213} (\bibinfo {year} {1973})}\BibitemShut {NoStop}%
\bibitem [{\citenamefont {Giamarchi}(2004)}]{Giamarchi2004}%
  \BibitemOpen
  \bibfield  {author} {\bibinfo {author} {\bibfnamefont {T.}~\bibnamefont
  {Giamarchi}},\ }\href {http://books.google.com/books?id=GVeuKZLGMZ0C} {\emph
  {\bibinfo {title} {Quantum Physics in One Dimension}}},\ International series
  of monographs on physics\ (\bibinfo  {publisher} {Oxford University Press},\
  \bibinfo {year} {2004})\BibitemShut {NoStop}%
\bibitem [{\citenamefont {Nomura}\ and\ \citenamefont
  {Okamoto}(1993)}]{doi:10.1143/JPSJ.62.1123}%
  \BibitemOpen
  \bibfield  {author} {\bibinfo {author} {\bibfnamefont {Kiyohide}\
  \bibnamefont {Nomura}}\ and\ \bibinfo {author} {\bibfnamefont {Kiyomi}\
  \bibnamefont {Okamoto}},\ }\bibfield  {title} {\enquote {\bibinfo {title}
  {Phase diagram of {S}=1/2 antiferromagnetic {XXZ} chain with
  next-nearest-neighbor interactions},}\ }\href {\doibase 10.1143/JPSJ.62.1123}
  {\bibfield  {journal} {\bibinfo  {journal} {J. Phys. Soc. Jap.}\ }\textbf
  {\bibinfo {volume} {62}},\ \bibinfo {pages} {1123--1126} (\bibinfo {year}
  {1993})}\BibitemShut {NoStop}%
\bibitem [{\citenamefont {Haldane}(1982)}]{PhysRevB.25.4925}%
  \BibitemOpen
  \bibfield  {author} {\bibinfo {author} {\bibfnamefont {F.~D.~M.}\
  \bibnamefont {Haldane}},\ }\bibfield  {title} {\enquote {\bibinfo {title}
  {Spontaneous dimerization in the {$S=\frac{1}{2}$} {Heisenberg}
  antiferromagnetic chain with competing interactions},}\ }\href {\doibase
  10.1103/PhysRevB.25.4925} {\bibfield  {journal} {\bibinfo  {journal} {Phys.
  Rev. B}\ }\textbf {\bibinfo {volume} {25}},\ \bibinfo {pages} {4925--4928}
  (\bibinfo {year} {1982})}\BibitemShut {NoStop}%
\bibitem [{\citenamefont {Tomaras}\ and\ \citenamefont
  {Kehrein}(2011)}]{0295-5075-93-4-47011}%
  \BibitemOpen
  \bibfield  {author} {\bibinfo {author} {\bibfnamefont {C.}~\bibnamefont
  {Tomaras}}\ and\ \bibinfo {author} {\bibfnamefont {S.}~\bibnamefont
  {Kehrein}},\ }\bibfield  {title} {\enquote {\bibinfo {title} {Scaling
  approach for the time-dependent {Kondo} model},}\ }\href
  {http://stacks.iop.org/0295-5075/93/i=4/a=47011} {\bibfield  {journal}
  {\bibinfo  {journal} {EPL (Europhysics Letters)}\ }\textbf {\bibinfo {volume}
  {93}},\ \bibinfo {pages} {47011} (\bibinfo {year} {2011})}\BibitemShut
  {NoStop}%
\bibitem [{\citenamefont {Bukov}\ \emph
  {et~al.}(2015{\natexlab{c}})\citenamefont {Bukov}, \citenamefont
  {D'Alessio},\ and\ \citenamefont
  {Polkovnikov}}]{doi:10.1080/00018732.2015.1055918}%
  \BibitemOpen
  \bibfield  {author} {\bibinfo {author} {\bibfnamefont {Marin}\ \bibnamefont
  {Bukov}}, \bibinfo {author} {\bibfnamefont {Luca}\ \bibnamefont {D'Alessio}},
  \ and\ \bibinfo {author} {\bibfnamefont {Anatoli}\ \bibnamefont
  {Polkovnikov}},\ }\bibfield  {title} {\enquote {\bibinfo {title} {Universal
  high-frequency behavior of periodically driven systems: from dynamical
  stabilization to {Floquet} engineering},}\ }\href {\doibase
  10.1080/00018732.2015.1055918} {\bibfield  {journal} {\bibinfo  {journal}
  {Advances in Physics}\ }\textbf {\bibinfo {volume} {64}},\ \bibinfo {pages}
  {139--226} (\bibinfo {year} {2015}{\natexlab{c}})},\ \Eprint
  {http://arxiv.org/abs/https://doi.org/10.1080/00018732.2015.1055918}
  {https://doi.org/10.1080/00018732.2015.1055918} \BibitemShut {NoStop}%
\bibitem [{\citenamefont {Weinberg}\ and\ \citenamefont
  {Bukov}(2017)}]{SciPostPhys.2.1.003}%
  \BibitemOpen
  \bibfield  {author} {\bibinfo {author} {\bibfnamefont {Phillip}\ \bibnamefont
  {Weinberg}}\ and\ \bibinfo {author} {\bibfnamefont {Marin}\ \bibnamefont
  {Bukov}},\ }\bibfield  {title} {\enquote {\bibinfo {title} {{QuSpin: a Python
  Package for Dynamics and Exact Diagonalisation of Quantum Many Body Systems
  part I: spin chains}},}\ }\href {\doibase 10.21468/SciPostPhys.2.1.003}
  {\bibfield  {journal} {\bibinfo  {journal} {SciPost Phys.}\ }\textbf
  {\bibinfo {volume} {2}},\ \bibinfo {pages} {003} (\bibinfo {year}
  {2017})}\BibitemShut {NoStop}%
\bibitem [{\citenamefont {Sandvik}(2010)}]{Sandvik2010}%
  \BibitemOpen
  \bibfield  {author} {\bibinfo {author} {\bibfnamefont {Anders~W.}\
  \bibnamefont {Sandvik}},\ }\bibfield  {title} {\enquote {\bibinfo {title}
  {Computational studies of quantum spin systems},}\ }\href {\doibase
  http://dx.doi.org/10.1063/1.3518900} {\bibfield  {journal} {\bibinfo
  {journal} {AIP Conference Proceedings}\ }\textbf {\bibinfo {volume} {1297}},\
  \bibinfo {pages} {135--338} (\bibinfo {year} {2010})}\BibitemShut {NoStop}%

\bibitem [{\citenamefont {Zhang}\ \emph
  {et~al.}(2017{\natexlab{b}})\citenamefont {Zhang}, \citenamefont {Pollmann},
  \citenamefont {Sondhi},\ and\ \citenamefont {Moessner}}]{Zhang2017}%
  \BibitemOpen
  \bibfield  {author} {\bibinfo {author} {\bibfnamefont {Carolyn}\ \bibnamefont
  {Zhang}}, \bibinfo {author} {\bibfnamefont {Frank}\ \bibnamefont {Pollmann}},
  \bibinfo {author} {\bibfnamefont {S.~L.}\ \bibnamefont {Sondhi}}, \ and\
  \bibinfo {author} {\bibfnamefont {Roderich}\ \bibnamefont {Moessner}},\
  }\bibfield  {title} {\enquote {\bibinfo {title} {Density-matrix
  renormalization group study of many‐body localization in floquet
  eigenstates},}\ }\href {\doibase 10.1002/andp.201600294} {\bibfield
  {journal} {\bibinfo  {journal} {Ann. Phys. (Leipzig)}\ }\textbf {\bibinfo
  {volume} {529}},\ \bibinfo {pages} {1600294} (\bibinfo {year}
  {2017}{\natexlab{b}})}\BibitemShut {NoStop}%

\bibitem [{\citenamefont {De~Raedt}\ and\ \citenamefont
  {De~Raedt}(1983)}]{PhysRevA.28.3575}%
  \BibitemOpen
  \bibfield  {author} {\bibinfo {author} {\bibfnamefont {Hans}\ \bibnamefont
  {De~Raedt}}\ and\ \bibinfo {author} {\bibfnamefont {Bart}\ \bibnamefont
  {De~Raedt}},\ }\bibfield  {title} {\enquote {\bibinfo {title} {Applications
  of the generalized {Trotter} formula},}\ }\href {\doibase
  10.1103/PhysRevA.28.3575} {\bibfield  {journal} {\bibinfo  {journal} {Phys.
  Rev. A}\ }\textbf {\bibinfo {volume} {28}},\ \bibinfo {pages} {3575--3580}
  (\bibinfo {year} {1983})}\BibitemShut {NoStop}%

\bibitem [{\citenamefont {Polizzi}(2009)}]{PhysRevB.79.115112}%
  \BibitemOpen
  \bibfield  {author} {\bibinfo {author} {\bibfnamefont {Eric}\ \bibnamefont
  {Polizzi}},\ }\bibfield  {title} {\enquote {\bibinfo {title}
  {Density-matrix-based algorithm for solving eigenvalue problems},}\ }\href
  {\doibase 10.1103/PhysRevB.79.115112} {\bibfield  {journal} {\bibinfo
  {journal} {Phys. Rev. B}\ }\textbf {\bibinfo {volume} {79}},\ \bibinfo
  {pages} {115112} (\bibinfo {year} {2009})}\BibitemShut {NoStop}%

\bibitem[{\citenamefont{Birkhoff and Rota}(1989)}]{birkhoff1989ordinary}
\bibinfo{author}{\bibfnamefont{G.}~\bibnamefont{Birkhoff}} \bibnamefont{and}
\bibinfo{author}{\bibfnamefont{G.}~\bibnamefont{Rota}},
\enquote{\bibinfo{title}{Ordinary Differential Equations}},
\href{https://books.google.com/books?id=YBjEQgAACAAJ}{(\bibinfo{publisher}{Wiley}, \bibinfo{year}{1989}), ISBN
	\bibinfo{isbn}{9780471860037}}.

\bibitem[{\citenamefont{Rodriguez-Vega
		et~al.}(2018)\citenamefont{Rodriguez-Vega, Lentz, and
		Seradjeh}}]{1367-2630-20-9-093022}
\bibinfo{author}{\bibfnamefont{M.}~\bibnamefont{Rodriguez-Vega}},
\bibinfo{author}{\bibfnamefont{M.}~\bibnamefont{Lentz}}, \bibnamefont{and}
\bibinfo{author}{\bibfnamefont{B.}~\bibnamefont{Seradjeh}},
\enquote{\bibinfo{title}{Floquet perturbation theory: formalism and application to low-frequency limit}},
\href{http://stacks.iop.org/1367-2630/20/i=9/a=093022}{\bibinfo{journal}{New Journal of Physics} \textbf{\bibinfo{volume}{20}}, \bibinfo{pages}{093022} (\bibinfo{year}{2018})}.




\bibitem[{\citenamefont{{Vogl} et~al.}(2019)\citenamefont{{Vogl}, {Laurell},
		{Barr}, and {Fiete}}}]{2019arXiv190207237V}
\bibinfo{author}{\bibfnamefont{M.}~\bibnamefont{{Vogl}}},
\bibinfo{author}{\bibfnamefont{P.}~\bibnamefont{{Laurell}}},
\bibinfo{author}{\bibfnamefont{A.~D.} \bibnamefont{{Barr}}},
\bibnamefont{and} \bibinfo{author}{\bibfnamefont{G.~A.}
	\bibnamefont{{Fiete}}},
\enquote{\bibinfo{title}{Analogue of Hamilton-Jacobi theory for the time-evolution operator }}, \href{https://arxiv.org/abs/1902.07237}{\bibinfo{journal}{arXiv:} \bibinfo{eprint}{1902.07237}
	(\bibinfo{year}{2019})}.

\end{thebibliography}

\newpage

%%%%%%%%%%%%%
%Appendix %
%%%%%%%%%%%%%
\appendix

\section{Spin in a rotating magnetic field}
\label{app:spin_in_rotfield}
The first terms of the effective Hamiltonian corresponding to Eq.\eqref{Rabi-Ham} in a Magnus expansion are given as,
\begin{equation}
	H_\mathrm{eff}\approx \left(
	\begin{array}{cc}
	B_z-\frac{B_p^2}{\omega } & -\frac{2 B_p B_z}{\omega } \\
	-\frac{2 B_p B_z}{\omega } & \frac{B_p^2}{\omega }-B_z \\
	\end{array}
	\right).
\end{equation}
Next, we calculate the Hamiltonian in the rotating frame using Eq.\eqref{approx_floweq}.
We make the ansatz,
 \begin{equation}
H(s,t)=\begin{pmatrix}
B_2(s)&B_0(s)-iB_1(s)\\
B_0(s)+iB_1(s)&-B_2(s)
\end{pmatrix},
\end{equation}
and find the flow equations
\begin{equation}
\begin{aligned}
\frac{d B_2(s)}{d s}&=-\frac{2 B_p (B_0(s) (\cos (\omega t )-1)+B_1(s) \sin (\omega t ))}{\omega }, \\
\frac{d B_0(s)}{d s}&=\frac{2 B_p B_2(s) (\cos (\omega t )-1)}{\omega }-B_p \cos (\omega t ),\\
\frac{d B_1(s)}{d s}&=\frac{2 B_p B_2(s) \sin (\omega t )}{\omega }-B_p \sin (\omega t ),
\end{aligned}
\end{equation} 
with initial conditions,
\begin{equation}
\begin{aligned}
&B_2(0)=B_z;\quad B_0(0)=B_p\cos(\omega t),\\
& B_1(0)=B_p\sin(\omega t).
\end{aligned}
\end{equation}
The solutions to the flow equations at $s=1$ given our boundary conditions at $s=0$ are now given by,
\begin{equation}
\begin{aligned}
B_2=&\frac{1}{4} (4 B_z-\omega ) \cos \left(\frac{4 B_p \sin \left(\frac{t \omega }{2}\right)}{\omega }\right)\\&- B_p \sin \left(\frac{t \omega }{2}\right) \sin \left(\frac{4 B_p \sin \left(\frac{t \omega }{2}\right)}{\omega }\right)+\frac{\omega}{4},\\
B_0=&\frac{1}{4} \sin \left(\frac{t \omega }{2}\right) (\omega -4 B_z) \sin \left(\frac{4 B_p \sin \left(\frac{t \omega }{2}\right)}{\omega }\right),\\
&- B_p \sin \left(\frac{t \omega }{2}\right) \cos \left(\frac{4 B_p \sin^2 \left(\frac{t \omega }{2}\right)}{\omega }\right),\\
B_1=&\frac{1}{2} B_p \sin (t \omega ) \cos \left(\frac{4 B_p \sin \left(\frac{t \omega }{2}\right)}{\omega }\right),\\
&-\frac{1}{4} (\omega -4 B_z) \cos \left(\frac{t \omega }{2}\right) \sin \left(\frac{4 B_p \sin \left(\frac{t \omega }{2}\right)}{\omega }\right).
\end{aligned}
\end{equation}

After taking an average over one period we end up with the effective time independent Hamiltonian
\begin{equation}
\begin{aligned}
&H_\mathrm{eff}=\left(
\begin{array}{cc}
B_2& B_0 \\
B_0 & -B_2 \\
\end{array}
\right)\\
&B_2=\frac{1}{4} \left(\omega +(4 B_z-\omega ) J_0\left(\frac{4 B_p}{\omega }\right)-4 B_p J_1\left(\frac{4 B_p}{\omega }\right)\right)\\
&B_0=B_p J_2\left(\frac{4 B_p}{\omega }\right)-B_z J_1\left(\frac{4 B_p}{\omega }\right).
\end{aligned}
\end{equation}

\section{Effective Hamiltonians}
\label{app:Heff}

For the lowest order Magnus expansion the effective Hamiltonian is,
\begin{equation}
H_\mathrm{eff}=\frac{1}{T}\int_0^t dt\left(H(t)+i\int_0^tdt_1\comm{H(t_1)}{H(t)}\right).
\end{equation} 
This will be a reference point for our flow equation approach.

\subsection{Flow equation approach for the XY-spin chain with anti-symmetric exchange}
We find that at each flow-step of Eq.\eqref{approx_floweq} the Hamiltonian $H(s)$ retains the form
\begin{equation}
\begin{aligned}
H&(s,t)=\sum_i\left(c_1(s,t)S_i^xS_{i+1}^x+c_2(s,t)S_i^yS_{i+1}^y\right.\\
&\left.+c_3(s,t)S_i^xS_{i+1}^y+c_4(s,t)S_i^yS_{i+1}^x+c_5(s,t)S_i^z \right),
\end{aligned}
\end{equation}
where our initial Hamiltonian,  Eq.\eqref{hinit}, tells us that we have the initial conditions
\begin{equation}
\begin{aligned}
c_1(0,t)&=J_x,\\
c_2(0,t)&=J_y,\\
c_3(0,t)&=D,\\
c_4(0,t)&=-D,\\
c_5(0,t)&=h_0+h(t).
\end{aligned}
\end{equation}

Defining $h_I(t):=\int_0^t dt' h(t')$ we can compactly write the flow equations for the coefficients as
\begin{equation}
\begin{aligned}
\frac{dc_1(s,t)}{ds}&=h_I(t)\cdot(c_3(s,t)+c_4(s,t)),\\
\frac{dc_2(s,t)}{ds}&=-h_I(t)\cdot(c_3(s,t)+c_4(s,t)),\\
\frac{dc_3(s,t)}{ds}&=h_I(t)\cdot(c_2(s,t)-c_1(s,t)),\\
\frac{dc_4(s,t)}{ds}&=h_I(t)\cdot(c_2(s,t)-c_1(s,t)),\\
\frac{dc_5(s,t)}{ds}&=-h(t).\\
\end{aligned}
\end{equation}
The solution at $s=1$ is found as
\begin{equation}
\begin{aligned}
c_1(1,t)&=J+\frac{\Delta J}{2}\cos(2h_I(t)),\\
c_2(1,t)&=J-\frac{\Delta J}{2}\cos(2h_I(t)),\\
c_3(1,t)&=D-\frac{\Delta J}{2}\sin(2h_I(t)),\\
c_4(1,t)&=-D-\frac{\Delta J}{2}\sin(2h_I(t)),\\
c_5(1,t)&=h_0.
\end{aligned}
\end{equation}

Taking the explicit form $h(t)=h\sin(\omega t)$, we can take a time average over a period of the Hamiltonian and find the approximate Hamiltonian at stroboscopic times as
\begin{equation}
\begin{aligned}
	H_\mathrm{eff}&=\sum_i(J_x^{(f)}S_i^xS_{i+1}^x+J_y^{(f)} S_i^yS_{i+1}^y+D_+^{(f)}S_i^xS_{i+1}^y\\
	&+D_-^{(f)}S_i^yS_{i+1}^x+h_0S_i^z),\\
	J_x^{(f)}&:=J+\frac{\Delta J}{2}  \cos \left(\frac{2 h}{\omega }\right) J_0\left(\frac{2 h}{\omega }\right),\\
	J_y^{(f)}&:=J-\frac{\Delta J}{2}  \cos \left(\frac{2 h}{\omega }\right) J_0\left(\frac{2 h}{\omega }\right),\\
	D_{\pm}^{(f)}&:=\pm D-\frac{\Delta J}{2}\sin \left(\frac{2 h}{\omega }\right) J_0\left(\frac{2 h}{\omega }\right).
\end{aligned}
\end{equation}
\label{appXYflow}

\subsection{Flow equation approach for the \texorpdfstring{$J_1-J_2$}{J1-J2} model with time dependent magnetic field in \texorpdfstring{$x$}{x}-direction}
We find that at each flow-step [using Eq.\eqref{approx_floweq}] the Hamiltonian has the form,
\begin{equation}
\begin{aligned}
H(s,t)=&\sum_n\sum_i\left(C_{xx}^n(s,t)S_i^xS_{i+n}^x+C_{yy}^n(s,t)S_i^yS_{i+n}^y\right.\\
&\left.+C_{zz}^n(s,t)S_i^zS_{i+n}^z+C_{yz}^n(s,t)(S_i^zS_{i+n}^y+S_i^yS_{i+n}^z)\right)\\
&+\sum_i\tilde h(s,t)S_i^z,
\end{aligned}
\end{equation}
where our initial Hamiltonian,  Eq.\eqref{hinit2}, gives us the boundary conditions
\begin{equation}
\begin{aligned}
C_{xx}^n(0,t)=J_n,\\
C_{yy}^n(0,t)=J_n,\\
C_{zz}^n(0,t)=\Delta J_n,\\
C_{yz}^n(0,t)=0,\\
\tilde h(0,t)=h(t).
\end{aligned}
\end{equation}
Defining $h_I(t):=\int_0^t dt' h(t')$ we find that an infinitesimal step implies the flow equations
\begin{equation}
\begin{aligned}
\frac{dC_{xx}^n(s,t)}{ds}&=0,\\
\frac{dC_{yy}^n(s,t)}{ds}&=2h_I(t)\cdot C_{yz}^n(s,t),\\
\frac{dC_{zz}^n(s,t)}{ds}&=-2h_I(t)\cdot C_{yz}^n(s,t),\\
\frac{dC_{yz}^n(s,t)}{ds}&=-h_I(t)\cdot (C_{yy}^n(s,t) - C_{zz}^n(s,t)), \\
\frac{d\tilde h(s,t)}{ds}&=-h(t).
\end{aligned}
\end{equation}
The solution at $s=1$ is found as
\begin{equation}
\begin{aligned}
C_{xx}^n(1,t)&=J_n,\\
C_{yy}^n(1,t)&=\frac{J_n}{2}\left(2+2J_n^z-1)\sin^2(h_I(t))\right),\\
C_{zz}^n(1,t)&=\frac{J_n}{2}\left(2+2(J_n^z-1)\cos^2(h_I(t))\right),\\
C_{yz}^n(1,t)&=\frac{J_n}{2}(J_n^z-1)\sin(2h_I(t)),\\
\tilde h(1,t)&=0
\end{aligned}
\end{equation}
For the special case of 
\begin{equation}
h(t)=B\begin{cases}
1;\quad  2n\pi<\omega t<2n\pi+\pi\\
-1;\quad  2n\pi+\pi< \omega t<2(n+1)\pi
\end{cases};\quad n\in\mathbb{Z}
\end{equation}
we find
\begin{equation}
\begin{aligned}
H_\mathrm{eff}&=\sum_{n=1}^2\sum_i(J_nS_i^xS_{i+1}^x+J_n^y S_i^yS_{i+n}^y+J_n^zS_i^zS_{i+n}^z\\
&+\Gamma_n(S_i^zS_{i+n}^y+S_i^yS_{i+n}^z) ),\\
J_n^y&:=\frac{J_n }{4} \left(2 J_n^z -\frac{(J_n^z -1) \omega  \sin \left(\frac{2 \pi  B}{\omega }\right)}{\pi  B}+2\right),\\
J_n^z&:=\frac{J_n \left((J_n^z -1) \omega  \sin \left(\frac{2 \pi  B}{\omega }\right)+2 \pi  (J_n^z +1) B\right)}{4 \pi  B},\\
\Gamma_n&:=\frac{(J_n^z -1) J_n \omega  \sin ^2\left(\frac{\pi  B}{\omega }\right)}{2 \pi  B}.
\end{aligned}
\end{equation}
\label{appJ1J2flowmagfield}

\section{Time evolution operator for the \texorpdfstring{$XY$}{XY} model with time-dependent magnetic field in \texorpdfstring{$z$}{z}-direction}
The equation for the time evolution operator in this case has the form
\begin{equation}
i\partial_t U(t)=\sum_kH_kU(t),
\label{timeev_eq}
\end{equation}
with $\comm{H_k}{H_{k^\prime}}=0$, $\forall k\neq k^\prime$.  One may make a separation of variables ansatz
\begin{equation}
U=\prod_k U_k,
\end{equation}
where we assume that $\comm{U_k}{U_{k^\prime}}=0$, $\forall k\neq k^\prime$.  Inserting this ansatz into Eq.\eqref{timeev_eq} gives
\begin{equation}
i\sum_k(\partial_t U_k)\left[\prod_{k^\prime\neq k}U_{k^\prime}\right]=\sum_kH_kU_k\left[\prod_{k^\prime\neq k}U_{k^\prime}\right].
\end{equation}
It is a sufficient condition for this equation to be fulfilled that it is satisfied term by term. This yields
\begin{equation}
i(\partial_t U_k)=H_k(t)U_k.
\label{reduced_time_ev_eq}
\end{equation}
In our case generically $H_k(t)$ has the form
\begin{equation}
H_k(t)=(c_k^\dag,c_{-k})\begin{pmatrix}
H^k_{11}(t)&H^k_{12}(t)\\
H^k_{21}(t)&H^k_{22}(t)
\end{pmatrix}\begin{pmatrix}
c_k\\c_{-k}^\dag
\end{pmatrix}.
\end{equation}
An ansatz for $U_k$ therefore is
\begin{equation}
\begin{aligned}
U_k(t)=&A^k_0(t)+A^k_1(t)c_k^\dag c_k+A^k_2(t)c_{-k}^\dag c_{-k}+A^k_3(t)c_{-k}c_k\\
&+A^k_4(t)c_{-k}^\dag c_k^\dag+A^k_5(t)c_{-k}^\dag c_k^\dag c_{-k}c_k,
\end{aligned}
\end{equation}
which indeed fulfills $\comm{U_k}{U_{k^\prime}}=0$, $\forall k\neq k^\prime$. Inserting the ansatz in Eq.\eqref{reduced_time_ev_eq} we find that it is consistent since no further terms appear. By equating coefficients we find
	\begin{widetext}
	\begin{equation}
	\frac{\partial \vect A^k(t)}{\partial t}=i \left(
	\begin{array}{cccccc}
	-H^k_{22}(t) & 0 & 0 & 0 & H^k_{21}(t) & 0 \\
	-H^k_{11}(t) & -H^k_{11}(t)-H^k_{22}(t) & 0 & 0 & -H^k_{21}(t) & 0 \\
	H^k_{22}(t) & 0 & 0 & 0 & -H^k_{21}(t) & 0 \\
	-H^k_{21}(t) & -H^k_{21}(t) & -H^k_{21}(t) & -H^k_{22}(t) & 0 & H^k_{21}(t) \\
	H^k_{12}(t) & 0 & 0 & 0 & -H^k_{11}(t) & 0 \\
	0 & -H^k_{22}(t) & H^k_{11}(t) & H^k_{12}(t) & -H^k_{21}(t) & -H^k_{11}(t) \\
	\end{array}
	\right)\vect A^k(t),
	\end{equation}
	\end{widetext}
	as a linear system of equations for the coefficients $A_i^k$ from the ansatz, where we defined
	\begin{equation}
		\vect A^k(t)=(A_1^k(t),A_2^k(t),A_3^k(t),A_4^k(t),A_5^k(t))^T.
	\end{equation}
The initial conditions for a time evolution operator in this notation are
	\begin{equation}
	\vect A^k(0)=(1,0,0,0,0)^T,
	\end{equation}
	where $T$ denotes the transpose here (not the period of the time-dependent Hamiltonian). From here the time evolution operator was evaluated numerically.
\label{time-ev-XY}

\section{Numerical calculation of time evolution operators using exact diagonalization}\label{app:ED}
To study the $J_1$-$J_2$ models in Eqs. \eqref{hinit2}, \eqref{hinit3} and their approximate counterparts in Eqs. \eqref{heff2}, \eqref{heff3} we employ exact diagonalization \cite{Sandvik2010}. We note that the time evolution of a given initial state is more efficiently calculated using Krylov subspace methods, as in Ref. \cite{Machado2017}, and that DMRG-based methods are more powerful in the Floquet-MBL regime \cite{Zhang2017}, being capable of reaching larger system sizes. Here we want, however, to compare the full time evolution operators using the most unbiased numerical method possible. To calculate this operator for a given (exact or approximate) Hamiltonian, we write
\begin{align}
	H\left( t\right)	&=	H_0 + V\left( t\right),
\end{align}
where $V(t)\equiv 0$ for the time-independent effective Hamiltonians. We next ``Trotterize'' the problem by introducting discrete time steps $t_j=jT/N_\mathrm{steps} \equiv j\delta t$, where $\delta t$ is chosen small enough not to affect the results. Here $T$ is the period of the time-dependent Hamiltonian. Then the time evolution operator over a period is given by
\begin{align}
	{U}(T,0)	&=	\prod_{j=0}^{N_\mathrm{steps}-1} {U}\left( t_{j+1}, t_j \right),
\end{align}
where, using a second-order Trotter-Suzuki decomposition \cite{PhysRevA.28.3575}, we write
\begin{align}
	  {U} \left( t+\delta t, t\right)	&=	\exp \left[ -\frac{i\delta t}{2\hbar} V \left( t+ \frac{\delta t}{2} \right) \right]	\exp \left[ -i\delta t H_0/\hbar \right]	\nonumber\\
					&\times \exp \left[ -\frac{i\delta t}{2\hbar} V \left( t+ \frac{\delta t}{2} \right) \right].
\end{align}
If the time-dependence of $V(t)$ factors out, i.e. $V(t)=f(t)V_0$, the problem of calculating $U(T,0)$ can be reduced to two matrix diagonalizations, of $H_0$ and $V_0$, respectively, and numerically efficient matrix-matrix multiplications. If one of the matrices is integrable the problem can be simplified further, as in Ref. \cite{DAlessio2015}, but we do not assume that here. We write $H_0$ and $V(t)$ in a basis implementing translational invariance, and, for Eqs. \eqref{hinit3}, \eqref{heff3} also magnetization conservation \cite{Sandvik2010}. The full diagonalization of the two matrices is achieved using the FEAST Eigenvalue Solver \cite{PhysRevB.79.115112}.

\section{Transverse Ising model: BCH, Flow and replica Hamiltonians}
In this section we summarize the treatment of the transverse Ising model.
\label{appBCHresum}
\subsection{Flow equations for the delta function model}
The flow equations for the $\delta$-function model,  Eq.\eqref{transverseIsingdelta}, in our approximation,  Eq.\eqref{approx_floweq}, are found as
\begin{equation}
\hspace*{-0.1cm}
	\begin{aligned}
&\frac{dC_x^{F,\delta}(s,t)}{ds}=- h_x (\delta (t)-1),\\
&\frac{dC_z^{F,\delta}(s,t)}{ds}=- h_z (\delta (t)-1),\\
&\frac{dC_{xx}^{F,\delta}(s,t)}{ds}=-4 h_z (t-1) C_{xy}^{F,\delta}(s,t),\\
&\frac{dC_{xy}^{F,\delta}(s,t)}{ds}=2 (t-1) [h_z (C_{xx}^{F,\delta}(s,t)-C_{yy}^{F,\delta}(s,t))-h_x C_{xz}^{F,\delta}(s,t)],\\
&\frac{dC_{yy}^{F,\delta}(s,t)}{ds}= 4(t-1) [h_z C_{xy}^{F,\delta}(s,t)-h_x C_{yz}^{F,\delta}(s,t)],\\
&\frac{dC_{xz}^{F,\delta}(s,t)}{ds}=2 (t-1) [h_x C_{xy}^{F,\delta}(s,t)-h_z C_{yz}^{F,\delta}(s,t)],\\
&\frac{dC_{yz}^{F,\delta}(s,t)}{ds}=2(t-1) [h_z C_{xz}^{F,\delta}(s,t)+h_x (C_{yy}^{F,\delta}(s,t)-C_{zz}^{F,\delta}(s,t))],\\
&\frac{dC_{zz}^{F,\delta}(s,t)}{ds}= 4h_x (t-1) C_{yz}^{F,\delta}(s,t),
	\end{aligned}
\end{equation}
with initial conditions
\begin{equation}
	\begin{aligned}
	&C_x^{F,\delta}(0,t)=h_x \delta (t),\\
	&C_z^{F,\delta}(0,t)=h_z \delta (t),\\
	&C^{F,\delta}_{zz}(0,t)=J_z,\\
	&C^{F,\delta}_{xx}(0,t)=C^{F,\delta}_{xy}(0,t)=0,\\
	&C^{F,\delta}_{yy}(0,t)=C^{F,\delta}_{xz}(0,t)=C^{F,\delta}_{yz}(0,t)=0.
	\end{aligned}
\end{equation}
The solution at $s=1$ gives coefficients
\begin{equation}
\hspace*{-0.1cm}
\begin{aligned}
&C_{x/z}^{F,\delta}=h_{x/z},\\
&C_{xx}^F=\frac{h_x^2 h_z^2 J_z (12 h-8 \sin (2 h)+\sin (4 h))}{8 h^5},\\
&C_{xy}^{F,\delta}=\frac{h_x^2 h_z J_z \sin ^4(h)}{h^4},\\
&C_{xz}^{F,\delta}=\frac{h_x h_z J_z \left( h \left(2h_z^2- h_x^2\right)+ \sin (2 h) \left(h_x^2-h_z^2\right)-\frac{h_x^2\sin (4 h)}{4}\right)}{2 h^5},\\
&C_{zz}^{F,\delta}=\frac{J_z \left(4 h \left(h_x^4+2 h_z^4\right)+h_x^4 \sin (4 h)+8 h_x^2 h_z^2 \sin (2 h)\right)}{8 h^5},\\
&C_{yz}^{F,\delta}=\frac{h_x J_z \sin ^2(h) \left(h_x^2 \cos (2 h)+h_x^2+2 h_z^2\right)}{2 h^4},\\
&C_{yy}^{F,\delta}=-\frac{h_x^2 J_z (\sin (4 h)-4 h)}{8 h^3},\\
&C_{xzz}^{F,\delta}=0.
\end{aligned}
\end{equation}

\subsection{Flow equations for the Heaviside \texorpdfstring{$\theta$}{theta}-function model}
The flow equations for the Heaviside $\theta$-function model,  Eq.\eqref{transverseIsingsign}, in our approximation,  Eq.\eqref{approx_floweq}, are found to generate an infinite amount of terms. This means an exact solution of \eqref{approx_floweq} is impossible in turn this also means that a rotating frame approximation is impossible because matrix exponentials and also the rotation of operators induced by it cannot be calculated. Our method allows to truncate terms and therefore find an approximate rotating frame transformation. The terms that appear in Eq.\eqref{transverseIsingeffham} are generated quickly when using an ansatz that starts with the form of the original Hamiltonian, and subsequently adding the new terms that appear to that ansatz. This motivates one to include as many terms from the Hamiltonian Eq.\eqref{transverseIsingeffham} as possible while still allowing for a compact analytical result. We choose the ansatz Hamiltonian,
\begin{equation}
\begin{aligned}
	H_{\mathrm{Ansatz}}^{F,\theta}(s)=&\sum_i\Bigg[ C_x^{F,\theta}\sigma_i^x+C_z^{F,\theta}\sigma_i^z+C_{yy}^{F,\theta}\sigma_i^y\sigma_{i+1}^y\\
	&+C_{zz}^{F,\theta}\sigma_i^z\sigma_{i+1}^z+C_{xz}^{F,\theta}(\sigma_i^x\sigma_{i-1}^z+\sigma_i^x\sigma_{i+1}^z)\\
	&+C_{yz}^{F,\theta}(\sigma_i^y\sigma_{i-1}^z+\sigma_i^y\sigma_{i+1}^z)\\
	&+C_{xzz}^{F,\theta}\sigma_{i}^x\sigma_{i-1}^z\sigma_{i+1}^z\Bigg].
	\end{aligned}
\end{equation}
The flow equations,  Eq.\eqref{approx_floweq}, give us the following equations for the coefficients
\begin{equation}
		\begin{aligned}
	&\frac{dC_x^{F,\theta}(s,t)}{ds}=- (4J_z C_{yz}^{F,\theta}(s,t) f_I(t)+h_x f(t),)\\
	&\frac{dC_z^{F,\theta}(s,t)}{ds}=- h_z f(t),\\
	&\frac{dC_{yy}^{F,\theta}(s,t)}{ds}= 4h_x C_{yz}^{F,\theta}(s,t) f_I(t),\\
	&\frac{dC_{xz}^{F,\theta}(s,t)}{ds}=2 h_z C_{yz}^{F,\theta}(s,t) f_I(t),\\
	&\frac{dC_{yz}^{F,\theta}(s,t)}{ds}=2f_I(t) ( J_z C_{x}^{F,\theta}(s,t)- h_z C_{xz}^{F,\theta}(s,t),\\
	&\hspace*{2cm}+J_z C_{xzz}^{F,\theta}(s,t)- h_x \text{Cyy}(s,t)+ h_x \text{Czz}(s,t)),\\
	&\frac{dC_{zz}^{F,\theta}(s,t)}{ds}= (J_z f(t)-4h_x C_{yz}^{F,\theta}(s,t) f_I(t)),\\
	&\frac{dC_{xzz}^{F,\theta}(s,t)}{ds}=- 4J_z C_{yz}^{F,\theta}(s,t) f_I(t),
	\end{aligned}
\end{equation}
where $f(t)=\theta\left(t-\frac{1}{2}\right)$, $f_I(t)=t+(1-2 t) \theta \left(t-\frac{1}{2}\right)$ and $\theta$ the Heaviside function.  

The initial conditions are
\begin{equation}
\begin{aligned}
&C_x^{F,\theta}(0,t)= h_x (f(t)+1),\\
&C_z^{F,\theta}(0,t)=h_z (f(t)+1),\\
&C^{F,\theta}_{zz}(0,t)=- J_z (f(t)-1),\\
&C^{F,\theta}_{xx}(0,t)=C^{F,\theta}_{xy}(0)=0,\\
&C^{F,\mathrm{sgn}}_{yy}(0,t)=C^{F,\mathrm{sgn}}_{xz}(0,t)=C^{F,\mathrm{sgn}}_{yz}(0,t)=0.
\end{aligned}
\end{equation}

The solution at $s=1$ implies that the coefficients are
\begin{equation}
%\hspace*{-0.5cm}
\begin{aligned}
&C_{x}^{F,\mathrm{sgn}}=h_x-\frac{ 4 h_x J_z^2}{\gamma ^2}\left(1-\frac{\sin (\gamma )}{\gamma }\right),\\
&C_{z}^{F,\mathrm{sgn}}=h_z,\\
&C_{xx}^F=C_{xy}^{F,\mathrm{sgn}}=0,\\
&C_{xz}^{F,\mathrm{sgn}}=\frac{ 2 h_x h_z J_z}{\gamma ^2}\left(1-\frac{\sin (\gamma )}{\gamma }\right),\\
&C_{zz}^{F,\mathrm{sgn}}=J_z-\frac{ 4 h_x^2 J_z}{\gamma ^2}\left(1-\frac{\sin (\gamma )}{\gamma }\right),\\
&C_{yz}^{F,\mathrm{sgn}}=\frac{2 h_x J_z}{\gamma ^2}(1-\cos (\gamma )),\\
&C_{yy}^{F,\mathrm{sgn}}=\frac{ 4 h_x^2 J_z}{\gamma ^2}\left(1-\frac{\sin (\gamma )}{\gamma }\right),\\
&C_{xzz}^{F,\mathrm{sgn}}=\frac{ 8 h_x J_z^2}{\gamma ^2}\left(\frac{\sin (\gamma )}{\gamma }-1\right),
\end{aligned}
\end{equation}
where $\gamma=\sqrt{4h_x^2+h_z^2+4J_z^2}$.

\subsection{Result for the BCH idenity}
For the BCH idenity one finds coefficients
\begin{equation}
\begin{aligned}
&C_x^\mathrm{BCH}=h_x,\\
&C_z^\mathrm{BCH}=h_z,\\
&C_{zz}^\mathrm{BCH}=J_z,\\
&C_{yz}^\mathrm{BCH}=h_xJ_z,\\
&C_{xx}^\mathrm{BCH}=C_{xy}^\mathrm{BCH}=C_{xz}^\mathrm{BCH}=C_{yy}^\mathrm{BCH}=C_{xzz}^\mathrm{BCH}=0.
\end{aligned}
\end{equation}

\subsection{Result for the replica approximation}
The coefficients for the replica case were taken from Ref.[\onlinecite{PhysRevLett.120.200607}] as
\begin{equation}
\begin{aligned}
&C_x^R=h_x\left(J_z\cot (2J_z)+\frac{1}{2}\right),\\
&C_z^R=h_z,\\
&C_{zz}^R=J_z,\\
&C_{yz}^R=\frac{1}{2}h_xJ_z,\\
&C_{xzz}^R=h_x\left(J_z\cot (2J_z)-\frac{1}{2}\right),\\
&C_{xx}^R=C_{xy}^R=C_{xz}^R=C_{yy}^R=0.
\end{aligned}.
\end{equation}
\end{document}